\documentstyle[12pt,aps,floats,epsf]{revtex}

\catcode`\@=11
\long\def\@makefntext#1{ 
\protect\noindent \hbox to 3.2pt {\hskip-.9pt
$^{{\ninerm\@thefnmark}}$\hfil}#1\hfill} 

\def\thefootnote{\fnsymbol{footnote}}
 \def\@makefnmark{\hbox to 0pt{$^{\@thefnmark}$\hss}}  

\def\ps@myheadings{\let\@mkboth\@gobbletwo
\def\@oddhead{\hbox{} %
\rightmark\hfil\ninerm\thepage}
\def\@oddfoot{}\def\@evenhead{\ninerm\thepage\hfil %
\leftmark\hbox{}}\def\@evenfoot{}
\def\sectionmark##1{}\def\subsectionmark##1{}}

\begin{document}

\newcommand{\symbolfootnote}{\renewcommand{\thefootnote}
    {\fnsymbol{footnote}}}
\renewcommand{\thefootnote}{\fnsymbol{footnote}}
\newcommand{\alphfootnote}
    {\setcounter{footnote}{0}
     \renewcommand{\thefootnote}{\tenrm\alph{footnote}}}

\newcommand{\centeron}[2]{{%
   \setbox0=\hbox{#1}\setbox1=\hbox{#2}%
   \ifdim\wd1>\wd0\kern.5\wd1\kern-.5\wd0\fi
   \copy0\kern-.5\wd0\kern-.5\wd1\copy1%
   \ifdim\wd0>\wd1\kern.5\wd0\kern-.5\wd1\fi
}}
\newcommand{\st}[1]{\centeron{$#1$}{$/$}}

\newcommand {\Asy}{\mathop{{\rm Asy}}}
\newcommand
   {\ltap}{\;\centeron{\raise.35ex\hbox{$<$}}{\lower.65ex\hbox{$\sim$}}\;}
\newcommand
   {\gtap}{\;\centeron{\raise.35ex\hbox{$>$}}{\lower.65ex\hbox{$\sim$}}\;}
\newcommand \gsim{\mathrel{\gtap}}
\newcommand \lsim{\mathrel{\ltap}}

\newcommand{\LB}{\hbox{\bf($\!$(}}
\newcommand{\RB}{\hbox{\bf)$\!$)}}
\newcommand{\LBB}{\hbox{\bf\Big($\!\!$\Big(}}
\newcommand{\RBB}{\hbox{\bf\Big)$\!\!$\Big)}}

\newcommand{\iep}{\mbox{$i \epsilon $}}
\newcommand{\lvac}{\mbox{$\langle 0|$}}   
\newcommand{\rvac}{\mbox{$|0\rangle $}}   
\newcommand{\leftx}{\mbox{$\langle x|$}}   
\newcommand{\righx}{\mbox{$|x\rangle $}}   
\newcommand{\leftz}{\mbox{$\langle z|$}}   
\newcommand{\righz}{\mbox{$|z\rangle $}}   
\newcommand{\lpatp}{\mbox{$\langle p|$}}   
\newcommand{\rpatp}{\mbox{$|p\rangle $}}   
\newcommand{\lplus}{\mbox{$\langle p+\lambda e_{1}|$}}   
\newcommand{\rplus}{\mbox{$|p+\lambda e_{1}\rangle $}}   
\newcommand{\lN}{\mbox{$\langle N|$}}   
\newcommand{\rN}{\mbox{$|N\rangle $}}   
\newcommand{\ie}{\mbox{$i \epsilon $}}
\newcommand{\ieo}{\mbox{$i \epsilon x_{0}$}}
\newcommand{\xsl}{\mbox{$ \st x$}}
\newcommand{\xx}{\mbox{$x^{2}$}}
\newcommand{\veck}{\vec{k}}
\newcommand{\np}{\st p}
\newcommand{\nP}{\st{\cal P}}
\newcommand{\nk}{\st k}
\newcommand{\nq}{\st q}
\newcommand{\nx}{\st x}
\newcommand{\ny}{\st y}
\newcommand{\nz}{\st z}
\newcommand{\lan}{\langle }
\newcommand{\ran}{\rangle }
\newcommand{\rran}{\rangle \!\rangle }
\newcommand{\llan}{\langle \!\langle }
\def \p{_{\perp }}
\def \bu{_{\bullet}}
\def \bubu{_{\bullet\bullet}}
\def \ci{_{\circ }}
\def \cici{_{\circ \circ }}
\def \sr{_{\ast}}
\def \di{_{\diamond}}
\def \cip{_{\circ \perp }}
\def \csr{_{\circ \star}}
\def \srp{_{\ast\perp }}
\def \sbu{_{\ast\bullet}}
\def \sbus{_{\ast\bullet\sigma }}
\def \sci{_{\ast\circ }}
\def \O{^{\Omega }}
\def \da{^{\dagger }}
\def \ze{^{\zeta }}
\def \zo{^{\zeta _{0}}}
\def \an{^{(0)}_{A}}
\newcommand{\iy}{\infty }
\newcommand{\Repa}{\mathop{\rm Re}}         
\newcommand{\Impa}{\mathop{\rm Im}}         
\newcommand{\ch}{\mathop{\rm ch}}                
\newcommand{\sh}{\mathop{\rm sh}}                
\newcommand{\PV}{\mathop{\rm PV}}                
\newcommand{\Tr}{\mathop{\rm Tr}}                
\newcommand{\deter}{\mathop{\rm det}}                
\newcommand{\al}{\alpha }
\newcommand{\be}{\beta }
\newcommand{\ga}{\gamma }
\newcommand{\la}{\lambda }
\newcommand{\om}{\omega }
\newcommand{\eps}{\epsilon }
\newcommand{\rg}{renormalization group}
\newcommand{\Uz}{{U^{\zeta }}}
\newcommand{\Udz}{U^{\dagger \zeta }}
\newcommand{\Uo}{{U^{\sigma _{1}}}}
\newcommand{\Udo}{U^{\dagger \sigma _{1}}}
\newcommand{\UdO}{U^{\dagger \Omega }}
\newcommand{\Ut}{U^{\zeta _{0}}}
\newcommand{\Udt}{U^{\dagger \zeta _{0}}}
\newcommand{\Uhz}{{\hat{U}^{\zeta }}}
\newcommand{\Uhdz}{\hat{U}^{\dagger \zeta }}
\newcommand{\Uho}{{\hat{U}^{\sigma _{1}}}}
\newcommand{\Uhdo}{\hat{U}^{\dagger \sigma _{1}}}
\newcommand{\Uht}{\hat{U}^{\sigma _{2}}}
\newcommand{\Uhdt}{\hat{U}^{\dagger \sigma _{2}}}
\newcommand{\Us}{{U^{\sigma }}}
\newcommand{\Uds}{U^{\dagger \sigma }}
\newcommand{\Uhd}{\hat{U}^{\dagger }}
\newcommand{\Ud}{U^{\dagger }}
\newcommand{\Uh}{\hat{U}}
\newcommand{\Vd}{V^{\dagger }}
\newcommand{\as}{\frac {\alpha _{s}}{2\pi ^{2}}}
\newcommand{\si}{\sigma }
\newcommand{\ms}{\frac {m^{2}}{s}}
\newcommand{\pas}{\frac {p_{A}^{2}}{s}}
\newcommand{\pbs}{\frac {p_{B}^{2}}{s}}
\newcommand{\qcd}{quantum chromodynamics}   
\newcommand{\eea}{$e^{+}e^{-}$ annihilation}
\newcommand{\ipp}{inclusive particle production}
\newcommand{\dis}{deep inelastic scattering}
\newcommand{\ff}{fragmentation function}
\newcommand{\stf}{structure function}
\newcommand{\so}{string operator}
\newcommand{\ope}{operator product expansion}
\newcommand{\lico}{light-cone}
\newcommand{\mael}{matrix element}
\newcommand{\crsc}{cross section}
\newcommand{\cofu}{coefficient function}
\newcommand{\half}{\frac {1}{2}}
\newcommand{\xhalf}{\frac {x}{2}}
\newcommand{\ubar}{\bar{u}}

\def\abstracts#1{{
    \centering{\begin{minipage}{30pc}\tenrm\baselineskip=12pt\noindent
    \centerline{\tenrm ABSTRACT}\vspace{0.3cm}
    \parindent=0pt #1
    \end{minipage} }\par}}

\newcommand{\bibit}{\it}
\newcommand{\bibbf}{\bf}
\renewenvironment{thebibliography}[1]
    {\begin{list}{\arabic{enumi}.}
    {\usecounter{enumi}\setlength{\parsep}{0pt}
\setlength{\leftmargin 1.25cm}{\rightmargin 0pt}
     \setlength{\itemsep}{0pt} \settowidth
    {\labelwidth}{#1.}\sloppy}}{\end{list}}

\topsep=0in\parsep=0in\itemsep=0in
\parindent=1.5pc

\newcounter{itemlistc}
\newcounter{romanlistc}
\newcounter{alphlistc}
\newcounter{arabiclistc}
\newenvironment{itemlist}
        {\setcounter{itemlistc}{0}
     \begin{list}{$\bullet$}
    {\usecounter{itemlistc}
     \setlength{\parsep}{0pt}
     \setlength{\itemsep}{0pt}}}{\end{list}}

\newenvironment{romanlist}
    {\setcounter{romanlistc}{0}
     \begin{list}{$($\roman{romanlistc}$)$}
    {\usecounter{romanlistc}
     \setlength{\parsep}{0pt}
     \setlength{\itemsep}{0pt}}}{\end{list}}

\newenvironment{alphlist}
    {\setcounter{alphlistc}{0}
     \begin{list}{$($\alph{alphlistc}$)$}
    {\usecounter{alphlistc}
     \setlength{\parsep}{0pt}
     \setlength{\itemsep}{0pt}}}{\end{list}}

\newenvironment{arabiclist}
    {\setcounter{arabiclistc}{0}
     \begin{list}{\arabic{arabiclistc}}
    {\usecounter{arabiclistc}
     \setlength{\parsep}{0pt}
     \setlength{\itemsep}{0pt}}}{\end{list}}

\newcommand{\fcaption}[1]{
        \refstepcounter{figure}
        \setbox\@tempboxa = \hbox{\tenrm Fig.~\thefigure. #1}
        \ifdim \wd\@tempboxa > 6in
           {\begin{center}
        \parbox{6in}{\tenrm\baselineskip=12pt Fig.~\thefigure. #1 }
            \end{center}}
        \else
             {\begin{center}
             {\tenrm Fig.~\thefigure. #1}
              \end{center}}
        \fi}

\newcommand{\tcaption}[1]{
        \refstepcounter{table}
        \setbox\@tempboxa = \hbox{\tenrm Table~\thetable. #1}
        \ifdim \wd\@tempboxa > 6in
           {\begin{center}
        \parbox{6in}{\tenrm\baselineskip=12pt Table~\thetable. #1 }
            \end{center}}
        \else
             {\begin{center}
             {\tenrm Table~\thetable. #1}
              \end{center}}
        \fi}

\def\@citex[#1]#2{\if@filesw\immediate\write\@auxout
    {\string\citation{#2}}\fi
\def\@citea{}\@cite{\@for\@citeb:=#2\do
    {\@citea\def\@citea{,}\@ifundefined
    {b@\@citeb}{{\bf ?}\@warning
    {Citation `\@citeb' on page \thepage \space undefined}}
    {\csname b@\@citeb\endcsname}}}{#1}}

\newif\if@cghi
\def\cite{\@cghitrue\@ifnextchar [{\@tempswatrue
    \@citex}{\@tempswafalse\@citex[]}}
\def\citelow{\@cghifalse\@ifnextchar [{\@tempswatrue
    \@citex}{\@tempswafalse\@citex[]}}
\def\@cite#1#2{{$\null^{#1}$\if@tempswa\typeout
    {IJCGA warning: optional citation argument
    ignored: `#2'} \fi}}
\def\@cite#1#2{{%
   \if@cghi
      $\null^{#1}$
    \else
      #1
    \fi
    \if@tempswa\typeout
       {IJCGA warning: optional citation argument
       ignored: `#2'} \fi
}}
\newcommand{\citeup}{\cite}

\def\fnm#1{$^{\mbox{\scriptsize #1}}$}
\def\fnt#1#2{\footnotetext{\kern-.3em
    {$^{\mbox{\sevenrm #1}}$}{#2}}}

\font\twelvebf=cmbx10 scaled\magstep 1
\font\twelverm=cmr10 scaled\magstep 1
\font\twelveit=cmti10 scaled\magstep 1
\font\elevenbfit=cmbxti10 scaled\magstephalf
\font\elevenbf=cmbx10 scaled\magstephalf
\font\elevenrm=cmr10 scaled\magstephalf
\font\elevenit=cmti10 scaled\magstephalf
\font\bfit=cmbxti10
\font\tenbf=cmbx10
\font\tenrm=cmr10
\font\tenit=cmti10
\font\ninebf=cmbx9
\font\ninerm=cmr9
\font\nineit=cmti9
\font\eightbf=cmbx8
\font\eightrm=cmr8
\font\eightit=cmti8

\begin{flushright}
   MIT-CTP-2470\\
   hep-ph/9509348\\
   September1995
\end{flushright}

\centerline{\tenbf OPERATOR EXPANSION FOR HIGH-ENERGY SCATTERING}
\baselineskip=16pt
\vspace{0.4cm}
\centerline{\tenrm I.BALITSKY\footnote{On leave of absence from
St.Petersburg Nuclear
Physics Institute, 188350 Gatchina, Russia}}
\baselineskip=13pt
\centerline{\tenit Center for Theoretical Physics,
Laboratory for Nuclear Science}
\baselineskip=12pt
\centerline{\tenit Department of Physics, MIT, Cambridge 02139}
\vspace{0.4cm}
\abstracts{I demonstrate that the leading logarithms for high-energy
scattering can be
obtained as a result of evolution of the nonlocal operators - straight-line
ordered gauge factors - with respect to the slope of the straight line.}
\vfil
\rm\baselineskip=14pt

\alphfootnote


\section{Introduction }
\label{sec:intro}

The rapid increase of the structure function $F_{2}(x,Q^{2})$ at small $x$
that is observed in DESY at HERA (see e.g. \cite{eksp}) has revived interest
in the problem of the
high-energy behavior of QCD amplitudes.
In the leading logarithmic approximation it is governed by
BFKL equation \cite{fkl}\cite{bl}\cite{l} leading to
a $\sim x^{-0.5}$ behavior
of $F_{2}(x)$ which is not far from the  experimental curve.
Unfortunately, there are theoretical problems with
the BFKL answer which make it difficult, if not impossible,  to
use these leading logarithmic as a description of real high-energy
processes. First and foremost, the BFKL answer violates unitarity
and therefore it is at best some
kind of preasymptotic behavior which can
be reliable only at some intermediate energies. (The true
high-energy asymptotics would correspond to the unitarization
of the leading logarithmic results but this is a problem where
nobody has succeeded in 20 years and not because of lack of
effort.)

Moreover, even at those moderately high energies where
unitarization is not important, the BFKL results
in QCD are not completely rigorous.  Even if we start from the
scattering of hard objects such as heavy quarks, then already in
the leading logarithm approximation we obtain considerable contributions
from the region of small momenta (large distances) where
perturbative QCD is not applicable \cite{bl}\cite{l}. In other words, the hard
pomeron which is believed to describe the observed small-$x$ growth of
structure function $F_{2}$ interacts strongly with
the soft "old" pomeron made from non-perturbative gluons. Therefore,
it is highly desirable to have a  method of separation of
small- and large-distance contributions to high-energy amplitudes,
and the starting point here must be a properly gauge-invariant formalism
for the BFKL equation.

In present paper we suggest a kind of gauge-invariant
operator expansion for high-energy amplitudes. The relevant
operators are gauge factors ordered along (almost)
light-like lines stretching from minus to plus infinity.
These ``Wilson-line'' gauge factors
correspond to very fast quarks moving along the lines (see e.g. \cite{korch}).
It turns out that the small-$x$ behavior of structure functions
is governed by the evolution of these operators with respect
to deviation of the Wilson lines from the light cone; this
deviation
serves as a kind of ``renormalization point" for these operators.
In this language the BFKL equation is simply the evolution
equation for the  Wilson-line operators with respect to the
slope of the line. The gauge-invariant generalization of the
BFKL equation turns out to be a nonlinear equation which
contains more information than the usual BFKL equation --- for example,
it describes also the triple vertex of hard pomerons in QCD (cf. \cite{bart}).

Asymptotic expansions (in large momentum limits) play a vital
role in QCD.  Cross sections (or amplitudes) in these limits
simplify drastically, and one is thereby enabled to do
calculations that would otherwise be impossible.  The best
established of these expansions is Wilson's operator product
expansion for the $T$-product of two electromagnetic currents:
\begin{equation}
   T{j_{\mu }(x)j_{\nu }(0)}= \sum c_{n}(x) O_{n}(0).
\label{Wil}
\end{equation}
Here the coefficients $c_{n}$ contain all the singularities at
$x=0$, and the operators $O_{n}$ have no dependence on $x$.
Taking the expectation value of Eq.~(\ref{Wil}) in a nucleon state
and then Fourier transforming gives integer moments of the
factorization theorem for deep-inelastic structure functions:
\begin{equation}
   F_{2}(x_{B},Q^{2}) = \sum _{i} C_{i} (x_{B},Q^{2}/\mu ^{2} ,
\alpha _{s}(\mu
^{2}))\otimes
  f_{i} (x_{B},\mu ^{2}, \alpha _{s} (\mu ^{2}))+ \dots .
\end{equation}
Here the parton densities $f_{i} (x_{B},\mu ^{2}, \alpha _{s} (\mu ^{2}))$
are matrix elements of \lico\ operators. The dots stand for the
contributions of higher twist terms, i.e., terms
damped by extra powers of $1/Q^{2}$. $x_{B}$ is the
Bjorken scaling variable $x_{B}=Q^{2}/2p\cdot q$, and $\mu $ is the
renormalization scale.

Both Wilson's \ope\ and the factorization theorem can be
expressed in terms of coefficient functions and operator matrix
elements. This implies that a precise definition can be given to
the quantities involved. In particular, there are contributions
to the cross sections that come from the non-perturbative domain
of large distances.  The matrix element factors include these
contributions, and their definitions include non-perturbative
contributions.

The renormalization scale $\mu $ has the qualitative effect of
separating ``hard'' and ``soft'' contributions to the cross
section.  Integrals over soft momenta, those much less than $\mu $,
give suppressed contributions to the coefficient functions.
Integrals over hard momenta, those much greater than $\mu $, give
suppressed contributions to the matrix elements.  {\em Roughly}
speaking the coefficient functions are given by integrals over
large momenta $Q^{2}>p^{2}>\mu ^{2}$, while the matrix elements are given by
integrals over small momenta $p^{2}<\mu ^{2}$.  The crucial property that
enables calculations to be done easily is that the $\mu $ dependence
is given by the renormalization group equations.  One can set
$\mu =O(Q)$ in the coefficient functions.  Both these and the
kernel of the renormalization group equation can then be
calculated perturbatively, in powers of $\alpha _{s}(Q)$.

\begin{figure}[htb]
\mbox{
\epsfxsize=7cm
\epsfysize=5cm
\hspace{4cm}
\epsffile{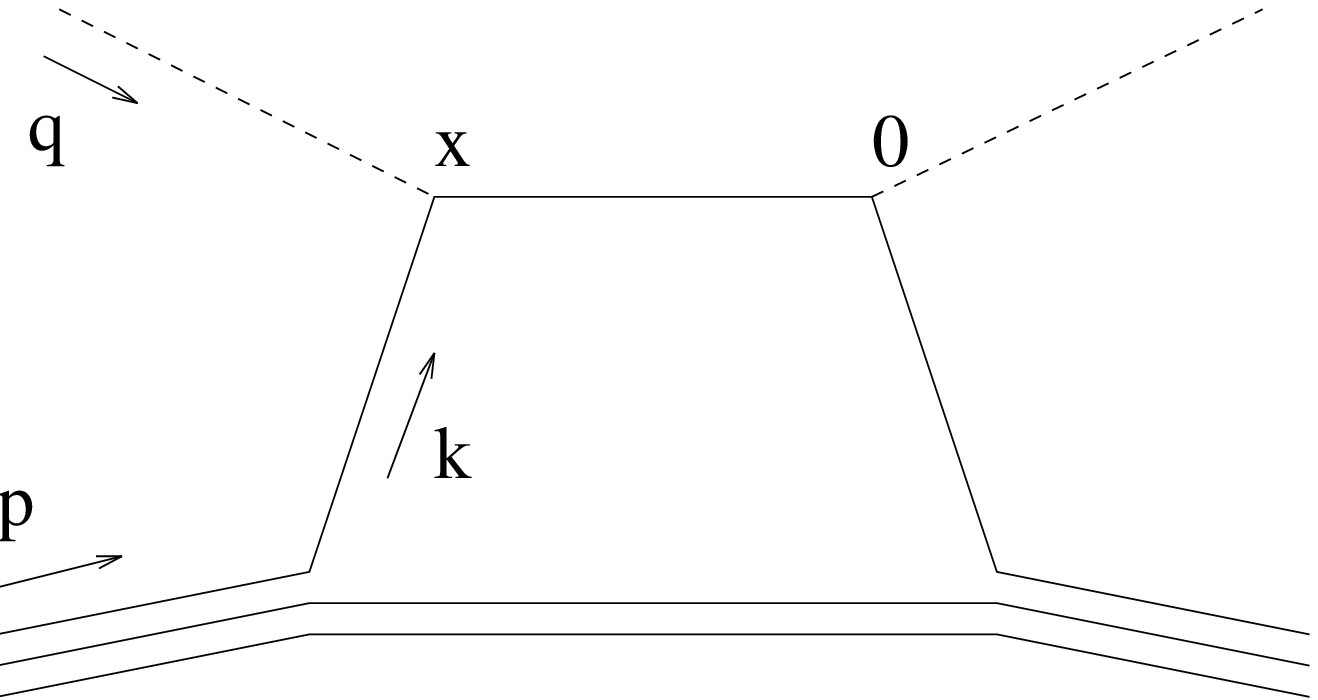}}
\vspace{0.1cm}
{\caption{\label{fig:dis} A typical diagram for \protect\dis.}}
\end{figure}

Let us recall how the usual Wilson expansion helps us to
find the $Q^{2}$ dependence of the moments of structure
functions of deep inelastic scattering. The essence is that instead of the
dependence of the physical amplitude on $Q^{2}$
(in the Euclidean region, which
corresponds to the moments of structure functions), we study the dependence of
\mael s of local operators on the renormalization point $\mu $.
Consider the simplest Feynman diagram shown in Fig.\ \ref{fig:dis}.

At large $q$ we can
expand the current quark propagator in inverse powers of $q$
\begin{equation}
\frac {1}{\nq+\nk}=\frac {1}{\nq}-\frac {1}{\nq}\nk\frac {1}{\nq}+
\frac {1}{\nq}\nk\frac {1}{\nq}\nk\frac {1}{\nq}+...
\label{2.1}
\end {equation}
where $n^{th}$ term of the expansion corresponds to the $n^{th}$ moment of the
structure function.
Unfortunately, after the expansion the loop integrals over $k$ become UV
divergent so we must modify our Taylor series (\ref{2.1}) somehow. To this end,
we note that each term on the right-hand side
of Eq.\ (\ref{2.1}) corresponds to the matrix
element of a certain quark operator of the type
$\bar{\psi }(\partial _{\mu })^{n}\gamma _{\nu }\psi $; the UV divergence
reflects
merely the large dimensions of these operators. It is well known how to
regularize these UV divergences --- we must
introduce the regularized operators
normalized at some point $\mu $ and expand our physical amplitudes in a series
of these regularized operators. Roughly speaking,
each term on the right of Eq.\ (\ref{2.1})
will be integrated over $k$ only up to
$k=\mu $. The dependence on $\mu $ will be cancelled : in the next order in
$\alpha _{s}$ the coefficients of the Taylor expansion will be modified
also ---
they will contain terms $\sim \alpha _{s} \ln{Q^{2}\over\mu ^{2}}$
which will cancel the
dependence of matrix elements of the (renormalized)
local operators on $\mu ^{2}$.
In the leading logarithmic approximation we can simply take $\mu ^{2}=Q^{2}$
and the
dependence of moments of structure functions on $Q^{2}$ will reflect the
dependence of the matrix elements of the operators
$\bar{\psi }\nabla _{\mu _{1}}...\nabla _{\mu _{n-1}}\gamma _{\mu _{n}}\psi $
on the
normalization point.  This dependence
is given by renormalization-group equation (see e.g. the book \cite{collbook}).

Summarizing, in order to find the dependence of the structure functions
of \dis\ at large $Q^{2}$ we perform the following steps: (i) formally expand
in
inverse powers of $Q^{2}$, (ii) regularize the obtained UV divergent matrix
elements of local operators, and (iii) write down (and solve) the evolution
equation with respect to the normalization point $\mu $. In the original
Feynman
diagrams for \stf s of \dis\ the photon virtuality $Q^{2}$ plays the
role of a
``physical" cutoff for the integrals over loop momenta. After expansion in
powers of $1/Q^{2}$ these integrals became UV divergent; by adding counterterms
in the usual way we introduce an ``artificial" cutoff $\mu ^{2}$ for these loop
integrals. Now, instead of studying of the $Q^{2}$ behavior of the original
Feynman diagrams, we should trace the dependence of the matrix
elements of the
operators on the cutoff $\mu ^{2}$.  This is a lot easier thing to do because
it
is governed by \rg .

Now, we would like to generalize these ideas for high-energy
scattering.  In order to find the high-energy behavior of a
certain physical amplitude (say, the \stf\ of \dis\ at very
small $x$), we will perform the same three steps: (i) formally
expand the amplitude at large energy $s$ --- after that we will
have the divergences in the longitudinal integrals, (ii)
regularize these longitudinal divergences by introducing a
certain cutoff, and (iii) find (and hopefully solve) the evolution
equations with respect to this cutoff. As in the case of Wilson
expansion, the dependence of the relevant matrix elements on the
cutoff determines the high-energy behavior of the original amplitude.
In the subsequent three Sections we will perform these steps.
In the Appendices we present the shock-wave picture of high-energy
scattering in the virtual photon frame.

\section{High Energy Limit}
\label{sec:HELimit}

\begin{figure}[htb]
\vspace{-6cm}
\mbox{
\epsfxsize=7cm
\epsfysize=14cm
\hspace{4cm}
\epsffile{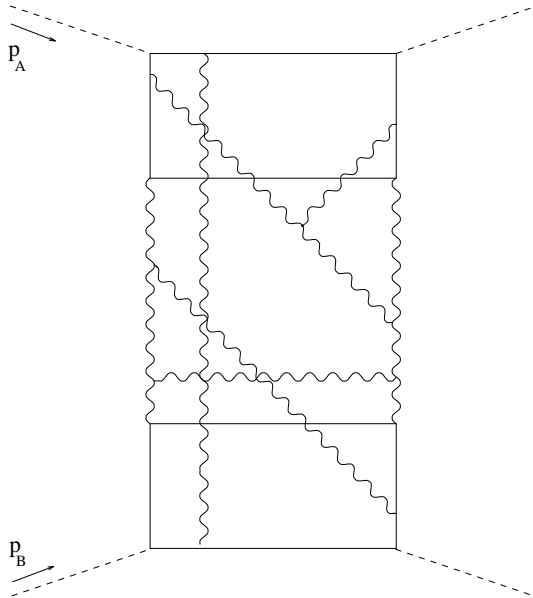}}
\vspace{0.1cm}
{\caption{\label{fig:his} High-energy scattering of virtual photons.}}
\end{figure}

As an example, let us consider the high-energy behavior of the
forward
scattering  amplitude for virtual photons in the region where
$s=(p_A+p_B)^2\gg p_A^2,p_B^2$:
\begin{equation}{\cal A}(p_A,p_B)=-ie^A_{\mu}e^A_{\nu}e^B_{\xi}e^
B_{\eta}\int\!d^4x\,d^4y\,d^4z\,e^{ip_A\cdot x+ip_B\cdot y}\lvac
T\{j^{\mu}(x+z)j^{\nu}(z)j^{\xi}(y)j^{\eta}(0)\}\rvac.\label{defA}
\end{equation}
(Only the connected part of the Green function is used, and
the vectors $e^A_{\mu}$ and $e^B_{\mu}$ are the polarizations of the
photons.)
For simplicity, we assume that the virtualities of photons are
negative, since
then our amplitude (\ref{defA}) will have only one discontinuity corresponding
to the total cross section of virtual photon scattering.
(At $s\gg p_{A}^{2}\gg
p_{B}^{2}$ this will be the \crsc\ of \dis\ from
a virtual photon at small $x$.)
A typical graph is shown in Fig.\ \ref{fig:his}.
Our aim is to
obtain the leading contribution (in powers of $s$) from graphs
for the amplitude, and it is well known that the leading
large-energy asymptotic behavior ($s$ times logarithms of $s$)
corresponds to diagrams with gluon exchanges.

To orient the reader in our subsequent technical treatment, we first
explain the qualitative features of the process in coordinate space.
Suppose that we view the process as one of the incoming photons, $A$
say, travelling through the color field due to the other.  Because of
time dilation and Lorentz contraction, each beam particle may be
viewed as a collection of fast-moving, point-like objects distributed
over the transverse plane.  The probability of a large momentum
transfer $Q$ is of order $1/Q^2$, so that the partons are to be
regarded as travelling along straight lines while they are crossing
the non-trivial part of the field.  This is, of course, the parton
model.  In a lowest order approximation, such as Fig.\ \ref{fig:his},
the partons in question are given by a single quark-antiquark pair.
The photon has fluctuated to a state of a quark-antiquark pair, and
this state is almost unchanged while the pair traverses the field,
since the time-scale for the evolution of the state is much longer
than the time to cross the field.

Another view of the same situation can be obtained in the rest frame
of the beam.  The field of the other particle is Lorentz contracted,
time dilated (and intensified).  It looks like a shock wave of width
$L\sim |p_B^2|/s$, and the quarks of the beam cannot make a significant
movement in the transverse direction during that time.  Here, $L$ is
the typical length scale of the field.

{}From either viewpoint, we see that the situation is one in which some
version of the eikonal approximation is valid.  That is, the effect of
the field on the state of the quark-antiquark pair is given by
integrals over the gluon field along the (straight-line) trajectories
of the partons.  Indeed, we will see that what we need are exactly
path-ordered exponentials of the gluon field.  Unlike the simple
eikonal approximation, the field affects both the phase and the color
orientation of the state.

\subsection{Amplitude as integral over gluon field}

As usual, at high energies it is convenient to use a
decomposition in Sudakov variables.
For the momenta we write the standard formula
\begin{equation}p^{\mu}~=~\al_pp_1^{\mu}+\be_pp_2^{\mu}+p\p^{\mu}
,\label{sudm}\end{equation}
where
$p_1^{\mu}\simeq p_A^{\mu}-\frac {p_A^2}sp_B^{\mu}$ and
$p_2^{\mu}\simeq p_B^{\mu}-\frac {p_B^2}sp_A^{\mu}$ are
light-like vectors close to $p_A$ and $p_B$ respectively.
(Then $g_{\mu\nu}=\frac 2s(p_{1\mu}p_{2\nu}+p_{1\nu}p_{2\mu})+g^{
\perp}_{\mu\nu}$.)  These variables
are essentially identical to light-front coordinates, $\al=p_+/\sqrt{s},
{}~\be=p_-/\sqrt{s}$.
For the coordinates we use a slightly different form
\begin{equation}
z^{\mu}~=~\frac 2sz_{*}p_1^{\mu}+\frac 2sz_{\bu}p_
2^{\mu}+z\p^{\mu},\label{sudc}
\end{equation}
where $z_{\bu}\equiv z_{\mu}p_1^{\mu}$ and $z_{*}\equiv z_{\mu}p_
2^{\mu}$\footnote{ Sometimes, however, we shall use also
"covariant" coordinates - Sudakov variables:
\begin{equation}
z^{\mu}~=~up_1^{\mu}+vp_
2^{\mu}+z\p^{\mu},\label{sudc1}
\end{equation}}.
One advantage of these coordinates is their simple
scaling properties when we take the high energy limit,
as in Eq.~(\ref{lim}), below.  The factors $2/s$ in the
formula for the components, Eq.~(\ref{sudc}), avoid extra
factors of $s$ in the combination $p\cdot z$
$=$$\alpha_pz_{\bu}+\beta_pz_{*}-p_{\perp}\cdot z_{\perp}$.

The Jacobian of the
transition to Sudakov variables is $s/2$ so that
\begin{equation}
\int \! d^{4}z={2\over s}\int \! dz\bu dz_{*} d^{2}z\p~,~~~\int \!
d^{4}p={s\over 2}\int \!
d\al_{p} d\be_{p} d^{2}p\p .
\end{equation}
To put the scattering amplitude (\ref{defA}) in a form symmetric
with respect the top and bottom photons, we make a shift of the
coordinates in the currents by $(z\bu,0,0\p)$ and then reverse
the sign of $z\bu$. This gives:
\begin{eqnarray}
   {\cal A}(p_{A},p_{B}) &=& -ie^{A}_{\mu }e^{A}_{\nu }e^{B}_{\xi }
e^{B}_{\eta }{2\over s}\int \!
d^{2}z\p dz\bu dz_{*} \int \! d^{4}x \,
d^{4}y \,
   e^{ip_{A}\cdot x+ip_{B}\cdot y}
\nonumber\\
\qquad
   && \hspace*{-1cm} \lvac T\{j^{\mu }(x\bu,x_{*}+z_{*},x\p+z\p)
             j^{\nu  }(0,z_{*},z\p)
             j^{\xi }(y\bu+z\bu,y_{*},y\p)
             j^{\eta }(z\bu,0,0\p)\}
    \rvac .
\end{eqnarray}
(We remind the reader that only the connected part of this Green
function is taken.)

It is convenient to
start with the upper part of the diagram,
i.e., to study how fast quarks
move in an external gluonic field.
After that, functional integration
over the gluon fields will reproduce us the Feynman diagrams of the
type of Fig. \ref {fig:his}:
\begin{eqnarray}
   {\cal A}(p_{A},p_{B})&=&
   -ie^{A}_{\mu }e^{A}_{\nu }e^{B}_{\xi }e^{B}_{\eta }{s\over 2}
   \int \! d^{2}z\p{\cal N}^{-1}
   \int {\cal D}A \, e^{iS(A)} {\rm det}(i\nabla )
\nonumber\\
&& \qquad
   \left\{{2\over s}\int \! dz_{*} \int \!d^{4}x
           \, e^{ip_{A}\cdot x}
           \left\lan T j^{\mu }(x\bu,x_{*}+z_{*},x\p+z\p)
                j^{\nu }(0,z_{*},z\p)
           \right\ran_{A}
    \right\}
\nonumber\\
&& \qquad
   \left\{{2\over s}\int \! dz\bu\int \!
          d^{4}y \, e^{ip_{B}\cdot y}
          \left\lan
              T j^{\xi }(y\bu+z\bu,y_{*},y\p)j^{\eta }(z\bu, 0, 0\p)
          \right\ran_{A}
    \right\},
\label{A.fnl.int}
\end{eqnarray}
where
\begin{equation}
\lan T j_{\mu }(x)j_{\nu }(y)\ran_{A}\equiv
   \frac {\int {\cal D}\psi {\cal D}\bar{\psi }e^{iS(\psi , A)}
j_{\mu }(x)j_{\nu  }(y)}{\int {\cal D}\psi {\cal D}\bar{\psi }e^{iS(\psi , A)}}
   - \mbox{Same at $A=0$} ,
\label{2.7}
\end{equation}
and $S(A)$ and $S(\psi ,A)$ are the gluon and quark-gluon parts of
the QCD action respectively.  ${\rm det}(i\nabla )$ is the determinant of
Dirac operator in the external gluon field; it gives the effect
of quark loops in Fig.\ \ref{fig:his}.  The subtraction in Eq.\
(\ref{2.7}) removes the disconnected graph.

The arrangement of the integrals in Eq.\ (\ref{A.fnl.int}) arises
from a choice to construct amplitudes that have all
momentum conservation delta functions removed.
The integrals over $x$ and $y$ set the momenta of the outgoing
photons to be $p_{A}$ and $p_{B}$.  The integral over $z_{*}$ sets the
$\beta $ component of the incoming photon momentum on the top
bubble to be equal to the corresponding component of the outgoing
momentum, $\beta _{p_{A}}\simeq p_{A}^{2}/s$.
The $\alpha $ component of the incoming
photon to the top bubble is the corresponding component for the
outgoing photon minus whatever $\alpha $ component of momentum the
external field happens to provide.
A similar statement applies to the $z\bu$
integral.
The $z\p$ integral enforces zero transverse momentum
transfer at one end, and we leave it as the outermost integral in
order to emphasize that we wish to treat transverse momenta
symmetrically between the upper and lower quark loops.
There remains the functional integral over the
gluon field, after which momentum is conserved.  Before this is
performed, there is no conservation of momentum, since the gluon
field is position dependent.

\subsection{Fast moving photon in external gluon field}

{}From Eq.\ (\ref{A.fnl.int}), it is clear that the amplitude for
the upper part of the diagram in Fig.\ \ref{fig:his}, describing
a virtual photon with momentum $p_{A}$ flying through the external
gluon field $A_{\mu }$, is given by the following expression
(see Fig.\ \ref{fig:ext}):
\begin{figure}[htb]
\vspace{-6cm}
\mbox{
\epsfxsize=7cm
\epsfysize=14cm
\hspace{4cm}
\epsffile{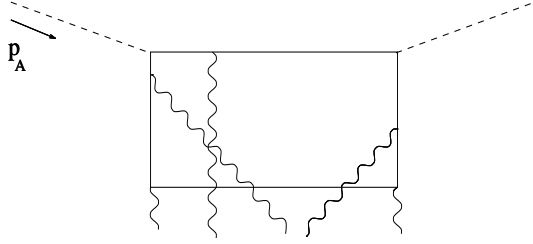}}
\vspace{-4cm}
{\caption{\label{fig:ext} High-energy scattering of virtual
photon from external field.}}
\end{figure}
\begin{eqnarray}
   \lefteqn{ \frac {2}{s} \int \! dz_{*} \int d^{4}x \, e^{ip_{A}\cdot x}
             \lan
             T\{j_{\mu }(x\bu,x_{*}+z_{*},x\p+z\p)
             j_{\nu  }(0,z_{*},z\p)\}\ran_{A}
           }
\hspace*{1.2in}
\nonumber\\
&=&
   \frac {2}{s} \sum _{i} e_{i}^{2}
   \int \! dz_{*} d^{4}x e^{ip_{A}\cdot x}
    \, {\rm Tr} \, \ga_{\mu }
     \LBB x\bu,x_{*}+z_{*},x\p+z\p \Big| \frac {i}{\nP}
        \Big| 0,z_{*},z\p
     \RBB
\nonumber\\
&& \hspace*{1.3in}
   \ga_{\nu } \LBB 0,z_{*},z\p \Big| \frac {i}{\nP}
            \Big| x\bu,x_{*}+z_{*},x\p+z\p
        \RBB
\nonumber\\
&&
   - \mbox{Same at $A=0$} ,
\label{Upper.z}
\end{eqnarray}
where we have used Schwinger's notations  for the propagators
in an external field, and $e_{i}$ are the quark charges.

In Schwinger's notations we write down formally the quark propagator in the
external gluon field $A_{\mu }(x)$ as a matrix element of the inverse Dirac
operator
\begin{equation}
G(x,y) = \LBB x \Big| \frac {i}{\nP} \Big| y\RBB =
         \LBB x \Big| \frac {i}{\st p + g\st A}
         \Big| y\RBB ,
\label{sch}
\end{equation}
where
\begin{equation}
   \LB x|y \RB =\delta ^{(4)}(x-y),
   {}~~~~ \LB x | p_{\mu } | y \RB =
   -i\frac {\partial }{\partial y^{\mu }}\delta ^{(4)}(x-y),~~~~
   \LB x | A_{\mu } | y \RB
   =A_{\mu }(x)\delta ^{(4)}(x-y) .
\label{2.10}
\end{equation}
Here $|x\RB$ are the eigenstates of the coordinate operator
${\cal X}|x\RB =x|x\RB$ (normalized according to the second
line in the above
equation). From Eq.\ (\ref{2.10}) it is also easy
to see that the eigenstates of
the free momentum operator $p$ are the plane waves
$|p\RB=\int d^4x\,e^{-ip\cdot x}|x\RB$.
It should be clear from the context whether the vectors
are eigenstates of momentum or position.  Note that the
states $|x\RB$ or $|p\RB$ are functions of {\em four\/}-vectors ($x^{\mu}$ or
$p^{\mu}$), unlike the actual quantum mechanical state vectors
of the field theory.

Thus,
for example, the first term of the expansion of
the propagator (\ref{sch}) in powers of external field is the
free propagator
\begin{equation}
   \LBB x \Big| \frac {i}{\st p} \Big| y\RBB
   = \int \frac {d^{4}p}{16\pi ^{4}} e^{-ip\cdot (x-y)  }\frac {i}{\st p} .
\label{bare}
\end{equation}

We are treating the gluon field as a matrix in the fundamental
representation of SU(3):
\begin{equation}
   A_{\mu }(x) = \sum _{\alpha } A^{\alpha }_{\mu } t_{\alpha } .
\end{equation}
Then the quark propagator, Eq.~(\ref{sch}), is a matrix in
both color and spinor space; the $\st p = \gamma ^{\mu }p_{\mu }$ is implicitly
multiplied by a unit color matrix.

Now let us Fourier transform Eq.\ (\ref{Upper.z}) over $z\p$, so
that it is a function of $q\p$ instead of $z\p$.
Going to Sudakov variables (\ref {sudm}), we have:
\begin{eqnarray}
\lefteqn{\int\!d^4x\int\!d^4z\,\delta (z\bu)e^{-i(q,z)\p}e^{ip_A\cdot
x}\ \lan T\{j_{\mu}(x+z)j_{\nu}(z)\}\ran_A}\hspace{0.6in}\nonumber\\
&=&\frac {s^2}4\sum_ie_i^2\int\frac {d^4k}{16\pi^4}\frac {d^4p}{1
6\pi^4}\frac {d^4p'}{16\pi^4}2\pi\delta (\be_p-\be'_p)4\pi^2\delta ^{(2)}
(p\p-p'\p-q\p)\nonumber\\
&\hspace{0.3in}&\Tr\,\bigg\{\gamma_{\mu}\LBB k\Big|\frac 1{\nP}\Big
|k-p\RBB\gamma_{\nu}\LBB k-p_A-p'\Big|\frac 1{\nP}\Big|k-p_A\RBB\bigg
\}\nonumber\\
&&-\mbox{\rm Same at $A=0$}.\label{quark.loop}\end{eqnarray}
Here, $(ab)\p$ denotes a scalar product of transverse components of
vectors $a$ and $b$.
In this expression, the quark-loop momentum is
$k^{\mu } = \al_{k} p_{1}^{\mu } + \be_{k} p_{2}^{\mu } + k\p^{\mu }$,
while
$p^{\mu } = \al_{p} p_{1}^{\mu } + \be_{p} p_{2}^{\mu } + p\p^{\mu }$ and
$p'^{\mu } = \al_{p'} p_{1}^{\mu } - \be_{p} p_{2}^{\mu } - p\p^{\mu } +
q\p^{\mu }$
are the momenta entering the two quark lines from the external
field.
Notice that the quark propagators do not conserve
momentum.
We have enforced conservation of the $\beta$ and the
transverse components of momentum by our choice of
external momentum, while conservation of the $\alpha$
component of the momenta will only be true after
the functional integral over the
external gluon field to form the complete amplitude, as
in Eq.~(\ref {A.fnl.int}).

\subsection{Regge limit}

Now, we must formally take the limit $s\rightarrow\iy$ in this
expression.  We will do this for a fixed external field.
The Regge limit $s\rightarrow\iy$ with $p_A^2$ and $p_B^2$ fixed corresponds
to the following rescaling of the virtual photon
momentum:
\begin{equation}p_A=\la p_1^{(0)}+\frac {p_A^2}{2\la p^{(0)}_1\cdot
p_2}p_2,\end{equation}
with $p_B$ fixed.  This is equivalent to
\begin{equation}
   p_{1}=\la p_{1}^{(0)}, \ \ \ p_{2}=p_{2}^{(0)} ,
\label{lim}
\end{equation}
where $p^{(0)}_1$ and $p_2^{(0)}$ are fixed light-like vectors so that
$\la$ is a large parameter associated with the
center-of-mass energy
($s=2\la p_1^{(0)}\cdot p_2^{(0)}$).\footnote{%
   The method we are using is a version of the
   methods used in \protect\cite{AV,Verlinde}.
}

Next, let us look at how the Sudakov
variables in Eq.\ (\ref{2.14}) scale with $\la$.
In general when treating an asymptotic limit of some Feynman
graphs, there will be a number of different regions of
loop-momentum space that contribute.  It is quite a
complicated problem to disentangle these.
However, we have chosen to start with the asymptotics for a {\em
fixed} external field.  So initially we do not have to conern
ourselves with the problem of multiple regions.  That problem
arises at a later stage of the argument when we perform the
the functional integral over the gauge field.

The limit we are taking is $\lambda \to \infty $ with the gauge field $A$
fixed.

First,
we shall see below
from the explicit form of integral that the important
values of the variables for the
quark loop attached to photon $A$ satisfy $\al_k\sim 1$ and
$\be_k\sim\frac 1{\la}$. Such momenta are obtained by boosting from
$\lambda =1$.
With this scaling, scalar products of quark
momenta, and the measure $d^4k$ are independent of $\lambda$;
this is a consquence of boost invariance.

Moreover, the important values of momenta transfered
from the gluon field obey
$\al_p\sim\frac 1{\la},\be_p\sim 1$,
since $\tilde{A}(\al_p,\be_p,p\p)=\tilde{A}^{(0)}(\la\al_p,\be_p,p\p)$
\begin{equation}
\tilde{A}^{(0)}(\al,\be,p\p)\equiv\int\!d^4x\,\,A(x)\,e^{
i\al p_1^{(0)}\cdot x+i\be p_2^{(0)}\cdot x-i(px)\p}
\label{azero}
\end{equation}
is a function independent of $\la$
\footnote{The scaling for $A$ applies before the functional
   integral over $A$.  After the integration over $A$, we will
   get contributions from $\alpha _{p} \sim 1$ and from $\alpha _{p}
\to  \infty $.
   The
   first region corresponds to higher-order corrections to the
   quark loop; these are just like higher order corrections to
   the Wilson expansion.  The second region corresponds to UV
   divergences in these same higher-order corrections.  In both
   cases subtractions must be applied; we treat this as a
   separate issue.  }.
(Hereafter the $\tilde{A}(p)$ denotes
the Fourier transform of the field $A(x)$)
So, we must compute the behavior of the quark loop
(\ref{quark.loop}) in the region
\begin{eqnarray}
   \al_{k}\sim 1, &
   {\displaystyle \be_{k}\sim {1\over \la}},
   & k\p^{2}\sim 1 ~(\cdot p_{A}^{2})
\nonumber\\
   \al_{p}\sim {1\over\la}, & \be_{p}\sim 1, & p\p^{2}\sim 1 ~
(\cdot p_{B}^{2})
{}.
\label{char}
\end{eqnarray}
(We shall see that $k\p^{2}\sim p_{A}^{2}$ from the explicit integral
(\ref{2.31})
and that the
characteristic $p\p^{2}$ of the external field is determined by the
characteristic scale of the source of this field, which is the
virtuality of the target photon $p_B^2$.)

\subsection{Quark propagator in external field}

\begin{figure}[htb]
\vspace{1cm}
\mbox{
\epsfxsize14cm
\epsfysize=3cm
\hspace{1cm}
\epsffile{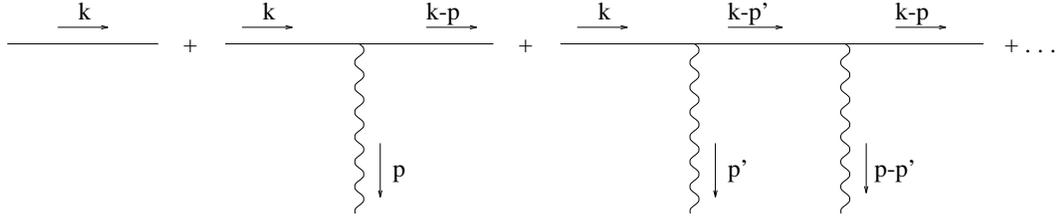}}
\vspace{1cm}
{\caption{\label{fig:exp}Quark propagator in the external field.}}
\end{figure}

As a first step, we will find the quark propagator in
the external field (see Fig.  \ref {fig:exp}).  In the limit
we are considering, we must recall the well known fact
that at high energy we can replace $g_{\mu\nu}$ for the gluon
propagators connecting quark lines with very different
rapidities by $\frac 2sp_{1\mu}p_{2\nu}$.  Thus, we can change the factors
$\ga^{\mu}A_{\mu}$ for the interaction to $\frac 2s\np_2A\bu$, correct to the
leading power of $\lambda$ (or $s$).  This gives:
\begin{eqnarray}
\lefteqn{\LBB k\Big|\frac 1{\st{{\cal P}}}\Big|k-p\RBB}\nonumber\\
&=&\frac {16\pi^4}{\nk}\delta^{(4)}(p)\nonumber\\
&&-g\frac 2s\frac {\nk}{k^2+\ie}\,\np_2A\bu(p)\,\frac {\nk-\np}{(
k-p)^2+\ie}\nonumber\\
&&+g^2\frac 2s\int\frac {d^4p'}{16\pi^4}\,\frac {\nk}{k^2+\ie}\,\np_
2A\bu(p')\frac {\nk-\np'}{(k-p')^2+\ie}\np_2A\bu(p-p')\,\frac {\nk
-\np}{(k-p)^2+\ie}\nonumber\\
&&+\dots,\label{2.14}\end{eqnarray}
where the dots stand for further terms in the
expansion in powers of the external field.

Let us start with the first nontrivial term $\sim A\bu$ shown in
Fig. \ref {fig:exp}b.
{}From the eq. (\ref{azero}) it follows that
\begin{equation}
 A\bu(\al_p,\be_p,p\p)=\la A^{(0)}\ci(\la\al_p,\be_p,p\p)
\end{equation}
(here $A\ci\equiv A_{\mu}p^0_{1\mu}$). Now it is easy to see that in the limit
$\la\rightarrow\iy$ the Fourier transform of the external field
$\tilde{A}(p)$ is proportional
to $\delta (\al_p)$ (we assume that the Fourier transform of
the external field $A^{(0)}(p)$ eq. (\ref{azero}) decreases at infinity).
The coefficient in front of the $\delta$-function can be figured out from
the following formula
\begin{equation}
ig\int\frac {d\al_p}{2\pi}\tilde{A}\bu(\al_p,\be_p,p\p)=i
g\int\!d^2x\p\,e^{-i(p,x)\p}\int\!du\,A\bu(up_1+x\p)e^{\frac i2s\be_
pu}.\label{A.lc.1}
\end{equation}
so the first two terms of the expansion of
the propagator (\ref {2.14}) reduce to
\begin{eqnarray}
\lefteqn{\LBB k\Big|\frac 1{\st{{\cal P}}}\Big|k-p\RBB}\nonumber\\
&=&\frac {16\pi^4}{\nk}\delta^{(4)}(p)\nonumber\\
&&-g\frac 2s\frac {\nk}{k^2+\ie}\,\np_22\pi\delta (\al_p)\left[\int\!d^
2x\p\,e^{-i(p,x)\p}\int\!du\,A\bu(up_1+x\p)
e^{\frac i2s\be_pu}\right]\,\frac {\nk
-\np}{(k-p)^2+\ie}
\label{2.1a}
\end{eqnarray}

 As we will see below,
the integrals in Eq.\ (\ref{quark.loop}) will force $\beta _{p}$ to be of
order $1/\lambda $, so that we should set $\beta _{p}=0$ in
Eq.~(\ref{A.lc.1}).
Eq.~(\ref{A.lc.1}) is the first order of the expansion of a path
ordered exponential whose precise definition is given in Eqs.\
(\ref{pexp}) and (\ref{U}) below.

Next, consider the third term on the right-hand side of
Eq.\ (\ref{2.14}) shown in Fig.\ \ref{fig:exp}c.
In the region (\ref{char}), the second propagator of
this term (times the $\st p_{2}$ factors on each side) reduces to
an eikonal denominator:
\begin{eqnarray}
  \np_{2} \,
  \frac {(\al_{k}-\al'_{p})\np_{1} + (\st k - \st p')\p}{(\al_{k}-\al'_{p})
(\be_{k} -\be'_{p})s-(k-p')\p^{2}+\ie}
  \, \np_{2}
   &=&
  \np_{2} \,
   \frac {(\al_{k}-\al'_{p})s}{(\al_{k}-\al'_{p})(\be_{k} -\be'_{p})s-
(k-p')\p^{2}+\ie}
\nonumber\\
  &\rightarrow&
  \np_{2} \,
   {1 \over -\be'_{p}+\ie\al_{k}} .
\end{eqnarray}
Furthermore, we can neglect $\al_{p},\al'_{p}$ as
compared to $\al_{k}$ in the free quark
propagators in Eq.\ (\ref{2.14}).

Again, since characteristic $\al_p\rightarrow 0$ at $\la\rightarrow\iy$
this contribution
will be proportional to $\delta (\al_p)$. In order to find the
coefficient in front of this $\delta$ function we shall integrate
the propagator over $\al_p$. As we will see below,
when we do the integral over $\beta_k$, both the quark and
antiquark lines are restricted to being forward moving,
i.e., $0<\alpha _{k}<1$.
Then the integrals of the gluon fields
in this third term reduce to the second order
term in the path-ordered gauge factor:
\begin{eqnarray}
\lefteqn{
   -ig^{2} \int \frac {d\al_{p}}{2\pi }
   \frac {d\al'_{p}}{2\pi } \frac {d\be'_{p}}{2\pi }
   \frac {d^{2}p'\p}{4\pi ^{2}}
   \tilde{A}\bu(\al'_{p}, \be'_{p},p'\p)
   \frac {1}{ -\be'_{p} + \ie}
   \tilde{A}\bu(\al_{p}-\al'_{p},\be_{p}-\be'_{p},p\p-p'\p)
}
\hspace*{0.5in}
\nonumber\\
&=&
   -g^{2}\int d^{2}x\p \, e^{-i (p,x)\p}
   \int dv \int du
      \Theta (v-u)
    A\bu (vp_{1}+x\p) A\bu (up_{1}+ x\p)
    \, e^{\frac {i}{2}s\be_{p}u}
\nonumber\\
&\to &
   -g^{2}\int d^{2}x\p \, e^{-i (p,x)\p}
   \int dv \int du
      \Theta (v-u)
    A\bu (vp_{1}+x\p) A\bu (up_{1}+ x\p)\,.
\label{2.19}
\end{eqnarray}
In the last line, we have again used the result, to be
demonstrated later, that $\beta _{p}$ is of order $1/\lambda $.
So, the expansion (\ref{2.14}) takes the form:
\begin{eqnarray}
\lefteqn{
   \LBB k
   \Big| {1\over \st{\cal P}}
   \Big| k-p
   \RBB
}
\nonumber\\
&=&
   {16\pi^4\over\nk} \delta ^{(4)}(p)
\nonumber\\
&&
   {}- g{2\over s} 2\pi\delta(\al_p)
     \frac {\nk}{k^{2}+\ie} \,
     \np_{2}\left[\int \! d^{2}x\p \, e^{-i(p,x)\p}
      \int \!du \, A\bu(up_{1}+x\p) e^{\frac {i}{2}s\be_{p}u}\right] \,
   \frac {\nk-\np}{(k-p)^{2}+\ie}
\nonumber\\
&&
   {}+g^{2} \frac {2}{s} 2\pi\delta(\al_p)
   \frac {\nk}{k^{2}+\ie}\np_2 \,
\nonumber\\
&&
  \left[\int d^{2}x\p \, e^{-i (p,x)\p}
   \int dv \int du
      \Theta (v-u)
    A\bu (vp_{1}+x\p) A\bu (up_{1}+ x\p)\right]\,
    \frac {\nk-\np}{(k-p)^{2}+\ie}
   {}+ \dots ,
\label{2.2a}
\end{eqnarray}

We now express these and all the higher terms of the expansion of the
right-hand-side of Eq.\ (\ref{2.14}) in terms of path ordered
exponentials of the gluon field.
Let us use $[x,y]$ to denote the path-ordered gauge factor
along the straight line connecting the points $x$ and $y$:
\begin{equation}
   [x,y]=Pe^{ig\int _{0}^{1} du \, (x-y)^{\mu }A_{\mu }(ux+(1-u)y)} .
\label{pexp}
\end{equation}
Then it can be demonstrated fairly easily
that further terms of the expansion of the
right-hand-side
of Eq.\ (\ref{2.14}) in powers of $A\bu$ will reproduce the
subsequent terms in the expansion (in powers of $A\bu$) of the following
gauge factors:\footnote{The form of our notations, $[U-1]$, $[\Ud-1]$,
reflects the fact that
at $\beta _{p}=0$, these gauge factors are simple path-ordered exponentials
along an infinite line, but that
the zeroth term in the expansion in powers of gauge field is missing.}:
\begin{eqnarray}
   [U-1](p\p) &=& U(p\p)-4\pi ^{2}\delta ^{(2)}(p\p),
\nonumber\\{}
   [\Ud-1](p\p) &=& \Ud(p\p)-4\pi ^{2}\delta ^{(2)}(p\p) ,
\end{eqnarray}
where $U(p\p)$ and $\Ud(p\p)$ are Fourier transforms of
gauge factors along lines extending to infinity in both
directions:
\begin{eqnarray}
   U(x\p )   &=& [\iy p_{1}+x\p ,-\iy p_{1} +x\p ] ,
\nonumber\\
   \Ud (x\p) &=& [-\iy p_{1}+x\p,\iy p_{1}+x\p] .
\label{U}
\end{eqnarray}
These correspond to quarks moving across
the external field with the speed of
light. The relevance of these gauge

Thus we finally have the propagator of a fast-moving
quark with $0<\alpha _{k}<1$:
\begin{eqnarray}
\lefteqn{
    \LBB k \Big| {1\over\nP} \Big| k-p \RBB
}\qquad
\nonumber\\
&=&
    16\pi ^{4} \delta ^{(4)}(p) {1\over\nk}
    + {4\pi i\over s}\delta (\al_{p})
      \frac {k}{k^{2}+\ie}
      \, \np_{2} \,
       [U-1](p\p)
      \frac {\nk - \np}{(k-p)^{2}+\ie} .
\label{qprop}
\end{eqnarray}
This is valid to the leading power of $\lambda $, when $\lambda $ is large and
the external gluon field $A$ is fixed.  A similar formula is
valid for the antiquark propagator.

\subsection {Impact factor}

Let us rewrite the expression (\ref{quark.loop})
for the upper part of the
diagram in Fig.\ \ref{fig:ext} using the above
formula for quark propagator (\ref{qprop}), and the corresponding
formula for the antiquark propagator. We obtain
\begin{eqnarray}
\lefteqn{
   - \sum _{i} e_{i}^{2} \int  \frac {d^{4}k\p}{16\pi ^{4}}\Theta(1>\al_k>0)
}
 \hspace*{0.2in}
 \nonumber\\
&&\Tr \Bigg\{
      i \,\gamma _{\mu }\,
      \frac {\nk}{ k^{2}+\ie}
      \, \gamma _{\nu } \,
\frac {[\nk-\np_A+\nq] \, \np_{2} \, [\nk-\np_A]}{[(k-p_A+q)\p^{2} + \ie] \,
[(k\-p_A)^{2}+\ie]}
   \left( [\Ud-1](q\p) \right)
\nonumber\\
&&
   {}-i \, \gamma _{\mu } \,
   \frac {\nk \, \np_{2} \, (\nk-\nq)}{[ k\p^{2}+\ie] [(k-q)^{2}+\ie]}
   \, \gamma _{\nu } \,
   \frac {\nk-\np_A}{(k-p_A)^{2}+\ie}
   \left( [U-1](q\p) \right)
\nonumber\\
&&
   {}-\int
   \frac {d^{4}p\p}{16\pi ^{2}} 2\pi\delta(\al_p)
 \, \gamma _{\mu } \,
\frac {\nk\np_2(\nk-\np)}{(k^{2}+\ie) [(k-p)^{2}+\ie]}
   \, \gamma _{\nu } \,
   \left([U-1](p\p)\right)
\nonumber\\
&&\hspace*{0.5in}
\frac {(\nk-\np_A-\np+\nq) \, \np_{2} \,\nk}{[(k-p_A-p+q)^{2}+\ie]
[(k-p_A)^{2}+\ie]}
   \left( [\Ud-1](q\p-p\p \right)
\Bigg\} ,
\label{2.25}
\end{eqnarray}
where we used the notation $\bar{\al}_{k}\equiv 1-\al_{k}$.
Now we can use contour integration to perform the integrals over
$\beta _{k}$ and $\beta _{p}$.  It is easy to verify that the dominant
contribution arises when both these variables are of order a
squared transverse momentum divided by $s$, i.e., of order $1/\lambda $.
It is easy to see that the linear terms in $U$ and $\Ud$ cancel in
Eq.~(\ref{2.25}) so after some algebra one obtains the final answer in the
following form\footnote{A more careful analysis performed in
   Appendix \ref{app:ExtField} shows
   that the Wilson lines $U$ and $\Ud$ are connected by gauge factors
   at infinity so
\begin{eqnarray}
\lefteqn{\Tr{U(x\p)\Ud(y\p)}}\\
\label{endf}
&\rightarrow&\lim_{\la\rightarrow\iy}
\Tr[-\la p_1+x\p,\la p_1+x\p][\la p_1+x\p,\la p_1+y\p]
[\la p_1+y\p,-\la p_1+y\p][-\la p_1+y\p
,-\la p_1+x\p].
\nonumber
\end{eqnarray}
   The gauge factors connecting the end points of the
   eikonals $U$ and $\Ud$ reduce at
   infinity the
   gauge factors made from pure gauge fields so the precise form
of the contour connecting
   the end points
   of Wilson lines does not matter.}
\begin{eqnarray}
\lefteqn{
   \int \! d^{4}x \int \! d^{4}z \, \delta (z\bu) e^{-i(qz)\p}
e^{ip_{A}\cdot  x}
   \lan T\{j_{\mu }(x+z)j_{\nu }(z)\}\ran_{A}
}
\hspace*{0.5in}
\nonumber\\
   &=& \sum e_{i}^{2}\int \frac {d^{2}p\p}{4\pi ^{2}}
   I_{\mu \nu }(p\p,q\p) \Tr\{ U(p\p)\Ud(q\p-p\p) \} ,
\label{2.30}
\end{eqnarray}
where $I_{\mu \nu }(p,q)$ is the so-called ``impact factor":
\begin{eqnarray}
   I_{\mu \nu }(p\p ,q\p) &=&
   \bar{I}_{\mu \nu }(p\p ,q\p)-\bar{I}_{\mu \nu }(0,q\p)
\nonumber\\
   \bar{I}_{\mu \nu }(p\p ,q\p) &=&
  - \int _{0}^{1} \frac {d\al}{2\pi }
   \int  \frac {d^{2}k\p}{4\pi ^{2}}
   \Tr \Bigg\{
      \gamma _{\mu } \,
      \frac {(\al\np_{1}+\nk\p) \, \np_{2} \,
[\al\np_{1}+(k-p)\p]}{s (k\p^{2}-p_{A}^{2}\al\bar{\al} )}
\nonumber\\
&&\hspace*{1.5in}
      \, \gamma _{\nu } \,
      \frac { [-\bar{\al}\np_{1}+(\nk-\np+\nq)\p] \, \np_{2} \,
(-\bar{\al}\np_{1}+ \nk\p)}{s [(k-p+q\al)\p^{2} +
(q\p^{2}-p_{A}^{2})\al\bar{\al} ]}
    \Bigg\} .
\label{2.31}
\end{eqnarray}
Contrary to appearances, the impact factor is independent of $s$
(and hence of $\lambda $).
The easiest way to see this is to observe that by a boost of the
coordinates used in Eq.\ (\ref{lim}) we may obtain the large $s$
limit by scaling $p_{2}$.  But in Eq.\ (\ref{2.31}) $p_{2}$ only
occurs in the combination $p_{2}/s$.

When the photon indices $\mu $ and $\nu $ are transverse, we
obtain the following explicit expression for $\bar{I}_{\mu \nu }$:
\begin{eqnarray}
   \bar{I}_{\mu \nu }(p\p,q\p)
   &=&
   -\half\int _{0}^{1} \frac {d\al}{2\pi }
   \int _{0}^{1} \frac {d\al'}{2\pi }
   \left\{
      P\p^{2}\al'\bar{\al'}+(q\p^{2}\al'-p_{A}^{2})\al\bar{\al}
   \right\}^{-1}
\nonumber\\
&& \qquad
   \Bigg\{
      (1-2\al\bar{\al}-2\al'\bar{\al'}+8\al\bar{\al}\al'\bar{\al'})
      P\p^{2} g_{\mu \nu }
      + 8\al\bar{\al}\al'\bar{\al'}P_{\perp \mu } P_{\perp \nu }
\nonumber\\
&& \qquad \
      +2g_{\mu \nu }q\p^{2}\al\bar{\al}(1-2\al')
      -4\al\bar{\al}(1-2\al)\bar{\al'}P_{\perp \mu }q_{\perp \nu }
   \Bigg\} ,
\label{2.32}
\end{eqnarray}
where $P\p\equiv p\p-q\p\al$. At $q\p=0$ this result agrees with \cite{XIV}.

Note that the limit $\la\rightarrow\iy$ enforces the vanishing of the total
$\be$ argument of the gauge factors $U$ (and $\Ud$) in Eq.\
(\ref{2.30}), whereas the individual
$A\bu$ fields forming this gauge factors may have nonvanishing $\be's$.
It is instructive to write down the final formula for the quark propagator
in this case:
\begin{eqnarray}
\lefteqn{
    \LBB k \Big| {1\over\nP} \Big| k-p \RBB
}\qquad
\nonumber\\
&=&
    16\pi ^{4} \delta ^{(4)}(p) {1\over\nk}~+
\nonumber\\
&&
    {4\pi i\over s}\delta (\al_{p})
      \frac {\nk\p\np_2}{k\p^{2}+\ie}
      \left( [U-1](p\p)
             \Theta (\al_{k})
             - [\Ud-1](p\p)
             \Theta (-\al_{k})
      \right)
      \frac {\nk - \np}{(k-p)\p^{2}+\ie} .
\label{qpro}
\end{eqnarray}
It is worth noting that the form of the answer (\ref{qpro})
--- ${\rm free~propagator} \otimes {\rm eikonal~factor}\otimes
{\rm free~propagator}$
---
is due to the shock-wave structure of the external field at large
energies (see Appendix \ref{app:ExtField}).

Let us also present the result (\ref{2.30}) in the transverse coordinate
representation. One has for the forward scattering:
\begin{eqnarray}
\lefteqn{
   \int \! d^{4}x \! \int \! d^{4}z \, \delta (z\bu) e^{ip_{A}\cdot x}
   \lan T\{j_{\mu }(x+z)j_{\nu }(z)\}\ran_{A}
}
\hspace*{0.5in}
\nonumber\\
   &=&
   \sum _{i}e_{i}^{2} \int \! d^{2}x\p\! \int \! d^{2}z\p \,
   I^{A}_{\mu \nu }(x\p)\Tr\{U(x\p+z\p)\Ud(z\p)\} ,
\label{2.34}
\end{eqnarray}
where the impact factor in coordinate representation has the form:
\begin{eqnarray}
   I^{A}_{\mu \nu }(x\p) &\equiv & \int \! \frac {d^{2}p\p}{4\pi ^{2}}
   e^{i(p,x)\p} I^{A}_{\mu \nu }(p\p)
\nonumber\\
   &=&
   \int _{0}^{1} \frac {d\al d\al'}{4\pi \al'\bar{\al}'}
   \sqrt {-\frac {p_{A}^{2}\al\bar{\al}}{x\p^{2}\al'\bar{\al}'}}
   \left\{ -2g_{\mu \nu }(1-2\al\bar{\al})(1-2\al'\bar{\al}')
      K_{1}\left( \sqrt {-p_{A}^{2}x\p^{2}
\frac {\al\bar{\al}}{\al'\bar{\al}'}} \right)
   \right.
\nonumber\\
&& \hspace*{0.2in}
   + \sqrt {-p_{A}^{2}x\p^{2}\frac {\al\bar{\al}}{\al'\bar{\al}'}}
   \left[
      g_{\mu \nu } (1-2\al\bar{\al}-2\al'\bar{\al}'+
8\al\bar{\al}\al'\bar{\al}')
      + 8\al\bar{\al}\al'\bar{\al}' \frac {x_{\mu }x_{\nu }}{x\p^{2}}
   \right]
\nonumber\\
&& \hspace*{0.4in}
   \left. K_{2}\left( \sqrt {-p_{A}^{2}x\p^{2}
\frac {\al\bar{\al}}{\al'\bar{\al}'}} \right)
   \right\},
\label{2.35}
\end{eqnarray}
where $I^{A}_{\mu \nu }(p\p)\equiv I^{A}_{\mu \nu }(p\p,0)$ and
$K_{n}(z)$ is the McDonald function.

Formula (\ref{2.34}) describes a quark and antiquark moving fast
through an external gluon field. After integrating over gluon
fields (in the functional integral) we obtain the virtual photon
scattering amplitude (\ref{2.7}). It is convenient to rewrite it
in the factorized form:
\begin{equation}
   {\cal A}(p_{A},p_{B})=i{s\over 2}
   \sum e_{i}^{2}
   \int \frac {d^{2}p\p}{4\pi ^{2}}I^{A}(p\p)
   \llan\Tr\{\hat U(p\p) \hat U^{\dagger }(-p\p)\}\rran .
\label{2.36}
\end{equation}
where $I^A(p\p)=e^A_{\mu}e^A_{\nu}I^A_{\mu\nu}(p\p)$.
The gluon fields in $U$ and $U^{\dagger }$ have been promoted to
operators, a fact we signal by replacing $U$ by $\hat U$, etc.
The reduced matrix
elements of the
operator $\Tr\{ \hat U(p\p) \hat U^{\dagger }(-p\p)\}$
between the ``virtual photon states" are defined as follows:
\begin{eqnarray}
   \llan \Tr\{ \hat U(p\p) \hat U^{\dagger }(-p\p)\}\rran
   &=&
   \int \! d^{2}x\p e^{-i(px)\p}
   \llan \Tr\{ \hat U(x\p) \hat U^{\dagger }(0) \} \rran
\nonumber\\
   \llan \Tr\{ \hat U(x\p) \hat U^{\dagger }(x'\p)\}\rran
   &\equiv &-\int \! d^{4}z \delta (z\sr) \int \! d^{4}y e^{ip_{B}\cdot y}
      e^{B}_{\xi }e^{B}_{\eta }
\nonumber\\
&&\quad
   \lvac T\{ \Tr \{ \hat U(x\p) \hat U^{\dagger }(x'\p) \}
         j^{\xi }(y+z)j^{\eta }(z)\}
   \rvac .
\label{2.37}
\end{eqnarray}

It is worth noting that for a real photon our definition of the
reduced matrix element can be rewritten as
\begin{equation}
   \lan \epsilon , p_{B} |
     \Tr \{ \hat U(x\p) \hat U^{\dagger }(x'\p) \}
   |\epsilon ', p_{B}+\be p_{B}\ran
   ~=~ 2\pi \delta (\beta ) \,
   \llan \Tr\{ \hat U(x\p) \hat U^{\dagger }(x'\p) \} \rran ,
\label{2.38}
\end{equation}
where $\epsilon $ and $\epsilon '$ represent the polarizations of the photon
states. The factor $2\pi \delta (\be)$ reflects the fact that the forward
matrix element of the operator $\hat U(x\p) \hat U^{\dagger }(x'\p)$
contains an unrestricted integration along $p_{1}$. Taking the
integral over $\be$ one easily reobtains Eq.\ (\ref{2.37}).

Our expression Eq.\ (\ref{2.30}) represents the upper part of the
graph as a numerical factor times a function of the gluon field.
The result is independent of what we chose to put in as the lower
part of the Green function.  Thus
we may say that this formula is correct in the operator sense:
\begin{eqnarray}
\lefteqn{
   \int \! d^{4}x\!\int \! d^{4}z
   \delta (z\bu)e^{-i(q,z)\p} e^{ip_{A}\cdot x} T\{j_{\mu }(x+z)j_{\nu }(z)\}
}\hspace*{2in}
\nonumber\\
&=&
   \sum _{{\rm flavors}} \! e_{i}^{2} \int \frac {d^{2}p\p}{4\pi ^{2}}
   I^{A}_{\mu \nu }(p\p,q\p)\Tr\{\hat{U}(p)\hat{\Ud}(q-p)\} ,
\label{2.24}
\end{eqnarray}
where the operators $\hat{U}$ and $\hat{\Ud}$ are given by the same
formulas (\ref{U}) with the substitution of the external field $A$
by the field operator $\hat{A}$. (We continue to use the $(~\hat{}~)$
notation for the operators in order to distinguish them from the
corresponding expressions constructed from external fields.)

This formula is a bit misleading, since the derivation assumes
that the gluon field only has Fourier components that obey
$\alpha _{p} \ll 1$ and $\beta _{p} \lsim 1$.  However, we expect that Fourier
components that do not obey this condition, in particular
$\alpha _{p} \sim 1$, will effectively give higher order corrections to
the coefficient $I^{A}$.

This is the first term in an expansion in powers of $\la$
at large $\la$.  In Eq.\ (\ref{2.24}), $I^{A}_{\mu \nu }$ has the same status
as a Wilson coefficient: it is a numerical coefficient that
multiplies an operator.  However, unlike the case of the Wilson
expansion, the coefficient is not a pure ultra-violet quantity;
we plan to express it as yet another operator matrix element.

Unfortunately, matrix elements of the operators
$\Tr\{\hat{U}(p)\hat{\Ud}(q-p)\}$
ordered along the light-like line will have a longitudinal divergence
in Feynman integrals.  This is rather like the Wilson OPE where
the matrix elements of the (unrenormalized) local operators will
have UV divergence in the integrals over loop virtualities. In the
next Section we will introduce ``regularized"
eikonal-line operators $U$ and $\Ud$ which will be the analogs of
the local renormalized operators for high-energy amplitudes.

\section{Regularized Wilson-line operators}
\label{sec:Reg.Wilson}

In the previous Section we have found that the formal high-energy
limit of the virtual photon scattering amplitude is described by
a matrix element of a Wilson-line operator (\ref{U}) ordered
along a light-like line. However, a matrix elements of such an
operator has a longitudinal divergence.

We will now explain how the divergence arises and how to treat
it, with the aid of a low order example.

\subsection{General structure; divergences, subtractions}
\label{sec:Gen.struct}

Originally we had graphs of the form of Fig.\ \ref{fig:his}.  For
the present part of our argument, let us choose all transverse
momenta to be of some given order of magnitude. Call this
magnitude $m$. (Finally we will see that all the integrals over
tranverse momenta converge on scales of order of photon virtualities).
Then the operator factorization Eq.~(\ref{2.36})
applies as it stands when all the gluons have $\alpha  \ll 1$.  The
longitudinal momenta are then restricted to give unsuppressed
contributions only when they are not too big: $|\alpha \beta s| \lsim m^{2}$.
(This last statement is actually in need of proof.)
We are using a Sudakov representation for the
momenta --- Eq.~(\ref{sudm}).

When we consider the integral over all the longitudinal gluon
momenta, we can partition the graph into factors ordered from top
to bottom.  The $\alpha $'s are strongly ordered between the different
factors, with the largest values at the top.  (We will not
present the proof that configurations with strong reverse
ordering between two factors are power-law suppressed.)

\begin{figure}[htb]
\vspace{-6cm}
\mbox{
\epsfxsize=7cm
\epsfysize=14cm
\hspace{4cm}
\epsffile{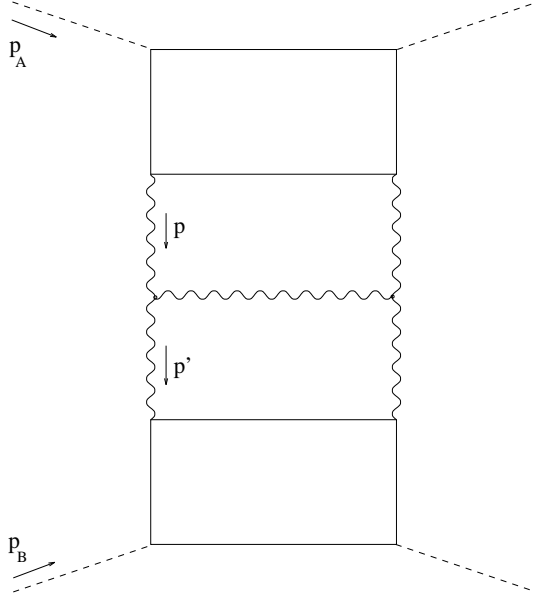}}
\vspace{1cm}
{\caption{\label{fig:1loopampl} First correction to a lowest order graph for
   high-energy scattering.}}
\end{figure}

Let us add one extra gluon rung to a lowest order graph,
so we have Fig.\ \ref{fig:1loopampl}, and let us write the graph as
\begin{equation}
   G = \int  d^{4}p \, I(p) ,
\end{equation}
where $p^{\mu }$ is the momentum flowing from one of the gluon lines
into the upper quark loop.
The procedure we explained in the previous section gives us the
asymptotics for the integrand when $m^{2}/s \lsim \alpha _{p} \ll 1$, where,
as we defined earlier, $m$ represents the typical scale of
transverse momenta.  So the factorized form is in an integral
\begin{equation}
   \int  d^{4}p \, \Asy_{\alpha _{p}\ll 1} I(p),
\end{equation}
with $\Asy I$ being of the form of the impact factor times the
integrand for a graph for
$\llan \Tr\{ \hat U(p\p) \hat U^{\dagger }(-p\p) \}\rran$.

Then we write the graph, Fig.\ \ref{fig:1loopampl}, as
\begin{equation}
   G = \int  d^{4}p \, \Asy_{\alpha _{p}\ll 1} I(p)
      + \int  d^{4}p \, \left[ I(p) - \Asy_{\alpha _{p}\ll 1} I(p) \right].
\label{R.G1}
\end{equation}
The first term is the lowest order impact factor times a
correction to the operator.  The second term has its $\alpha _{p} \ll 1$
behavior subtracted off, and so the dominant contribution to the
integral in this term is from $\alpha _{p}$ of order unity, or bigger.
This term should be treated as giving a higher order correction
to the impact factor, as we now explain.

Suppose that we have proved the factorized formula,
Eq.~(\ref{2.36}), in general.  Consider its expansion in
powers of the coupling.  The first term on the right of
Eq.~(\ref{R.G1}) is a contribution to
\begin{equation}
    \mbox{lowest order impact factor}
    \times  \mbox{next order matrix element} ,
\end{equation}
while the second term is a contribution to
\begin{equation}
    \mbox{next order impact factor}
    \times  \mbox{lowest order matrix element} .
\end{equation}

But, as we will see shortly, the integral over $\Asy_{\alpha _{p}\ll 1} I(p)$
has a divergence as $\alpha _{p}\to \infty $,
since the replacement of $I(p)$ by
$\Asy I(p)$ removes a convergence factor provided by the quark
loop.  (The approximations used to derive Eq.~(\ref{2.36}) are
only valid when $\alpha _{p}\ll 1$.)
We must therefore redefine the operator $\hat U(p\p) \hat
U^{\dagger }(-p\p)$ so that it has no divergence.  Ideally we would like to
do this by some kind of generalized renormalization procedure.
But for our discussion we will find it sufficient to change the
line along which the path ordered exponential is taken.

The structure of Eq.~(\ref{R.G1}) and the arguments that we will
need are completely analogous to those for the ordinary operator
product expansion.  However, it is important to realize that the
divergence we are concerned with is not a conventional
ultra-violet divergence.  Thus the methods used for the operator
product expansion need to be generalized.  (The operator indeed
has ultra-violet divergences, in certain graphs.  These are
associated with $p_{\perp }\to \infty $ behavior, and constitute a relatively
trivial problem.)

The decomposition of the amplitude into the impact factor times
matrix element has a very illuminating (although qualitative)
interpretation in terms of  functional intergral representation
for the amplitude.  It corresponds to the decomposition of the
functional integral into a product of two integrals - over the
(quark and gluon) fields with large light-cone fraction $\al\sim 1$
and over the fields with small $\al\sim {1\over\la}$ (which
corresponds to fields that are not scaled with $\la$). More
presicely, we choose $\si$ such as $\si\ll 1,~g^2\ln\si\ll 1$
($\si$ is independent of $\la$) and separate the functional
integration over the fields with lighcone fraction $\al$ either
greater or lesser than $\si$. First , we perform the integration
over the $\al>\si$ fields which scale with $\la$ and it yields
impact factors times the Wilson-line gauge factors constructed
from "external" fields with small $\al<\si$ and on the second
step the remaining integral over these
small-$\!\al$ fields will  give us the matrix elements of the
Wilson-line operators. (In the leading logarithmic approximation
these Wilson-line operators still correspond to the
slope $\parallel~p_A$ since we make no difference between $\ln{s\over m^2}$
and $\ln{s\si\over m^2}$). So, the impact factor is given by a functional
integral over the $\al\sim 1$ fields in the external small-$\al$ fields
which technically is a series of diagrams in the external field. The
leading-order impact factor calculated in Sect.2 is the simplest of such
diagrams. In the next order in coupling constant we will have more
complicated diagrams with large-$\al$ gluon fields as shown in Fig.5.
Unfortunately, there is no consistent quantitative decomposition of
the functional ontegral into product of $\al>\si$ and $\al<\si$ integrals
which goes beyond leading logarithmic approximation. So, at this point we
are forced to return to the original logic of the operator expansion and
define the impact factor as the coefficient function in front of
the (Wilson-line) operator by comparing the matrix elements of the
T-product of two currents and of the Wilson-line operators. If we
knew that the expansion goes in terms of Wilson lines beforehand
and our purpose was just to calculate the coeficients, it would be enough
to compare these matrix elements
between two (or four) real gluons. But since we want to prove that the
gluon operators assemble in Wilson lines we must compare these matrix
elements between an arbitrary number of real gluons - i.e., in external
gluon field. So, again the impact factors are given by the diagrams in
the external gluon field but the interpretation now is different - the
external gluon field is a convenient way to represent many-gluon states
between which we must take  the operator expansion in order to determine
the coefficient functions (impact factors). Maybe if the correct
gauge-invariant way to separate functional integrations over large and
small distances will appear some day it would possibly make the two
interpretations of the same diagrams in external fields equivalent.

\subsection{Loop corrections to Wilson line operators}

\begin{figure}[htb]
\vspace{-8cm}
\mbox{
\epsfxsize=7cm
\epsfysize=14cm
\hspace{4cm}
\epsffile{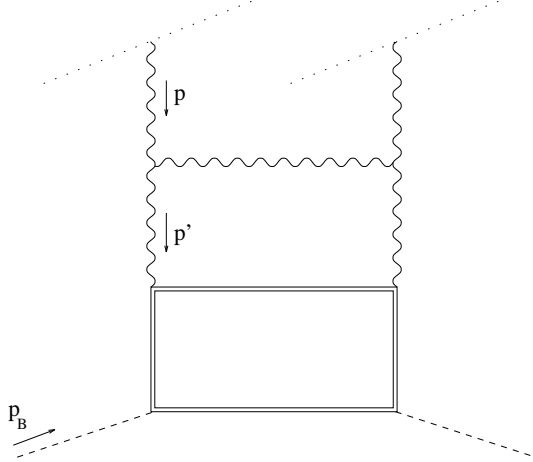}}
\vspace{1cm}
{\caption{\label{fig:mael}  Typical diagram for the matrix element of the
operator
         $\Tr\{\hat{U}\hat{\Ud}\}$ between ``virtual-photon"
         states.}}
\end{figure}

Consider the example of the one-rung ladder
diagram shown in Fig.\ \ref{fig:mael}.

\begin{figure}[htb]
\vspace{-9cm}
\mbox{
\epsfxsize=15cm
\epsfysize=12cm
\hspace{0cm}
\epsffile{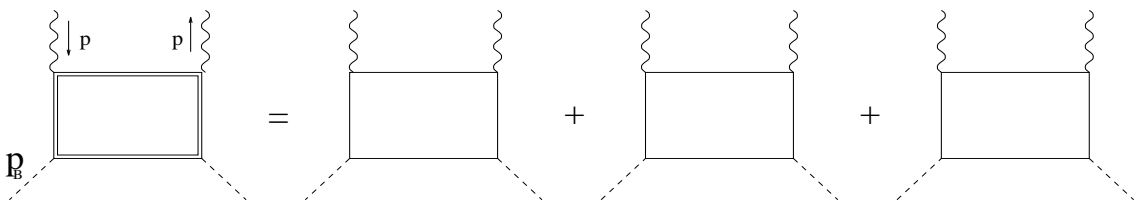}}
\vspace{1cm}
{\caption{\label{fig:bulb}  Quark bulb.}}
\end{figure}

The corresponding contribution to the matrix element
$\llan \Tr\{\Uh(p\p)\Uhd(-p\p)\}\rran$ has the form:
\begin{eqnarray}
\lefteqn{
   -{i\over 2}g^{6}
   \int \!\frac {d\al_{p}}{2\pi }  \frac {d\al'_{p}}{2\pi }
   \frac { d\be'_{p}}{2\pi } \frac {d^{2}p'\p}{4\pi ^{2}}
}\hspace*{1.2in}
\nonumber\\
&&
   \frac {{4\over  s^{2}}\Gamma \sbu{}^{\si}(p,-p')
\Gamma \sbus(p,-p')}{p\p^{4}(\al'_{p}\be'_{p}s-p^{'2}\p+\ie)^{2} \,
[-(\al_p-\al_p')\be'_{p}s - (p-p')\p^{2}+\ie]}
   \Phi ^{B}(p') ,
\label{3.1}
\end{eqnarray}
up to the trivial color factor $N_c(N_c^2-1)$.  Here the momenta are
defined in Fig.\ \ref{fig:mael},
$\Gamma _{\mu \nu \si}(p,-p',p'-p)=-(p+p')_{\si}g_{\mu \nu }+
(2p'-p)_{\mu }g_{\nu \si}+(2p-p')_{\nu }g_{\si\mu }$ is a
three-gluon vertex, and
\begin{eqnarray}
   \Phi ^{B}_{\xi \eta }(p) &=& e^{B}_{\xi }e^{B}_{\eta }
   \int \! \frac {d^{4}k}{16\pi ^{4}} \Tr{1\over\nk}\np_{1}{1\over\nk+\np}
\nonumber\\
&&
   \left[ \np_{1}{1\over\nk}\ga^{\eta }{1\over\nk-\np_{B}}
         \ga^{\xi }+\np_{1}{1\over\nk}\ga^{\xi }{1\over\nk+\np_{B}}
         \ga^{\eta }+\ga^{\eta }{1\over\nk+\np-\np_{B}}\np_{1}
         {1\over\nk-\np_{B}}\ga^{\xi }
   \right]
\label{3.2}
\end{eqnarray}
is the quark loop shown in Fig.\ \ref{fig:bulb}.
In this section we also omit for brevity the trivial
factors due to the electric charges of the quarks.

In writing Eq.~(\ref{3.1}), we have assumed that rapidity of the
gluon rung is much larger than that of the quark loop.  This
accounts for the indices on the 3-gluon vertices.
We used Feynman gauge and substituted $g_{\xi \eta }$ by ${2\over
s}p_{1\xi }p_{2\eta }$ which is valid for gluons connecting lines with very
different rapidities.

\subsection{Calculation of divergences}

It is easy to see that the integral over $\al$ in Eq.~(\ref{3.1}) is
logarithmically divergent. At first sight, the divergence
appears to be linear, since
\begin{equation}
   {4\over s^{2}}\Gamma \sbu{}^{\si}(p,-p')\Gamma \sbus(p,-p')
   =(\al_{p}-2\al'_{p})\be'_{p}s-(p+p')^{2}\p ,
\label{3.3}
\end{equation}
but careful analysis carried below shows that $\be'_{p}$ is
$\sim {1\over \al_{p}}$ at large $\al_{p}$.) The only gauge-invariant way
to regularize this divergence which we have found is to change
slightly the slope of the supporting line (as it was done in
ref.\cite{slope} for the case of Sudakov formfactor). We define:
\begin{equation}
   \hat{\Us} (x\p )= [\iy p\ze+x\p ,-\iy p\ze +x\p ]~ ,
   ~\hat{\Uds} (x\p)=[-\iy p\ze+x\p,\iy p\ze +x\p]
\label{3.4}
\end{equation}
where
\begin{equation}
   p\ze~=~ p_{1}~+~\zeta p_{2}
\label{3.5}
\end{equation}
so that at $\zeta \ll 1$ the operators (\ref{3.4}) are ordered
along a slightly non-light-like line.

Now let us demonstrate
on our example that changing of
the slope of the line according to Eq.\ (\ref{3.5}) does regularize the
longitudinal
divergence in \mael s of the operators $\hat{U}$.
It is easy to see that the changing of
the slope of the line according to Eq.\ (\ref{3.5}) leads to
the substitution $p=\al p_{1}+p\p\rightarrow p=\al p_{1}-\zeta \al p_{2}+p\p$
in the diagram in Fig.\ \ref{fig:mael}. Therefore, we obtain the
contribution  of this diagram to the \mael\ of the operator $\llan
\Tr\Uhz(p\p)\Uhdz(q\p-p\p)\rran$ in the following form:
\begin{eqnarray}
\lefteqn{
   -{i\over 2}g^{6}\int \! \frac {d\al_{p}}{2\pi }
\frac {d^{4}p'\p}{16\pi ^{4}}
}\hspace*{0.5in}
\nonumber\\
&&
  \frac {[(\al_{p}-2\al'_{p})\be'_{p}s-(p+p')^{2}\p]
\Phi ^{B}(p')}{(\zeta \al_{p}^{2}s+p\p^{2}-\ie)^{2} \, (\al'_{p}\be'_{p}s-
{p'}^{2}\p+\ie)^{2} \, [-(\al_{p}-\al')(\al_{p}\zeta +\be'_{p})s-
(p-p')\p^{2}+\ie]}
\label{3.6}
\end{eqnarray}

As we shall see below, the logarithmic contribution comes from the region
$\sqrt {m^{2}\over\zeta s} \gg \alpha _{p} \gg \alpha '_{p} \sim \ms$,
$1 \gg \beta _{p}' \gg \beta _{p} =
-\zeta \alpha _{p} \sim \sqrt {m^{2}\zeta \over s}$.
In this region one can perform the integration over $\be'_{p}$  by taking
the residue at the pole
$\left[ -(\al_{p}-\al')(\al_{p}\zeta +\be'_{p})s-(p-p')\p^{2}+
\ie \right]^{-1}$,
and the result is
\footnote
   {In the region we are investigating, we can neglect
   the $\beta '_{p}$ dependence  of the lower quark loop.}
\begin{eqnarray}
\lefteqn{
   \frac {g^{6}}{s} \int  \frac {d\alpha _{p}}{2\pi }
\frac {d\alpha '_{p}}{2\pi }
   \int \frac {d^{2}p'\p}{4\pi ^{2}}
   \left[ \Theta (\alpha _{p}>\alpha '_{p}>0 + \Theta (0>\alpha '_{p}>
\alpha _{p}) \right]
}\hspace*{1.1in}
\nonumber\\
&&
\frac {\left( p\p^{2}+ {p'}\p^{2} -\alpha _{p}^{2}\zeta s/2 \right) \,
\Phi ^{B}\left(\alpha _{p}' p_{1}-(\alpha _{p}\zeta +
{(p-p')\p^{2}\over \al_{p} s})p_{2}+p'\p \right)}{|\al_{p}-\al'_{p}| \,
(\zeta \al_{p}^{2}s+p\p^{2}-\ie)^{2} \,
[\frac {\alpha '_{p}}{\alpha _{p}}(p-p')\p^{2}+p^{'2}\p+\ie]^{2}}
\label{3.7}
\end{eqnarray}
Here we have used the approximation that $\alpha _{p} \gg \alpha '_{p}$:
The component $\be_{k}$ along the $p_{2}$ vector in the quark bulb is $\sim 1$
(similar to the case of upper quark bulb where the component $\al_{k}$
along the
vector $p_{1}$ is $\sim 1$, see eq. (\ref{2.31})).
Therefore, $\al'_{p} \sim \ms$ and
we see now that our integral over $\al_{p}$
in the region $\sqrt {m^{2}\over\zeta s} \gg \al_{p} \gg \al'_{p} \sim \ms$
is indeed logarithmic. The lower limit of logarithmical
integration is provided by the matrix
element itself (since $\be_{k}\sim 1$ in the lower quark bulb) while
the upper limit, at $\alpha _{p}^{2} \sim m^{2}/\zeta s$
is enforced by the non-zero $\zeta $ and the result has the form
\begin{eqnarray}
   \llan Tr\Uhz(p\p)\Uhdz(q\p-p\p)\rran_{{\rm Fig.~\ref{fig:mael}}}
   &=&
   {g^{6}\over 8\pi }
   \ln \left( {s\over m^{2}\zeta } \right)
   \int \!\frac {d^{2}p'\p}{4\pi ^{2}} \frac {p\p^{2} +
{p'}\p^{2}}{p\p^{4} {p'}\p^{4}}
   I^{B}(p'\p)
\label{3.8}
\end{eqnarray}
where
\begin{equation}
I^{B}(p'\p)={2\over s}\int \!\frac {d\al'_{p}}{2\pi }
\Phi ^{B}(\al'_{p} p_{1}+p'\p)
\label{3.9}
\end{equation}
is the impact factor for the lower quark bulb.
It is easy to
demonstrate that $I^{B}$ can be reduced to the double-integral form
(\ref{2.32})
(with the trivial change $p_{A}\rightarrow p_{B}$).

Let us compare now the matrix element (\ref{3.8}) with the
corresponding contribution to physical amplitude shown in
Fig.\ \ref{fig:1loopampl} which has the form:
\begin{eqnarray}
\lefteqn{
   {g^{6}\over 2} \int \!   \frac {d^{4}p\p}{16\pi ^{4}}
\frac {d^{4}p'\p}{16\pi ^{4}}
}\quad
\nonumber\\
&&
\frac {\Phi ^{A}_{\mu \nu }(p)[(\al_{p}-2\al'_{p})\be'_{p}s-(p+p')^{2}\p]
\Phi ^{B}_{\xi \eta }(p')}{(\al_{p}\be_{p}s-p\p^{2}+\ie)^{2}
(\al'_{p}\be'_{p}s-p^{'2}\p+\ie)^{2}[(\al_{p}-\al'_{p})(\be_{p}-
\be'_{p})s- (p-p')^{2}+\ie]}
,
\label{3.10}
\end{eqnarray}
where the upper quark bulb $\Phi ^{A}_{\mu \nu }$ is the same
as in Sec.\\ref{sec:Gen.struct}b.
This integral is rather similar to the one for the matrix element
of the operator, except that there is now a factor of the upper
quark bulb, and there are integrals over $\beta _{p}$ and $p_{\perp }$.

The previous arguments show that the logarithmic contribution
comes from the region $\al_{p}\gg\al'_{p}\sim\ms$,
$\ms\sim\be\ll\be'\ll 1$, and now the upper limit of the
logarithmic integral is set not by the regularized path-ordered
exponential, but by the upper quark bulb, at $\alpha _{p} \sim 1$.  Hence
we have
\begin{eqnarray}
\mbox{L.h.s. of Eq.~(\protect\ref{3.10})}
&\sim&
   i{g^{6}\over 4\pi } \ln \left( {s\over m^{2}} \right)
   \int \! \frac {d^{2}p\p}{4\pi ^{2}} \frac {d^{2}p'\p}{4\pi ^{2}}
   \frac {p\p^{2}+{p'}^{2}\p}{p\p^{4} {p'}^{4}\p} I^{A}(p\p) I^{B}(p'\p).
\label{3.11}
\end{eqnarray}
This agrees with the with estimate Eq.~(\ref{3.8}), if we set
$\zeta  = {p_{A}^{2}\over s}$.  This corresponds to making the line in the
path-ordered exponential have a finite rapidity relative to the
photon.

Thus, a more correct version of the factorization formula
Eq.\ (\ref{2.36}) or Eq.~(\ref{2.24}) has the operators
$\hat{U}$ and $\hat{\Ud}$ ``regularized'' at $\zeta  \sim {p_{A}^{2}\over s}$:
\begin{eqnarray}
\lefteqn{
   \int  d^{4}x \int  d^{4}z \,
   \delta (z\bu) e^{ip_{A}\cdot x} T\{j_{\mu }(x+z)j_{\nu }(z)\}
}\qquad
\nonumber\\
   &=&
   \sum _{i} e_{i}^{2}  \int  \frac {d^{2}p\p}{4\pi ^{2}}
   I^{A}_{\mu \nu }(p\p)\Tr\{\Uh^{\zeta ={m^{2}\over s}}(p)
   \Uh^{\dagger \zeta ={m^{2}\over s}}(-p)\}
\nonumber\\
   &&+ \mbox{term with higher order impact factors}.
\label{3.12}
\end{eqnarray}

\subsection{Next-to-leading order}

Let us outline the situation in the next-to-leading order in coupling constant.
At the level of the $O(g^{6})$ calculation that we are examining,
our results so far show that this formula captures all of the
contributions from the original graphs (as $s\to \infty $) except for
those from $\alpha _{p} \sim 1$.  As indicated earlier,
in Sect.\
\ref{sec:Gen.struct}, these are included if we define the $O(g^{4})$
impact factor suitably.  The reason why we have indicated the
higher order impact factor separately in Eq.~(\ref{3.12}) is that
it is more than a trivial higher order correction to the lowest
order impact factor $I^{A}_{\mu \nu }$, as we will now show.

Following the strategy indicated by Eq.~(\ref{R.G1}), we examine
the difference
\begin{eqnarray}
   &&\int  d^{4}x \int  d^{4}z \,
   \delta (z\bu) e^{ip_{A}\cdot x}
   T\{j_{\mu }(x+z)j_{\nu }(z)\}
\nonumber\\
   && -
   \sum _{i} e_{i}^{2} \int  \frac {d^{2}p\p}{4\pi ^{2}}
   I^{A}_{\mu \nu }(p\p)
   \Tr \left\{
      \Uh^{\zeta ={m^{2}\over s}}(p)
      \Uh^{\dagger \zeta ={m^{2}\over s}}(-p)
   \right\}
\label{3.13}
\end{eqnarray}
in an external field. We are assuming a calculation to $O(g^{4})$ in
Eq.~(\ref{3.13}).

\begin{figure}[htb]
\vspace{-6cm}
\mbox{
\epsfxsize=14cm
\epsfysize=10cm
\hspace{0cm}
\epsffile{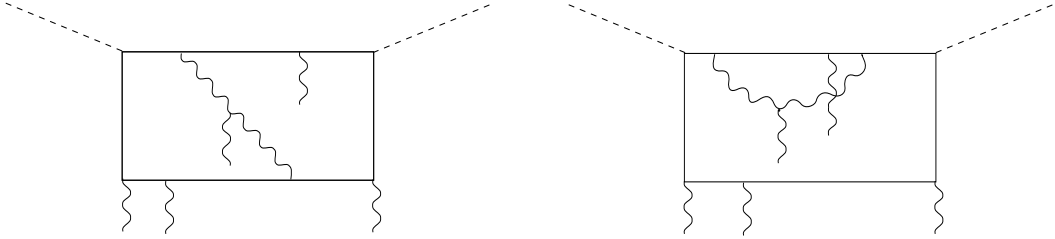}}
\vspace{0cm}
{\caption{\label{fig:coef} Impact factor in next-to-leading order in
$\al_{s}$.}}
\end{figure}

Typical diagrams are shown in Fig.\ \ref{fig:coef}. From the
shock-wave picture of external field (see Appendix
\ref{app:ExtField}) it is clear
that the general form of the answer for Eq.~(\ref{3.13}) is
\begin{eqnarray}
   &&
   g^2\int  d^{2}x\p d^{2}y\p d^{2}z\p
   J_{1}^{A}(x\p,y\p,z\p) \Tr \{t^a U(x\p+z\p)t^b \Ud(z\p) \}
                [U(y\p+z\p)]_{ab}
\nonumber\\
   &&+
   g^2\int  d^{2}x\p d^{2}z\p J_{2}^{A}(x\p) \Tr \{ U(x\p+z\p) \Ud(y\p) \} .
\label{3.14}
\end{eqnarray}
where $[U]_{ab}$ is the Wilson-line gauge factor (\ref{2.25}) in the
adjoint (gluon) representation. The first term corresponds to the case when
the shock wave hits two
quarks and a gluon and the second to when it only hits two quarks.
Without the subtraction term in Eq.~(\ref{3.13}),
the diagrams in Fig.\ \ref{fig:coef} would diverge logarithmically at
small $\al_{p}$.
But after the subtraction the result will converge; the integral
(to leading power) will be dominated by $\al\sim 1$.

Thus we obtain the coefficient functions (impact factors) $J_{1}$ and $J_{2}$.
So, the operator expansion up to the next-to-leading term has the form
\begin{eqnarray}
\lefteqn{
  \int  d^{4}x \int  d^{4}z \delta (z\bu) e^{ip_{A}\cdot x}
   \lan T\{j_{\mu }(x+z)j_{\nu }(z)\}\ran_{A}
}
\hspace*{1in}
\nonumber\\
   &=&
   \sum e_{i}^{2} \int  d^{2}x\p \int  d^{2}z\p
   I^{A}_{\mu \nu }(x\p) \Tr\{ \Uhz(x\p+z\p) \Uhdz(z\p) \}
\nonumber\\
   &+&
   g^2\int  d^{2}x\p d^{2}y\p d^{2}z\p
   J_{1}^{A}(x\p,y\p) \Tr\{t^a \Uhz(x\p+z\p)t^b \Uhdz(z\p) \}
   [ \Uhz(y\p+z\p)]_{ab}
\nonumber\\
   &&+
   g^2\int  d^{2}x\p d^{2}z\p
      J_{2}^{A}(x\p) \Tr\{ \Uhz(x\p+z\p) \Uhdz(y\p) \}
\nonumber\\
   && + O(g^{4}) ,
\label{3.15}
\end{eqnarray}
where the operators $U$ in the $O(g^{2})$ term must be also regularized at
$\zeta ={p_{A}^{2}\over s}$ in order to simulate a proper cutoff
for the logarithms
$\sim g^{4} \ln{s\over m^{2}}$ as well. In principle, this procedure may
be repeated
many times yielding the coefficient functions (impact factors) in any given
order of perturbation theory just as for the usual Wilson expansion.

It is worth noting that the above procedure of separating the Feynman integrals
into the contributions coming from large and small components of the momentum
$p_{A}$ can be repeated for the bottom part of the diagram with the result
being
the separation of loop integrals into contributions of large and small
components along the $p_{B}$. In the leading order in $\alpha _{s}$ the result
will
have the same form as Eq.\ (\ref{3.12}):
\begin{eqnarray}
\lefteqn{
  \int  d^{4}x \int  d^{4}z
  \, \delta (z\sr) e^{-i(q,z)\p} e^{ip_{B}\cdot x}T\{j_{\mu }(x+z)j_{\nu }(z)\}
}\hspace*{1in}
\nonumber\\
  &=&
   \sum e_{i}^{2}\int \frac {d^{2}p\p}{4\pi ^{2}}
   I^{B}_{\mu \nu }(p\p,q\p)\Tr\{\hat{U'}(p)\hat{U'^{\dagger }}(q-p)\} .
\label{3.16}
\end{eqnarray}
Here
\begin{equation}
   \hat{U'}(x\p )= [\infty p_{B}+x\p ,-\infty p_{B} +x\p ]~ ,
   {}~\hat{U'^{\dagger }}(x\p)=[-\infty p_{B}+x\p,\infty p_{B} +x\p]
\label{3.17}
\end{equation}
are gauge factors ordered along a straight line approximately
in the direction of
motion of the lower quarks ($p_{B}$) and the impact factor will be given by the
same expression (\ref{2.32}) save the trivial change $p_{A}^{2}\leftrightarrow
p_{B}^{2}$.
Therefore, the amplitude of scattering of virtual photons at high energy
(\ref{defA}) can be represented as a product of two impact factors times the
vacuum expectation value of four Wilson line operators representing the gluon
ladders (see Fig.~\ref{fig:Ladder}):
\begin{figure}[htb]
\vspace{-4cm}
\mbox{
\epsfxsize=14cm
\epsfysize=12cm
\hspace{0cm}
\epsffile{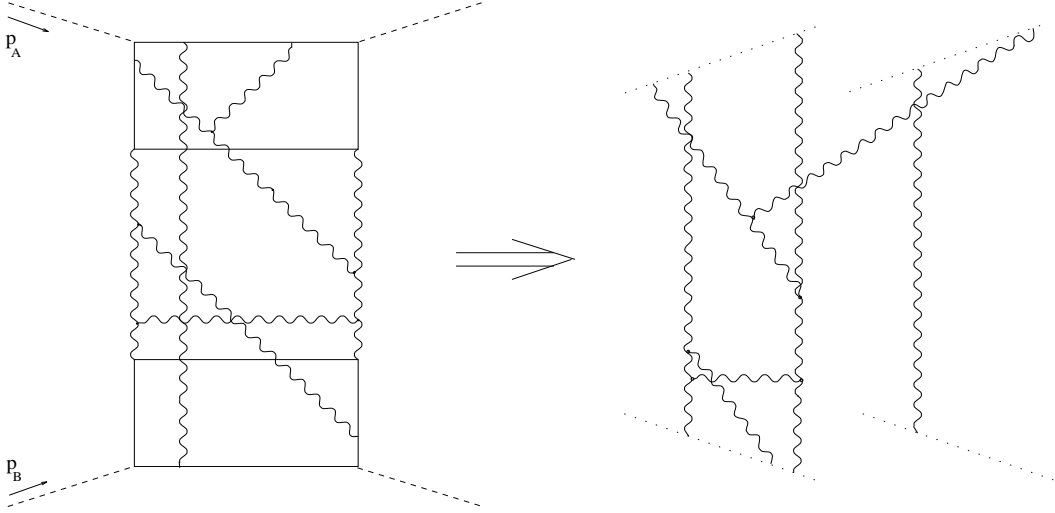}}
\vspace{1cm}
{\caption{\label{fig:Ladder} Factorization of gluon ladders from
the high-energy amplitude of virtual photon scattering.}}
\end{figure}

\begin{eqnarray}
\lefteqn{
   4\pi ^{2} \delta ^{(2)}(q\p)A(p_{A},p_{B},q)
   =-i{s\over 2}(\sum e_{i}^{2})^{2}
   \int \frac {d^{2}p\p}{4\pi ^{2}} \frac {d^{2}p'\p}{4\pi ^{2}}
   I^{A}(p\p,q\p)I^{B}(p'\p,0)
}\hspace*{2in}
\nonumber\\
&&
   \lvac \Tr\{\hat{U}(p\p)\hat{\Ud}(q\p-p\p)\}
         \Tr\{\hat{U'}(p'\p) \hat{U'^{\dagger }}(-p'\p)\}
    \rvac .
\end{eqnarray}
The operators $\hat{U}$ are ``normalized" in such a way that they are ordered
along
the slightly non-light-like line collinear to $p_{A}$ and $\hat{U'}$
along a line collinear to $p_{B}$.

\section{One-loop evolution of eikonal-line operators}
\label{sec:OneLoopEvol}

In previous sections we demonstrated that at large energies the
scattering amplitude of virtual photons can be reduced to the
matrix element (between the "virtual photon" states, see
definition (\ref{2.24})) of the two Wilson-line operators defined on a
near-light-like line collinear to $p_{A}$. Now we must study the
dependence of these matrix elements on energy which reveals
itself through the dependence on the slope of the supporting line.
 So, we must find\footnote{In this Section we omit the trivial
gauge end factors (\ref{2.31}) which will be restored in next Section}
\begin{equation}
\la{\partial \over\partial \la}\Tr\{\Uh(x\p)\Uhd(y\p)\}=
-2\zeta {\partial \over\partial \zeta }\Tr\{\Uh(x\p)\Uhd(y\p)\}
\label{4.1}
\end{equation}
at large $\la$. The operators $\Uh$ and $\Uhd$ are defined on the
lines collinear to $p\ze\equiv p_{1}+\zeta p_{2}$ where
$\zeta ={p_{A}^{2}\over s}$ is a small parameter which determines
the
deviation of the supporting line from the light cone. This derivative can be
expressed as
\begin{eqnarray}
&\zeta {\partial \over\partial \zeta }\Tr\{\Uh(x\p)\Uhd(y\p)\}=\nonumber\\
&ig\zeta \int \! udu
\left(\Tr\{[\iy,u]_xF\sbu (up\ze+x\p)[u,-\iy]_x\Uhd(y\p)\}\right.\nonumber\\
&\left.-\Tr\{\Uh(x\p)ig\zeta \int \! udu [-\iy,u]_yF\sbu
(up\ze+y\p)[u,\iy]_y\}\right)
\label{4.2}
\end{eqnarray}
where $[u,v]_x\equiv [up\ze+x\p,vp\ze+x\p]$. So,the derivative of the
two Wilson-line operator has reduced to a more complicated operator and
therefore, in general, we can extract no information on the behavior
with respect to $\la$. But in the case of large $\la$ (small $\zeta$)
we can expand the complicated operator in r.h.s.  of eq. (\ref{4.2})
in inverse powers of $\la$ as it was done for the T-product of quark
currents in the previous Section and we will show in this Section that
the result will have a similar form
\begin{eqnarray}
&ig\zeta \int \! udu\nonumber\\
&\left(\Tr\{[\iy, u]_x\hat{F}\sbu (up\ze+x\p)[u,-\iy]_x\Uhd(y\p)\}-
\Tr\{\Uh(x\p)[-\iy,u]_y\hat{F}\sbu
(up\ze+y\p)[u,\iy]_y\}\right)=\nonumber\\
&-{g^{2}\over 16\pi ^{3}}
\int \! dz\p\nonumber\\
&\left(\Tr\{\Uhz(x\p)\Uhdz(z\p)\}
\Tr\{\Uhz(z\p)\Uhdz(y\p)\}-
N_{c}\Tr\{\Uhz(x\p)\Uhdz(y\p)\}\right)\frac {
(x\p-y\p)^{2}}{(x\p-z\p)^{2}(z\p-y\p)^{2}}\nonumber\\
&~+~O(g^2)~+~O(\zeta)
\label{4.2a}
\end{eqnarray}
So, at large energies ($\la$) the derivative of the Wilson-line operator
do reduce to another operator constructed from Wilson lines. Unfortunately,
 the number of Wilson lines is not conserved since the equation is
non-linear. However, we shall see in Sect. 5 that in some important
cases the eq. (\ref{4.2a}) reduces to linear BFKL equation. In the
rest of this Section we shall derive this equation which is one of
the main results of this paper.

In order to establish the operator equation (\ref{4.2a}) let us
compare matrix elements of the l.h.s and r.h.s. If we knew that
the expansion goes in terms of Wilson lines beforehand, it would
be enough to compare these matrix elements
between two (or four) real gluons. But since we want to prove
that the gluon operators in r.h.s of eq. (\ref{4.2a}) assemble
in Wilson lines we must compare these matrix elements between
an arbitrary number of real gluons - i.e., in external gluon
field (see the discussion in Sect. 3). So, we must find the \mael\
of the operator in l.h.s. of eq. (\ref{4.2a}) in the external gluon
field at large $\la$ (small $\zeta$):
\begin{eqnarray}
&ig\zeta \int \! udu
\lan\Tr\{[\iy,u]_x\hat{F}\sbu (up\ze+x\p)[u,-\iy]_x\Uhd(y\p)\}-\nonumber\\
&\Tr\{\Uh(x\p)ig\zeta \int \! udu [-\iy,u]_y\hat{F}\sbu
(up\ze+y\p)[u,\iy]_y\}\ran_A
\label{4.2b}
\end{eqnarray}
 In the lowest order in coupling constant we obtain zero since
$\zeta\rightarrow 0$ and the external field is independent of $\zeta$.
But already in the first order in $\al_s$ the limit of the matrix
element (\ref{4.2b}) is nonvanishing due to the longutudinal
divergencies. (As we shown in previous Section, some of the
contributions to the matrix element of the operator
$TrU(x\p)\Ud(y\p)$ contain $\ln\zeta $ which means that the derivative
is $\sim {1\over\zeta}$ so the r.h.s. of eq. (\ref{4.2}) is actually
nonvanishing
at $\zeta \rightarrow 0$). Let us calculate the matrix element
(\ref{4.2b}) in the one-loop approximation.
It is convenient to use the light-like gauge $A\sr=0$
with the vector $n$ directed along $p_{B}$ (although all the calculations can
be repeated in the background-Feynman gauge with the same results, since we
have checked that the contributions due to gauge terms $\sim n_{\mu }$
in the propagator in the axial
gauge cancel).

\subsection{Calculation of the diagram in Fig. 10a}

\begin{figure}[htb]
\vspace{0cm}
\mbox{
\epsfxsize=15cm
\epsfysize=3cm
\hspace{0cm}
\epsffile{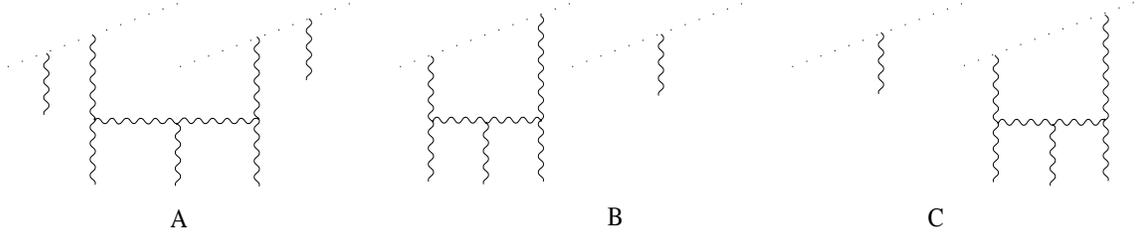}}
\vspace{0cm}
{\caption{\label{fig:1loop} One-loop diagrams for the evolution of the
   two-Wilson-line operator.}}
\end{figure}

In the first
order in $\alpha _{s}$ there are two one-loop diagrams for the matrix
element of
operator (\ref{4.2}) in external field (see Fig. \ref{fig:1loop}).

 We shall start with the diagram shown in Fig. \ref{fig:1loop}a. The
calculation is quite similar to the calculation of impact factor considered in
previous Section. For the sake of future applications we shall calculate
the derivative
$\zeta \frac {\partial }{\partial \zeta }\Uh(x\p)\otimes\Uhd(y\p)$
where
$\Uh(x\p)\otimes\Uhd(y\p)\equiv \{\Uh(x\p)\}^{i}_{j}\{\Uhd(y\p)\}^{k}_{l}$
is a product of Wilson-line operators with the non-convoluted color indices.
First, using the expression for the axial-gauge gluon propagator in the
external field from Appendix C we obtain\footnote{It can be
demonstrated that further terms in
expansion in
powers of gluon propagator (\ref{c5}) beyond those given in eq. (\ref{c6})
do not contribute in
the limit $\zeta \rightarrow 0$}:
\begin{eqnarray}
\lefteqn{ig\zeta \int \! udu
\lan[\iy,u]_x\hat{F}\sbu (up\ze+x\p)[u,-\iy]_x
\otimes\Uhd(y\p)\ran_A\nonumber}\\
&=&g\zeta \int \! udu \int\! du\nonumber\\
&&\left(\Theta(u-u')
[\iy,u]_xF\sbu (up\ze+x\p)[u,u']_xt^a[u',-\iy]_x+\right.\nonumber\\
&&\left.\Theta(u'-u)
[\iy,u']_xt^a[u',u]_xF\sbu (up\ze+x\p)[u,-\iy]_x\right)\nonumber\\
&&\otimes\int dv [-\iy,v]_yt^{b}[v,\iy]_y\nonumber\\
&&\LBB up\ze+x\p\Big|(p\ze_{\xi }-{\cal
P}\bu\frac {p_{2\xi }}{p\cdot p_{2}}){\cal O}^{\xi\eta}(p\ze_{\eta }
-\frac {p_{2\eta }}{p\cdot p_{2}}{\cal
P}\bu)\Big| vp\ze+y\p\RBB_{ab}\nonumber\\
&+&g\zeta\int du [\iy,u]_xt^{a}[u,-\iy]_x\otimes
\int dv [-\iy,v]_yt^{b}[v,\iy]_y\nonumber\\
&&
\LBB up\ze+x\p\Big|p\sr(p\ze_{\xi }-{\cal
P}\bu\frac {p_{2\xi }}{p\cdot p_{2}}){\cal O}^{\xi\eta}(p\ze_{\eta }
-\frac {p_{2\eta }}{p\cdot p_{2}}{\cal
P}\bu)\Big| vp\ze+y\p\RBB_{ab}
\label{4.3}
\end{eqnarray}
where the operator ${\cal O}$ has the form:
\begin{equation}
{\cal O}_{\mu \nu }= 4\frac {1}{{\cal P}^{2}}F^{\xi }_{\mu }\frac { 1}{{\cal
P}^{2}}F_{\xi \nu }\frac {1}{{\cal P}^{2}}-\frac {1}{{\cal
P}^{2}}(D^{\al}F_{\al\mu }\frac {p_{2\nu }}{p\cdot p_{2}}
+\frac {p_{2\mu }}{p\cdot p_{2}}
D^{\al}F_{\al\nu }-\frac {p_{2\mu }}{2p\cdot p_{2}}{\cal
P}^{\be}D^{\al}F_{\al\be}\frac {p_{2\nu }}{2p\cdot p_{2}})\frac {1}{{\cal
P}^{2}}
\label{4.4}
\end{equation}
(With our accuracy, multiplication by $p\ze$ coincides with
multiplication by $p_{1}$). We may drop the
terms
proportional to ${\cal P}\bu$ in the parenthesis since they lead to the terms
proportional to the integrals of total derivatives, namely
\begin{eqnarray}
&\int du [\iy,u]t^{a}[u,-\iy]p\ze_{\mu }(D^{\mu }\Phi
(up\ze,...))_{ab}=\nonumber\\
&\int du\frac {d}{du}\{[\iy,u]t^{a}[u,-\iy](\Phi (up\ze,...))_{ab}\}=0
\label{4.5}
\end{eqnarray}
and similarly for the total derivative with respect to $v$. Therefore we may
rewrite eq.(\ref{4.3}) as:
\begin{eqnarray}
\lefteqn{ig\zeta \int \! udu
\lan[\iy,u]_x\hat{F}\sbu (up\ze+x\p)[u,-\iy]_x
\otimes\Uhd(y\p)\ran_A}\nonumber\\
&=&g\zeta \int \! udu \int\! du'\nonumber\\
&&\left(\Theta(u-u')
[\iy,u]_xF\sbu (up\ze+x\p)[u,u']_xt^a[u',-\iy]_x\right.\nonumber\\
&+&\left.\Theta(u'-u)
[\iy,u']_xt^a[u',u]_xF\sbu (up\ze+x\p)[u,-\iy]_x\right)\nonumber\\
&&\otimes\int dv [-\iy,v]_yt^{b}[v,\iy]_y\nonumber\\
&&\LBB up\ze+x\p\Big|
{\cal O}\bubu\Big| vp\ze+y\p\RBB_{ab}\nonumber\\
&+&g\zeta\int du [\iy,u]_xt^{a}[u,-\iy]_x
\otimes\int dv [-\iy,v]_yt^{b}[v,\iy]_y\nonumber\\
&&\LBB up\ze+x\p\Big|p\sr{\cal O}\bubu\Big| vp\ze+y\p\RBB_{ab}
\label{4.6}
\end{eqnarray}

Now let us consider
the limit $\zeta \rightarrow 0$.  The Wilson lines made from external
fields are regular in this limit and the only singularity which can compensate
$\zeta $ in the numerator of l.h.s of eq. (\ref{4.6}) is ${1\over\zeta}$
coming from the differentiation
the gluon propagator in the external field which contains terms
$\sim\ln\zeta$. Therefore only the last term in eq. (\ref{4.6})
gives the nonvanishing result .
Adding the similar contribution from the second term in r.h.s. of
eq.(\ref{4.2}) we have:
\begin{eqnarray}
\lefteqn{\zeta {\partial \over\partial \zeta }
\lan\Uh(x\p)\Uhd(y\p)\ran_{A}}\hspace{1in}\nonumber\\
&=&
-g^{2}\int \! du [\iy,u]_xt^{a}[u,-\iy]_x\otimes
\int \! dv[-\iy,v]_yt^{b}[v,\iy]_y\nonumber\\
&&\LBB up_{A}+x\p\Big|up\sr{\cal O}\bubu-v{\cal O}\bubu
p\sr\Big| vp_{A}+y\p\RBB_{ab}
\label{4.7}
\end{eqnarray}
Let us at first neglect the gauge factors $[\iy,u]t^{a}[u,-\iy]$
and $[-\iy,v]t^{b}[v,\iy]$  - in other words, let us consider the trivial
zero-order term of expansion of these gauge factors in external field. We have
then
\begin{eqnarray}
&-ig^{2}\int \frac {d^4k}{16\pi ^{4}} \frac {d^4l}{16\pi ^{4}}
t^{a}\otimes t^{b}
(\frac {2}{s})^{2} 4\pi ^{2}\nonumber\\
&[\al_{k}\delta '(\be_{k}+\zeta \al_{k})\delta
(\be_{l}+\zeta \al_{l})+\al_{l}\delta '(\be_{l}+\zeta \al_{l})\delta
(\be_{k}+\zeta \al_{k})]e^{-i(k,x)\p+i(l,y)\p}
\LBB k\Big|{\cal O}\bubu \Big| l\RBB_{ab}
\label{4.8}
\end{eqnarray}
Note that as in the case of quark propagator we need the Green
functions integrated over the $\al$ component of external field.
The calculation of fast-moving gluon propagator in the external field
 mainly repeats the derivation of the formula
(\ref{qprop}) for the quark propagator and we will only sketch it here. In the
lowest order in expansion of the operator $\frac {1}{{\cal P}^{2}}$ in powers
of
$A_{\mu }$ in eq. (\ref{4.6}) one obtains:
\begin{eqnarray}
\lefteqn{
\LBB k\Big|{\cal O}\bubu\Big| k-p\RBB_{ab}=
}
\nonumber\\
&=&
{1\over \al_{k}}\frac
{1}{k^{2}+\ie}
\left\{-D^{\al}F_{\al\bu}(p)+
2\int\!\frac {d^4p'}{16\pi^4}F^{\xi }\bu(p')
\frac {1}{-\be'_{p}+\ie\al_{k}}
F_{\xi \bullet}(p-p')\right\}\frac {1}{(k-p)^{2}+\ie}
\label{4.9}
\end{eqnarray}
As can be seen from eq. (\ref{4.8}) our $\be_{p}$ are $\sim\zeta \al_{p}$ so
we can
neglect them in the arguments of external field.  Then the
expression in braces
is proportional to one of the operators:
\begin{eqnarray}
\lefteqn{[DF](x\p)+2i[FF](x\p)}
\nonumber\\
&=&\int \! du[\iy,u]_x
D^{\al}F_{\al\bullet }(up_{1}+x\p)[u,-\iy]_x+\nonumber\\
&&2i\int \! du\int \! dv\Theta (u-v)
[\iy,u]_x
F^{\xi }\bu(up_{1}+x\p)[u,v]_x
F_{\xi \bullet }(vp_{1}+x\p)[v,-\iy]_x
\label{4.10}
\end{eqnarray}
at $\al_{k}>0$ or
\begin{eqnarray}
\lefteqn{[DF\da](x\p)-2i[FF\da](x\p)}\nonumber\\
&=&\int \! du[-\iy,u]_x
D^{\al}F_{\al\bullet }(up_{1}+x\p)[u,\iy]_x+
\nonumber\\
&&2i\int \! du\int \! dv
\Theta (v-u) [-\iy,u]_x
F^{\xi }\bu(up_{1}+x\p)[u,v]_x
F_{\xi \bullet }(vp_{1}+x\p)[v,\iy]_x
\label{4.11}
\end{eqnarray}
at $\al_{k}<0$ in the lowest order in external field.  It can be demonstrated
that the subsequent terms of
the expansion of the operators $1/{\cal P}^{2}$ in eq.(\ref{4.6}) in
powers of external field "dress" lowest-order expressions for $[DF]$ and $[FF]$
by proper gauge factors according to eq. (\ref{4.10},\ref{4.11}) so we have
\begin{eqnarray}
&\LBB k\Big|4\frac {1}{{\cal P}^{2}}F_{\xi \bu}\frac {
1}{{\cal P}^{2}}F_{\xi \bu}\frac {1}{{\cal P}^{2}}-\frac {1}{{\cal
P}^{2}}(D_{\al}F_{\al\bu}\frac {s}{2p\cdot p_{B}}
+\frac {s\si_{1}}{2p\cdot p_{B}}
D_{\al}F_{\al\bu}-\frac {s}{2p\cdot p_{B}}{\cal
P}_{\be}D_{\al}F_{\al\be}\frac {s}{2p\cdot p_{B}})\frac {1}{{\cal
P}^{2}}\Big| k-p\RBB_{ab}~=\nonumber\\
&~\frac{2\pi i\delta(\al_p)}{\al_{k}}\frac { 1}{k^{2}+
\ie}\left(\Theta
(\al_k)\{i[DF](p\p)-2[FF](p\p)\}+
\Theta (-\al_k)\{i[DF\da](p\p)+2[FF\da](p\p)\}\right)
\frac {1}{(k-p)^{2}+\ie}\nonumber\\
&
\label{4.12}
\end{eqnarray}
It can be simplified even more if one notes that the operators in braces are in
fact the total derivatives of $U$ and $\Ud$ with respect to translations in the
perpendicular directions:
\begin{eqnarray}
&\partial \p^{2}U(x\p)\equiv \frac {\partial ^{2}}{\partial x_{i}\partial
x_{i}}U(x\p)=-i[DF](x\p)+2[FF](x\p)\nonumber\\
&\partial \p ^{2}U(x\p)\equiv \frac {\partial ^{2}}{\partial x_{i}\partial
x_{i}}\Ud(x\p)=i[DF](x\p)+2[FF](x\p)
\label{4.13}
\end{eqnarray}
(note that $\partial \p^{2}U=-\partial ^{2}U$). So, the eq. (\ref{4.12})
reduces to
the expression for the fast-moving gluon propagator in the external field in
the form:
\begin{equation}
\LBB k\Big|{\cal O}\bubu\Big| k-p\RBB=~\frac {-2\pi i\delta (\al_{p})}{\al_{k}}
\frac {\Theta (\al_{k})\partial \p^{2}U(p\p)-\Theta  (-\al_{k})\partial
\p^{2}\Ud(p\p)}{(k^{2}+\ie)[(k-p)^{2}+\ie]}
\label{4.15}
\end{equation}
This formula is valid in the region (\ref{char}) provided the $\be$ component
of the overall momentum transfer to external field is small (in our case
$\be_{p}\sim\zeta $, see eq. (\ref{4.7})). Substituting now eq.(\ref{4.15}) in
eq.
(\ref{4.7}) we have:
\begin{eqnarray}
&g^{2}t^{a}\otimes t^{b}
\int \frac {d^2k\p}{4\pi ^{2}}\frac {d^2l\p}{4\pi ^{2}}
e^{-i(k,x)\p+i(l,y)\p}
\int \frac {d\al_{k}}{2\pi } \nonumber\\
&\left(\frac {\zeta \al_{k}s}{(\zeta \al_{k}^{2}s+
k\p^{2})^{2}}(\Theta (\al)(\partial ^{2}U(k\p-l\p))-
\Theta (-\al)(\partial ^{2}\Ud(k\p-l\p))
\frac {1}{\zeta \al_{k}^{2}s+l^{2}\p}+ \right.\nonumber\\
&\left.\frac {1}{\zeta \al_{k}^{2}s+ k\p^{2}}
(\Theta (\al)(\partial ^{2}U(k\p-l\p))-\Theta (-\al)
(\partial ^{2}\Ud(k\p-l\p))
\frac {\zeta \al_{k}s}{(\zeta \al_{k}^{2}s+
l\p^{2})^{2}}\right)\nonumber\\
&={g^{2}\over 4\pi }\int \frac {dk\p}{4\pi ^{2}}
\frac {dp\p}{4\pi ^{2}}\frac {1}{k\p^{2}(k-p)\p^{2}}
[(\partial ^{2}U(p\p))_{ab}+(\partial
^{2}\Ud(p\p))_{ab}]e^{-i(k,x-y)\p-i(p,y)\p}
\label{4.16}
\end{eqnarray}
where the integral over $\al_{k}$ converges at $\al_{k}
\sim {k\p^{2}\over\zeta s}\sim 1$.

Now let us turn our attention to the omitted gauge factors
$[\iy,u]t^{a}[u,-\iy]$ and $[\iy,v]t^{b}[v,-\iy]$ in our starting
expression (\ref{4.6}). We demonstrate in
Appendix \ref{app:B} that they should be substituted
by $t^{a}[\iy,-\iy]\otimes
t^{b}[-\iy,\iy]$ or  $[\iy,-\iy]t^{a}\otimes [-\iy,\iy]t^{b}$ depending on the
sign of $\al_{k}$. (In the coordinate space it means that the transition
through the shock-wave "wall" can be before or after emission of quantum gluon
depending on the sign of $x\sr$ and $y\sr$, see Appendix \ref{app:ExtField}).
 After that our final result for the contribution
of the diagram in Fig.~\ref{fig:1loop}a reads
\begin{eqnarray}
\lefteqn{\zeta {\partial \over\partial \zeta }
\lan{\Uhz}(x\p){\Uhdz}(y\p)\ran_{A}~=}
\nonumber\\
&&
{g^{2}\over  4\pi }\left[\LBB x\p\Big|\frac {1}{p^{2}}(\partial ^{2}U)
\frac {1}{p^{2}}\Big| y\p\RBB_{ab}t^{a}U(x\p)\otimes t^{b}\Ud(y\p)
+\LBB x\p\Big|\frac {1}{p^{2}}
(\partial ^{2}\Ud)\frac {1}{p^{2}}\Big| y\p\RBB_{ab}
U(x\p)t^{a}\otimes \Ud(y\p)t^{b}\right]\nonumber\\
&
\label{4.17}
\end{eqnarray}
Using the identity
\begin{equation}
\partial ^{2}U^{(\dagger )}=-[p_{i}[p_{i},U^{(\dagger )}]]=
2p_{i}U^{(\dagger )}p_{i}-p^{2}U^{(\dagger )}-U^{(\dagger )}p^{2}
\label{4.18}
\end{equation}
it can be rewritten in the form:
\begin{eqnarray}
&\zeta {\partial \over\partial \zeta }\lan\{\Uhz(x\p)\}^{i}_{j}
\{\Uhdz(y\p)\}^{k}_{l}\ran_{A}~=\nonumber\\
&~
-{g^{2}\over 16\pi ^{3}}\int
dz\p\left[\{\Ud(z\p)U(x\p)\}^{k}_{j}\{U(z\p)\Ud(y\p)\}^{i}_{l}+
\{U(x\p)\Ud(z\p)\}^{i}_{l}\{\Ud(y\p)U(z\p)\}^{k}_{j}-\right.\nonumber\\
&
\left.\delta ^{k}_{j}\{U(x\p)\Ud(y\p)\}^{i}_{l}+\delta ^{i}_{l}
\{\Ud(y\p)U(x\p)\}^{k}_{j}\right]\frac {(x-z,y-z)\p}{(x-z) \p^{2}(y-z)\p^{2}}
\label{4.19}
\end{eqnarray}
where we have displayed the color indices explicitly.

\subsection{Calculation of the diagram in Fig. 10b}

\begin{figure}[htb]
\vspace{0cm}
\mbox{
\epsfxsize=10cm
\epsfysize=5cm
\hspace{3cm}
\epsffile{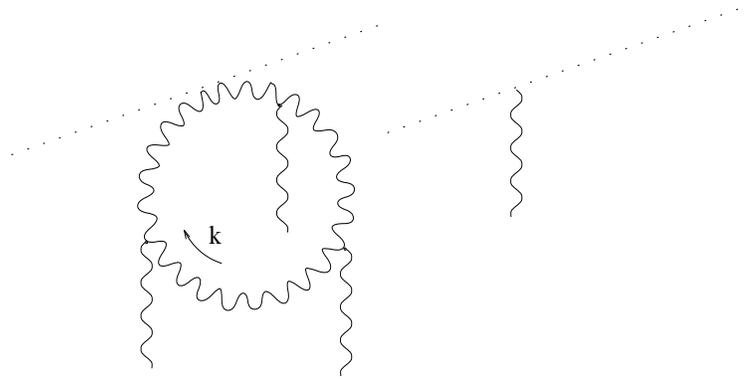}}
\vspace{0cm}
{\caption{\label{fig:Old8} Typical loop integral for the contribution of
the gauge
       terms $\sim n\parallel p_{2}$.}}
\end{figure}

The contribution of the diagram in Fig. \ref{fig:1loop}b is calculated in a
similar way. One starts with the expression (cf.\ eq.(\ref{4.4})):
\begin{eqnarray}
\lefteqn{ig\zeta \int \! udu
\lan[\iy,u]_x\hat{F}\sbu (up\ze+x\p)[up,-\iy]_x\otimes\Uhd(y\p)\ran_A}
\nonumber\\
&=&g\zeta \int \! udu \int\! du'\nonumber\\
&&\left(\Theta(v-v')\Theta(v'-u)[\iy,v]_x
t^a[v,v']_xt^b[v',u]_xF\sbu (up\ze+x\p)[u,-\iy]_x\right.\nonumber\\
&+&\left.
\Theta(v-v')\Theta(u-v)[\iy,u]_xF\sbu (up\ze+x\p)[u,v]_x
t^a[v,v']_xt^b[v',-\iy]_x\right.\nonumber\\
&+&\left.\Theta(v-u)\Theta(u-v')[\iy,v]_x
t^a[v,u]_xF\sbu (up\ze+x\p)[u,v']_xt^b[v',-\iy]_x
\right)\nonumber\\
&&\otimes \Ud(y\p)\nonumber\\
&&\LBB up\ze+x\p\Big|(p\ze_{\xi }-{\cal
P}\bu\frac {p_{2\xi }}{p\cdot p_{2}}){\cal O}^{\xi\eta}(p\ze_{\eta }
-\frac {p_{2\eta }}{p\cdot p_{2}}{\cal
P}\bu)\Big| vp\ze+x\p\RBB_{ab}\nonumber\\
&+&\nonumber\\
&&ig\zeta \int \! udu
[\iy,u]_x\hat{F}\sbu
(up\ze+x\p)[u,-\iy]_x\otimes
\nonumber\\
&&\int dv dv' \Theta(v-v')[\iy,v]_yt^{a}
[v,v']_yt^{b}[v',-\iy]_y
\nonumber\\
&&\LBB up\ze+x\p\Big|(p\ze_{\xi }-{\cal
P}\bu\frac {p_{2\xi }}{p\cdot p_{2}}){\cal O}^{\xi\eta}(p\ze_{\eta }
-\frac {p_{2\eta }}{p\cdot p_{2}}{\cal
P}\bu)\Big| vp\ze+x\p\RBB_{ab}\nonumber\\
&+&\nonumber\\
&&ig^{2}\int du dv\left(\Theta(u-v)
[\iy,u]_xt^{a}[u,v]_xt^{b}
[v,-\iy]_x\right.\nonumber\\
&+&\left.\Theta(v-u)
[\iy,v]_xt^{b}[v,u]_xt^{a}
[u,-\iy]_x\right)\otimes\Ud(y\p)
\nonumber\\
&&\LBB up\ze+x\p\Big|p\sr(p\ze_{\xi }-{\cal
P}\bu\frac {p_{2\xi }}{p\cdot p_{2}}){\cal O}^{\xi\eta}(p\ze_{\eta }
-\frac {p_{2\eta }}{p\cdot p_{2}}{\cal
P}\bu)\Big| vp\ze+x\p\RBB_{ab}
\label{4.20}
\end{eqnarray}
First, let us demonstrate that  the contribution of the terms $\sim {\cal
P}\bu$ in the parentheses in r.h.s. of eq.(\ref{4.20}) vanishes (cf.
eq.(\ref{4.4})). Indeed, using eq.(\ref{4.5}) and integrating by parts it is
easy to reduce these terms to the sum of the contributions of the type
\begin{eqnarray}
&g^{2}f_{abc}\int ^{\iy}_{-\iy}du [\iy,u]_xt^{c}[u,-\iy]_x\otimes\Ud(y\p)
\nonumber\\
&\LBB up\ze+x\p\Big|(p\ze_{\xi }-\half{\cal
P}\ci\frac {p_{2\xi }}{p\cdot p_{2}})\left\{\frac {1}{{\cal P}^{2}g_{\xi  \eta
}+2iF_{\xi \eta }}-\frac {1}{{\cal P}^{2}g_{\xi \lambda }+2iF_{\xi \lambda
}}(D_{\al}F_{\al\lambda }
\frac {p_{2\rho }}{p\cdot p_{2}}
+\right.
\nonumber\\
&\left.\frac {p_{2\lambda }}{p\cdot p_{2}}
D_{\al}F_{\al\rho }-
\frac {p_{2\lambda }}{p\cdot p_{2}}{\cal
P}_{\be}D_{\al}F_{\al\be}\frac {p_{2\lambda }}{p\cdot p_{2}})\frac {1}{{\cal
P}^{2}g_{\rho \eta }+2iF_{\rho \eta }}\right\}
\frac {p_{2\eta }}{p\cdot p_{2}}\Big| vp\ze+x\p\RBB_{ab}
\label{4.21}
\end{eqnarray}
This expression corresponds to diagram shown in Fig.~\ref{fig:Old8}.

 Let us consider the integration over $\be_{k}$ in the loop integral
corresponding to the momentum $k$.
As we shall see below, similarly to the case of diagram shown in Fig.
\ref{fig:1loop}a the contribution which survives in the limit
$\zeta\rightarrow 0$ has characteristic $\al_{k}\sim 1$
while all other $\tilde{\al}'s$ corresponding to
external field are $\sim {1\over\la}\sim \zeta $. This means that in
the contour integral over $\be_{k}$ all poles coming from the denominators
\begin{equation}
\{(\al_{k}+\tilde{\al})(\be_{k}+\tilde{\be})s-(k+\tilde{p})\p^{2}+\ie\}^{-1}
\label{422}
\end{equation}
lie to the one side of $X$ axis so that the resulting integral is zero. This
happened since we have cancelled the eikonal denominator in the original
diagram in Fig.~\ref{fig:1loop}b.
We shall see in a minute that this eikonal pole in $\be_{k}$
may lie to the opposite side of the $X$ axis thereby leading to the non-zero
contribution for the terms not proportional to ${\cal P}\bu$. Actually, the
cancellation of the terms proportional to longitudinal part of the gluon
propagator (in the external field) is a consequence of the gauge invariance of
the operator $U$.

So, we have reduced the contribution of the diagram in
Fig.~\ref{fig:1loop}b to
\begin{eqnarray}
\lefteqn{ig\zeta \int \! udu
\lan[\iy,u]_x\hat{F}\sbu (up\ze+x\p)[up,-\iy]_x\otimes\Uhd(y\p)\ran_A}
\nonumber\\
&=&g\zeta \int \! udu \int\! du'\nonumber\\
&&\left(\Theta(v-v')\Theta(v'-u)[\iy,v]_x
t^a[v,v']_xt^b[v',u]_xF\sbu (up\ze+x\p)[u,-\iy]_x\right.\nonumber\\
&+&\left.
\Theta(v-v')\Theta(u-v)[\iy,u]_xF\sbu (up\ze+x\p)[u,v]_x
t^a[v,v']_xt^b[v',-\iy]_x\right.\nonumber\\
&+&\left.(\Theta(v-u)\Theta(u-v')[\iy,v]_x
t^a[v,u]_xF\sbu (up\ze+x\p)[u,v']_xt^b[v',-\iy]_x
\right)\nonumber\\
&&\otimes \Ud(y\p)\nonumber\\
&&\LBB up\ze+x\p\Big|{\cal O}\bubu\Big| vp\ze+x\p\RBB_{ab}
\nonumber\\
&+&\nonumber\\
&&ig\zeta \int \! udu
[\iy,u]_x\hat{F}\sbu +
(up\ze+x\p)[u,-\iy]_x\otimes
\nonumber\\
&&\int dv dv' \Theta(v-v')[\iy,v]_yt^{a}
[v,v']_yt^{b}[v',-\iy]_y
\nonumber\\
&&\LBB up\ze+x\p\Big|{\cal O}\bubu\Big| vp\ze+x\p\RBB_{ab}\nonumber\\
&+&\nonumber\\
&&ig^{2}\int du dv\left(\Theta(u-v)
[\iy,u]_xt^{a}[u,v]_xt^{b}
[v,-\iy]_x\right.\nonumber\\
&+&\left.(\Theta(v-u)
[\iy,v]_xt^{b}[v,u]_xt^{a}
[u,-\iy]_x\right)\otimes\Ud(y\p)
\nonumber\\
&&\LBB up\ze+x\p\Big|p\sr{\cal O}\bubu\Big| vp\ze+x\p\RBB_{ab}
\label{4.23}
\end{eqnarray}
 As in the case of the diagram in Fig \ref{fig:1loop}a, a nonvanishing result
comes only from the last term where we differentiate propagator
in the external field  thus obtaining an extra factor ${1\over \zeta}\zeta=1$
in comparison to $\zeta\ln\zeta\rightarrow 0$ for the first three terms.
We have then
\begin{eqnarray}
\lefteqn{ig\zeta \int \! udu
\lan[\iy,u]_x\hat{F}\sbu (up\ze+x\p)[u,-\iy]_x\ran_A=}
\nonumber\\
&&
\hspace{-1cm}g^{2}\zeta \int ^{\iy}_{-\iy}du\int ^{u}_{-\iy}dv
[\iy,u]_xt^{a}[u,v]_xt^{b}[v,-\iy]_x
\LBB up_A+x\p\Big|up\sr{\cal O}\bubu-v{\cal O}\bubu
p\sr\Big| vp_A+x\p\RBB_{ab}
\label{4.24}
\end{eqnarray}
where we have omitted $\otimes \Ud(y\p)$ for brevity (it will be restored in
the final answer).

Again, let us neglect at first the gauge factors $[\iy,u]$,
$[u,v]$, and $[v,-\iy]$. We have then:
\begin{eqnarray}
&-g^{2}t^{a}t^{b}({2\over s})^{2}\int \frac {dk}{16\pi ^{4}}\int \frac {
dp}{16\pi ^{4}}
\frac {2\pi \delta '(\be_{p}+\zeta \al_{p})}{(\be_{k}+\zeta
\al_{k}-\ie)^{2}}\al_{k}
\LBB k\Big|{\cal O}\bubu\Big| k-p\RBB_{ab}
\label{4.25}
\end{eqnarray}
Now we can use our result for gluon propagator (\ref{4.15}) and obtain
\begin{eqnarray}
&&ig^{2}t^{a}t^{b}\int \frac {d\al_{k}}{2\pi \al_{k}}\frac {d\be_{k}}{2\pi }
\frac {dk\p}{4\pi ^{2}}
\frac {dp\p}{4\pi ^{2}}e^{-i(p,x)\p}\frac
{1}{(\be_{k}+\ze\al_{k}-\ie)^{2}}\frac { 1}{\al_{k}\be_{k}s-k\p^{2}+\ie}
\nonumber\\
&&[\Theta (\al)(\partial ^{2}U(p\p))-\Theta (-\al)(\partial ^{2}\Ud(p\p))]_{ab}
\frac {1}{\al_{k}(\be_{k} -\be_{p})s-(k-p)\p^{2}+\ie}
\nonumber\\
&&=~-{g^{2}\over4\pi }
t^{a}t^{b}\int \frac {dk\p}{4\pi ^{2}}\frac {dp\p}{4\pi ^{2}}
e^{-i(p,x)\p}{1\over k\p^{2}}\partial \p^{2}U(p\p){1\over
(k-p)\p^{2}}\nonumber\\
&&=~-{g^{2}\over 4\pi }t^{a}t^{b}\LBB x\p\Big|\frac {1}{p^{2}}(\partial \p
^{2}U)\frac {1}{p^{2}}\Big|x\p\RBB_{ab}
\label{4.26}
\end{eqnarray}
In the Appendix $\ref{app:B}$
it is demonstrated that the gauge factors of the type $[u,v]$
which we omitted  lead to the substitution
$[\iy,u]t^{a}[u,v]t^{b}
[v,-\iy]\rightarrow t^{a}U(x\p)t^{b}$ (in the coordinate space
it
means that the interaction with the shock-wave occurs between emission and
absorbtion of the quantum gluon). So we finally obtain:
\begin{eqnarray}
\lefteqn{\zeta {\partial \over\partial \zeta }\lan\hat{\Uz}(x\p)
\otimes\hat{\Udz}(y\p)\ran_{A}}\nonumber\\
&=&
-{g^{2}\over 4\pi } t^{a}U(x\p)t^{b}\otimes\Ud(y\p)
\LBB x\p\Big|\frac {1}{p^{2}}
(\partial ^{2}U)\frac {1}{p^{2}}\Big| x\p\RBB_{ab}=\nonumber\\
&=&-{g^{2}\over 16\pi ^{3}} \int
dz\p\left[U(z\p)\Tr\{U(x\p)\Ud(z\p)\}-N_{c}U(x\p)\right]\otimes\Ud(y\p)
\frac {1}{(x-z)\p^{2}}
\label{4.27}
\end{eqnarray}
where we have restored the omitted factor $\Ud(y\p)$. The contribution of the
remaining diagram in
Fig.~\ref{fig:1loop}c differs from eq.(\ref{4.27}) only in the
substitution $U\leftrightarrow \Ud$:
\begin{eqnarray}
\lefteqn{\zeta {\partial \over\partial \zeta }\lan\Uo(x\p)
\otimes\Udo(y\p)\ran_{A}}
\nonumber\\
&=&-{g^{2}\over 16\pi ^{3}}\int \! dz\p
U(x\p)\otimes\left[\Ud(z\p)\Tr\{U(z\p)\Ud(y\p)\}-N_{c}\Ud(y\p)\right]
\frac {1}{(y-z)\p^{2}}
\label{4.28}
\end{eqnarray}
Now we are in a position to write down the final answer for the one-loop
evolution of the operator $U(x\p)\Ud(y\p)$. Combining the expressions
(\ref{4.24}), (\ref{4.27}), and (\ref{4.28}) we obtain:
\begin{eqnarray}
\lefteqn{
\zeta {\partial \over\partial \zeta }\lan\{\Uhz(x\p)\}^{i}_{j}
\{\Uhdz(y\p)\}^{k}_{l}\ran_{A}}\nonumber\\
&=&
{g^{2}\over 16\pi ^{3}}\int \!dz\p
\left\{\left[\{\Ud(z\p) U(x\p)\}^{k}_{j} \{
U(z\p)\Ud(y\p)\}^{i}_{l} +
\{U(x\p)\Ud(z\p)\}^{i}_{l}\{\Ud(y\p)U(z\p)\}^{k}_{j}\right.\right.\nonumber\\
&&\left.\left.-\delta ^{k}_{j}
\{U(x\p)\Ud(y\p)\}^{i}_{l}-\delta ^{i}_{l}
\{\Ud(y\p)U(x\p)\}^{k}_{j}\right]\frac {(x-z,y-z)\p}{(x-z)\p^{2}(y-z)\p^{2}}-
\right.\nonumber\\
&&\left.\left[\{U(z\p)\}^{i}_{j}\Tr\{U(x\p)\Ud(z\p)\}-
N_{c}\{U(x\p)\}^{i}_{j}\right]
\Ud(y\p)^{k}_{l}\frac {1}{(x-z)\p^{2}}-\right.\nonumber\\
&&\left.
\{U(x\p)\}^{i}_{j} \left[\Ud(z\p)^{k}_{l}\Tr\{U(z\p)\Ud(y\p)\}-
N_{c}\{\Ud(y\p)\}^{k}_{l})\right]\frac {1}{(y-z)\p^{2}}\right\}
\label{4.29}
\end{eqnarray}
This is the one-loop result for the operator $\Uo(x\p)\otimes\Udo(y\p)$ in the
low-$\al$ external field A corresponding to the bottom part of the diagram in
Fig.2a. As we discussed in previous Section, the operator form of this one-loop
evolution is:
\begin{eqnarray}
\lefteqn{
\zeta {\partial \over\partial \zeta }\{\Uhz(x\p)\}^{i}_{j}
\{\Uhdz(y\p)\}^{k}_{l}}
\nonumber\\
&=&{g^{2}\over 16\pi ^{3}}
\int dz\p \left\{ \left[\{\Uhdz(z\p)\Uhz(x\p)\}^{k}_{j}
\{\Uhz(z\p)\Uhdz(y\p)\}^{i}_{l}+\{\Uhz(x\p)\Uhdz(z\p)\}^{i}_{l}\{\Uhdz(y\p)
\Uhz(z\p)\}^{k}_{j}\right.\right.
\nonumber\\
&&\left.\left.-\delta ^{k}_{j}
\{\Uhz(x\p)\Uhdz(y\p)\}^{i}_{l}-\delta ^{i}_{l}
\{\Uhdz(y\p)\Uhz(x\p)\}^{k}_{j}\right]
\frac {(x-z,y-z)\p}{(x-z)\p^{2}(y-z)\p^{2}}-
\right.\nonumber\\
&&\left.\left[\{\Uhz(z\p)\}^{i}_{j}\Tr\{\Uhz(x\p)\Uhdz(z\p)\}-
N_{c}\{\Uhz(x\p)\}^{i}_{j}\right]
\{\Uhdz(y\p)^{k}_{l}\}\frac {1}{(x-z)\p^{2}}-\right.\nonumber\\
&&\left.\{\Uhz(x\p)\}^{i}_{j} \left[\{\Uhdz(z\p)\}^{k}_{l}
\Tr\{\Uhz(z\p)\Uhdz(y\p)\}-N_{c}\{\Uhdz(y\p)\}^{k}_{l}\right]
\frac {1}{(y-z)\p^{2}}\right\}
\label{4.30}
\end{eqnarray}
where the operators $\Uhz$ and $\Uhdz$ are integrated along the line
collinear to $p\ze$ in order to impose cutoff $\al<\sqrt {m^{2}\over\zeta s}$
in
the matrix elements of these operators.

\section{Evolution of eikonal-line operators in leading log approximation}
\label{sec:LLAEvol}

\subsection{Linear evolution at large $N_{c}$}

Let us outline how to obtain the energy dependence of the amplitude using
the  expansion in eikonal-line operators. As we have discussed in
previous Sections after formal expansion at large energy we obtain in the
leading logarithmic approximation the operators $U$
and $\Ud$ "normalized" at the slope $\zeta =\pas$ times the impact factor:
\begin{equation}
\int \!\! dx\!\!\int \!\! dz
\delta (z\bu)e^{ip_{A}x}T\{j_{\mu }(x+z)j_{\nu }(z)\}=\!
\!\sum _{flavors}\!\!e_{i}^{2}\!\int \! dx\p dz\p
I^{A}_{\mu \nu }(x\p)\Tr\{U^{\zeta }(x\p+z\p)U^{\dagger \zeta }(z\p)\}+O(g^{2})
\label{5.1}
\end{equation}
where $I^{A}(x\p)$ ig given by eq. (\ref{2.35}) and the dots stand for the
next-to-leading term given in eq. (\ref{3.15}). (Hereafter we will wipe the
label
$(~\hat{}~)$ from the notation of the operators). Matrix element of this
operator $\ll U^{\zeta }(x\p)U^{\dagger \zeta }(y\p)\gg$ (see eq.(\ref{3.2})
for
the
definition) describes the gluon-photon scattering at large energies $\sim s$.
The behavior of this \mael\ with  energy is determined by the dependence on the
"normalization point" $\zeta $.
{}From the one-loop results for the
evolution of the operators $U$ and $\Ud$ it is easy to obtain evolution
equation:
\begin{eqnarray}
\lefteqn{\zeta \frac {\partial }{\partial \zeta }
\Tr\{\Uz(x\p)[x\p,y\p]_{-}\Udz(y\p)[y\p,x\p]_{+}\}}
\nonumber\\
&=&-{g^{2}\over 16\pi ^{3}}
\int \! dz\p(\Tr\{\Uz(x\p)[x\p,z\p]_{-}\Udz(z\p)[z\p,x\p]_{+}\}
\Tr\{\Uz(z\p)[z\p,y\p]_{-}\Udz(y\p)[y\p,z\p]_{+}\}-
\nonumber\\
&&
N_{c}\Tr\{\Uz(x\p)[x\p,y\p]_{-}\Udz(y\p)[y\p,x\p]_{+}\})\frac {
(x\p-y\p)^{2}}{(x\p-z\p)^{2}(z\p-y\p)^{2}}
\label{5.2}
\end{eqnarray}
where we have displayed the end gauge factors (\ref{endf}) explicitly. We see
that as a result of the evolution the evolution of the two-line operator
$\Tr\{U\Ud\}$ is the same operator (times the kernel) plus the four-line
operator $\Tr\{U\Ud\}\Tr\{U\Ud\}$. The result of the evolution of the four-line
operator will be the same operator times some kernel plus the six-line operator
of the type $\Tr\{U\Ud\}\Tr\{U\Ud\}\Tr\{U\Ud\}+\Tr\{U\Ud U\Ud\}\Tr\{U\Ud\}$ and
so on. Therefore it is instructive to consider at first the large-$N_{c}$ case
where, as we demonstrate below, the number of operators $U$ is always the same
during the evolution.

Let us consider the evolution equation (\ref{4.2}) in the large-$N_{c}$ limit.
It is convenient to rewrite it in terms of the operators
\begin{equation}
V(x\p,y\p)\equiv
\frac {1}{N_{c}}(x-y)\p^{-2}(\Tr\{U(x\p)[x\p,y\p]_{-}
\Ud(y\p)[y\p,x\p]_{+}\}-N_{c})
\label{V}
\end{equation}
so it reduces to:
\begin{eqnarray}
\lefteqn{\zeta \frac {\partial }{\partial \zeta } V(x\p ,y\p )}\nonumber\\
&=& -{g^{2}N_{c}\over 16\pi
^{3}}
\!\int \! dz\p \left\{\frac {V(x\p ,z\p )}{(z\p -y\p )^{2}}+
\frac {V(z\p ,y\p )}{(x\p-z\p)^{2}}-
\frac {V(x\p,y\p)(x\p -y\p )^{2}}{(x\p -z\p
)^{2}(z\p -y\p )^{2}}+V(x,z)V(z,y)\right\}
\label{5.3}
\end{eqnarray}
 It is easy to see that matrix elements of the operator $\llan V^{2}\rran$  are
$\sim {1\over N_{c}}$ in comparison to the matrix elements of the operator
$\llan
V\rran$ \footnote{The only exception is disconnected contributions of the type
$\llan V\rran\otimes\lvac V\rvac$ but they are $O(g^{2})$ corrections to the
matrix elements $\llan V\rran$ which we shall not consider in the leading
logarithmic approximation. In the next order in $g^{2}$ they must be taken into
account together with $O(g^{2})$ corrections to impact factor (\ref{2.35})}.

So, in the leading order in $N_{c}$ the evolution of the forward matrix element
of the two-line operator $V(x\p,y\p)$ is governed by linear BFKL equation
\footnote{ The connection between Wilson-line operators and the BFKL
equation was first discussed in Ref.\cite{collell}.}
\begin{equation}
\zeta \frac {\partial }{\partial \zeta } \llan V(x\p)\rran =
-{\al_{s}\over 4\pi
^{2}}
N_{c}\int \!
 dz\p\left\{\frac {\llan V(x-z\p)\rran}{z\p^{2}}+
\frac {\llan V(z\p)\rran}{(x\p-z\p)^{2}}-\frac {\llan V(x\p)\rran
x\p^{2}}{(x\p-z\p)^{2}z\p^{2}}\right\}
\label{5.4}
\end{equation}
where $\llan V(x\p)\rran\equiv \llan V(x\p,0)\rran$, see eq. (\ref{2.37}). The
eigenfunctions of this
equation
are powers $(x\p^{2})^{-\half+i\nu }$ and the eigenvalues are
$-{\alpha _{s}\over\pi }N_{c}\chi (\nu )$, where
$\chi (\nu )=-\Repa\psi (\half+i\nu )-C$. Therefore, the evolution of the
operator $V$ takes the form:
\begin{eqnarray}
&\llan V^{\zeta _{1}}(x\p)\rran=\int \! \frac {d\nu }{2\pi ^{2}}
(x\p^{2})^{-\half+i\nu }
\left(\frac {\zeta _{1}}{\zeta _{2}}\right)^{-{\alpha _{s}\over\pi }N_{c}\chi
(\nu )}
\int \! dz\p(z\p^{2})^{-\half-i\nu }\llan V^{\zeta _{2}}(z)\rran
\label{5.5}
\end{eqnarray}
We may proceed with this evolution as long as the upper limit of our
logarithmic integrals over $\al$ which is $\sqrt {p_{A}^{2}\over\zeta s}$ is
much
larger than the lower limit $\pbs$ which is determined by the lower quark bulb,
see the discussion in Sect. 3. (In other words, if we look on the derivation of
the evolution equation given in previuos Section we can see that it holds true
as long as we can neglect $\al_{p}\sim\pbs$ in comparison to
$\al_{k}\sim\sqrt {p_{A}^{2}\over\zeta s}$ in eq. (\ref{4.16})). It is
convenient to
stop evolution at a certain point $\zeta _{0}$ such as
\begin{equation}
\zeta _{0}=\si^{2}{s\over m^{2}},~~~\si\ll 1,~~~~g^{2}ln\si\ll 1
\label{5.6}
\end{equation}
Then the relative energy between the Wilson-line operator $V^{\zeta _{0}}$ and
lower virtual photon will be $s_{0}=m^{2}\si^{2}$ which is big enough to apply
our
usual high-energy approximations (such as pure gluon exchange and substitution
$g_{\mu \nu }\rightarrow {2\over s_{0}}p_{2\mu }p_{1\nu }$) but small in a
sence that
one does not need take into account the difference between $g^{2}\ln{s\over
m^{2}}$
and $g^{2}\ln{s\over m^{2}\si^{2}}$. Then finally the evolution (\ref{5.4})
takes the
form:
\begin{equation}
\llan V^{\zeta =\ms}(x\p)\rran=\int \! \frac {d\nu }{2\pi ^{2}}
(x\p^{2})^{-\half+i\nu }
(\frac {s}{m^{2}})^{{2\alpha _{s}\over\pi }N_{c}\chi (\nu )}
\int \! dz\p(z\p^{2})^{-\half-i\nu }\llan V^{\zeta _{0}}(z\p)\rran
\label{5.7}
\end{equation}
Now let us rewrite this evolution in terms of original operators $U\Ud$ in the
momentum representation. One has then:
\begin{eqnarray}
\lefteqn{\llan \Tr\{U^{\zeta =\ms}(p\p)
U^{\dagger \zeta =\ms}(-p\p)\}\rran}\nonumber\\
&=&\int \!
\frac {d\nu }{2\pi ^{2}}
(p\p^{2})^{-{1\over 2}-i\nu }
(\frac {s}{m^{2}})^{{2\alpha _{s}\over\pi }N_{c}\chi (\nu )}
\int \! dp'\p(p\p^{'2})^{\half+i\nu }\llan Tr\{U^{\zeta _{0}}(p'\p)
U^{\dagger \zeta _{0}}(-p'\p)\}\rran
\label{5.8}
\end{eqnarray}
where we omit for brevity the end factors (\ref{endf}). Since we
neglect the logarithmic corrections $\sim g^{2}\ln\si$ matrix element
of our operator $\Ut\Udt$ coincide with impact factor $I^{B}$ up
to $O(g^{2})$ corrections:
\begin{eqnarray}
\lefteqn{\llan Tr\{U^{\zeta _{0}}(p\p)
U^{\dagger \zeta _{0}}(-p\p)\}\rran}\nonumber\\
&=&g^{2}{N_{c}^{2}-1\over
2}\sum e_{i}^{2}\int \frac {d\al}{\pi s}\frac {\Phi ^{B}(\al_{p}p_{1}-\zeta
_{0}\al_{p}p_{2}+p\p)}{ (\zeta _{0}\al_{p}^{2}+p\p^{2})^{2}}\nonumber\\
&=&g^{2}{N_{C}^{2}-1\over 2}\sum e_{i}^{2}{1\over p\p^{4}}I^{B}(p\p)
\label{5.9}
\end{eqnarray}

Combining eqs. (\ref{2.36}),(\ref{5.8}), and (\ref{5.9}) we reproduce the usual
leading logarithmic result for virtual $\gamma \gamma $ scattering\cite{XIV}:
\begin{eqnarray}
\lefteqn{A(p_{A},p_{B})}\nonumber\\
&=&ig^{4}{s\over 2}(N_{c}^{2}-1)(\sum e_{i}^{2})^{2}\int \! d\nu
({s\over
m^{2}})^{{2\al_{s}\over\pi }N_{c}\chi (\nu )}
\int \! \frac {dp\p}{4\pi ^{2}} I^{A}(p\p)(p\p^{2})^{-{3\over 2}+i\nu }
\int \!\frac {dp'\p}{4\pi ^{2}} I^{B}(p'\p)(p\p^{2})^{-{3\over 2}-i\nu }
\label{5.10}
\end{eqnarray}
At $s\rightarrow 0$ the amplitude  (\ref{5.10}) is determined by the rightmost
singularity in the $\nu $ plane located at $\nu =0$ (in terms of complex
momenta
plane it corresponds to the position of the "bare pomeron" at
$j=1+{4\al_{s}\over\pi }N_{c}\ln2$) and the answer is
\begin{equation}
A(p_{A},p_{B})={i\over 2}sg^{4}\frac {N_{c}^{2}-1}{14\zeta
(3)N_{c}{\al_{s}\over\pi }}(\sum
e_{i}^{2})^{2}({s\over m^{2}})^{{4\al_{s}\over\pi }N_{c}\ln2}
\int \! \frac {dp\p}{4\pi ^{2}} I^{A}(p\p)(p\p^{2})^{-{3\over 2}}
\int \!\frac {dp'\p}{4\pi ^{2}} I^{B}(p'\p)(p\p^{2})^{-{3\over 2}}
\label{5.11}
\end{equation}
where $\zeta (3)\simeq 1.202$. In the case of small-x \dis\ the evolution of
the
matrix element \lN V\rN is the same as eq. (\ref{5.10}) with the only
difference that the lower impact factor $I^{B}$ should be substituted by the
nucleon impact factor $I^{N}$ determined by the matrix element of the operator
$U\Ud$ between the nucleon states \footnote{This is called "hard pomeron"
contribution to the structure functions of \dis\ since all the transverse
momenta in our calculations are large ($\sim Q^{2}$ or $\sim m_{N}^{2}$ which
is the
same in leading logarithmic approximation). However, due to the so-called
diffusion in transverse momenta the characteristic size of the $p\p^{2}$ in the
middle of gluon ladder is $e^{\sqrt {g^{2}ln s}}$ (see e.g.  Ref.\cite{lev}
for discussion) so at
very
small $x$ the region $p\p\sim \Lambda _{QCD}$ may become important. It
corresponds to the contribution of the so-called "soft" or "old" pomeron which
is constructed from non-perturbative gluons in our language and must be added
to the hard-pomeron result given by eq. (\ref{5.11})}:
\begin{equation}
\lan N,p_{B}|\Tr\{U^{\zeta _{0}}(x\p)U^{\dagger \zeta _{0}}(0)\}|N,p_{B}+\be
p_{2}\ran~=~2\pi \delta (\be)\int \!\frac {dp\p}{4\pi ^{2}}e^{i(px)\p}{1\over
p\p^{4}}I^{N}(p\p)
\label{5.12}
\end{equation}
where $2\pi \delta (\be)$ reflects the fact that matrix element of the operator
$U\Ud$ containes
unrestricted integration along $p^{\zeta _{0}}$, (cf. eq. (\ref{2.37})). The
nucleon impact factor $I^{B}(p\p)$ defined in (\ref{5.12}) is a
phenomenological
low-energy characteristic of the nucleon.  In the BFKL evolution it plays the
role similar to that of nucleon structure function at low normalization point
for GLAP evulution. In principle, it can be estimated using QCD sum rules or
phenomenological models of nucleon.

Let us discuss how the nucleon impact factor (\ref{5.12}) is related to the
gluon structure function of the nucleon which is defined as the matrix element
of the gluon light-cone operator:
\begin{equation}
{\cal D}(\omega ,\mu )=-{4\over s\omega }\int \! du e^{-i{s\over 2}\omega u}
\lN
\Tr\{F^{\xi }\bu(up_{1})[up_{1},0]F_{\xi \bullet}[0,up_{1}]\}\rN^{\mu }
\label{5.13}
\end{equation}
where $\mu $ is the normalization point for the light-cone operator. (The
unrenormalized operator $F(up_{1})F(0)$ is UV divergent so we regularize it by
counterterms just as for the local operator, see e.g. \cite{bb1}).The physical
meaning of $\mu $ is the resolution in the transverse size of the gluon: ${\cal
D}_{g}(\omega ,\mu )$ is the probability to find inside a nucleon the gluon
carrying the fraction $\omega $ of the nucleon momentum with the transverse
size
$\mu ^{-1}$. Formally,
\begin{equation}
{\cal D}_{g}(\om,\mu )\simeq\int \! dk\p \Theta (\mu ^{2}-k\p^{2}) {\cal
D}_{g}(\om,k\p)
\label{5.14}
\end{equation}
where ${\cal D}_{g}(\om,k\p)$ is the gluon distribution over transverse
momentum
$k\p$ and fraction of logitudinal momentum $\om p_{1}$:
\begin{eqnarray}
{\cal D}_{g}(\om,k\p)&=&\int \! dx\p e^{-i(kx)\p}
{\cal D}(\om,x\p),\nonumber\\
\om{\cal D}_{g}(\om,x\p)&=&-{4\over s}\int \! du e^{-i{s\over 2}\omega u} \lN
\Tr\{F^{\xi }\bu(up_{1}+x\p)\nonumber\\
&&[up_{1}+x\p,,0]F_{\xi \bullet}(0)[0,-\iy p_{1}+x\p]\}\rN
\label{5.15}
\end{eqnarray}
It is easy to relate impact factor to the gluon distribution - actually, it is
the same quantity with different regularization of longitudinal integrations.
Indeed,
\begin{eqnarray}
\lefteqn{{2\over p\p^{2}}I^{N}(p\p)
{}~=~2\frac {\lan N|\Tr\{\partial _{i}U^{\zeta _{0}}(p\p)
\partial _{i}U^{\dagger \zeta _{0}}(-p\p)\}|N\ran}{2\pi \delta (0)}}
\nonumber\\
&=&-{4\over s}g^{2}\int \!
du\lN \Tr\{[-\iy p\zo+x\p,up\zo+x\p]F^{\xi }\bu(up\zo+x\p)\nonumber\\
&&
[up\zo+x\p,-\iy p\zo+x\p][-\iy p\zo,0]F_{\xi \bullet}(0)[up\zo,-\iy p\zo]\}\rN
\label{5.16}
\end{eqnarray}
Now we see that the r.h.s.'s of eqs. (\ref{5.15}) and (\ref{5.16})
coincide up to a different cutoff in the longitudinal intergation in
\mael s: in the case of gluon distribtuion the integrals over the
$\al$ component are restricted from above by ${m^{2}\over\om s}$
whereas for the matrix element (\ref{5.16}) the cutoff is
$\sqrt {m^{2}\over\zeta _{0} s}={m^{2}\over s\si}$ so they coincide at
$\si=\om$
\footnote{For example, in the case of diagram in Fig. \ref{fig:mael} the
contribution to impact factor (\ref{5.16}) is (cf.eq. (\ref{3.6})):
\begin{eqnarray}
-{i\over 2}\int \!\frac {d\al_{p}}{2\pi }\frac {d\al'_{p}}{2\pi }
\frac {d\be'_{p}}{2\pi }\frac {dp'\p}{4\pi ^{2}}
\frac {g^{6}N_{c}(N_{c}^{2}-1)p\p^{2}\Gamma ^{\si}\sbu(p,-p')\Gamma
^{\si}\sbu(p,-p') \Phi ^{B}_{\xi \eta }(p')}{ (\zeta
\al^{2}s+p\p^{2}-\ie)^{2}(\al'_{p}\be'_{p}s-p^{'2}\p+\ie)^{2}[(\al-\al')
(\be-\be')s-(p-p')^{2}+\ie]}
\label{5.17}
\end{eqnarray}
whereas the contribution to the gluon distribution defined by r.h.s. of eq.
(\ref{5.15}) has the form:
\begin{eqnarray}
-{i\over 4}\int \!\frac {d\al_{p}}{2\pi }\frac {d\al'_{p}}{2\pi }
\frac {d\be'_{p}}{2\pi }\frac {dp'\p}{4\pi ^{2}}[\frac {1}{(\al\om
s-p\p^{2}-\ie)^{2}}+(\om\leftrightarrow -\om)]
\frac {g^{6}N_{c}(N_{c}^{2}-1)p\p^{2}\Gamma ^{\si}\sbu(p,-p')\Gamma
^{\si}\sbu(p,-p') \Phi ^{B}_{\xi \eta }(p')}{
(\al'_{p}\be'_{p}s-p^{'2}\p+\ie)^{2}[(\al-\al') (\be-\be')s-(p-p')^{2}+\ie]}
\label{5.18}
\end{eqnarray}
so we see that the only difference is in the cutoff for the logarithmical
integration over $\al$}. Therefore
\begin{equation}
{2\over p\p^{2}}I^{N}(p\p)=\si {\cal D}_{g}(\si,p\p) + O(g^{2})
\label{5.19}
\end{equation}
where the impact factor is determined by Wilson lines $U$ and $\Ud$
parallel to $p_{1}+{s\over m^{2}}\si p_{2}\parallel \si p_{2}+\ms p_{1}$ and
$\si m^{2}$ plays the role of the relative energy between Wilson lines
and nucleon.

\subsection{General case}

\begin{figure}[htb]
\vspace{0cm}
\mbox{
\epsfxsize=15cm
\epsfysize=8cm
\hspace{0cm}
\epsffile{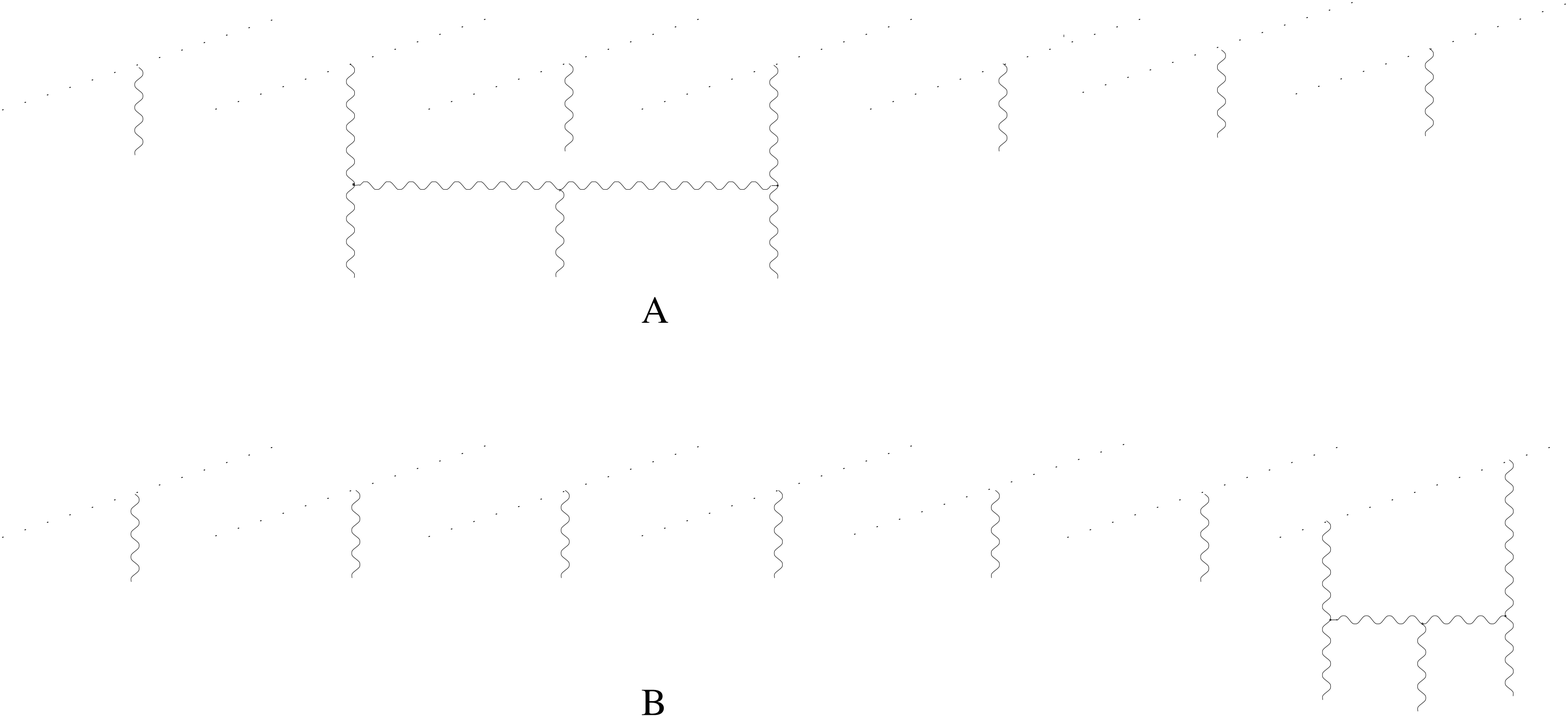}}
\vspace{0cm}
{\caption{\label{fig:OneLoopEvol} Typical diagrams for the one-loop
evolution of the
                   $n$-line operator.}}
\end{figure}

Unlike the linear evolution at large $N_{c}$, the general picture without
large-$N_{c}$ simplification is
very
complicated: not only the number of operators $U$ and $\Ud$ increase after each
evolution but they form more and more complicated structures like displayed
(and not displayed) in eq. (\ref{5.23}) below.
In the leading log approximation the evolution of the $2n$-line
operators such as
$\Tr\{U\Ud\}\Tr\{U\Ud\}...\Tr\{U\Ud\}$ come from either self-interaction
diagrams  or from the pair-interactions ones (see Fig.~\ref{fig:OneLoopEvol})

which we have already calculated in Sect.~\ref{sec:OneLoopEvol}
--- see
eqs.(\ref{4.17},\ref{4.27},\ref{4.28}). Therefore
the one-loop evolution equations for these operators can be constructed using
these rules which we shall list here for completeness in the explicit form
(in this Section we will use the notation $U_{x}\equiv U(x\p)$ etc):
\begin{eqnarray}
\lefteqn{\zeta \frac {\partial }{\partial \zeta }\{U_{x}\}^{i}_{j}
\{\Ud_{y}\}^{k}_{l}~=}\nonumber\\
&&
\hspace{-1cm}{g^{2}\over 16\pi ^{3}}\int dz\p(\{\Ud_{z}U_{x}\}^{k}_{j}
\{U_{z}\Ud_{y}\}^{i}_{l}+
\{U_{x}\Ud_{z}\}^{i}_{l}\{\Ud_{y}U_{z}\}^{k}_{j}-\delta ^{k}_{j}
\{U_{x}\Ud_{y}\}^{i}_{l}-\delta ^{i}_{l}
\{\Ud_{y}U_{x}\}^{k}_{j})\frac {(x-z,y-z)\p}{(x-z)\p^{2}(y-z)\p^{2}}
\nonumber\\
\lefteqn{\zeta \frac {\partial }{\partial \zeta }\{U_{x}\}^{i}_{j}
\{U_{y}\}^{k}_{l}~=}\nonumber\\
&&
\hspace{-1cm}-{g^{2}\over 16\pi ^{3}}
\int dz\p(\{U_{z}\}^{i}_{l}\{U_{y}\Ud_{z}U_{x}\}^{k}_{j}+
\{U_{x}\Ud_{z}U_{y}\}^{i}_{l}\{
U_{z}\}^{k}_{j}-\{U_{x}\}^{i}_{l}
\{U_{y}\}^{k}_{j}-\{U_{y}\}^{i}_{l}
\{U_{x}\}^{k}_{j})\frac {(x-z,y-z)\p}{(x-z)\p^{2}(y-z)\p^{2}}
\nonumber\\
\lefteqn{\zeta \frac {\partial }{\partial \zeta }\{U_{x}\}^{i}_{j}
\{\Ud_{y}\}^{k}_{l}~=}\nonumber\\
&&
\hspace{-1cm}-{g^{2}\over 16\pi ^{3}}
\int dz\p(\{\Ud_{z}\}^{i}_{l}\{\Ud_{y}U_{z}
\Ud_{x}\}^{k}_{j}+\{\Ud_{x}U_{z}\Ud_{y}\}^{i}_{l}\{
\Ud_{z}\}^{k}_{j}-\{\Ud_{x}\}^{i}_{l}
\{\Ud_{y}\}^{k}_{j}-\{\Ud_{y}\}^{i}_{l}
\{\Ud_{x}\}^{k}_{j})\frac {(x-z,y-z)\p}{(x-z)\p^{2}(y-z)\p^{2}}\nonumber\\
\label{5.20}
\end{eqnarray}
for the pair-interaction diagrams in Fig.~\ref{fig:OneLoopEvol}a and
\begin{eqnarray}
&\zeta \frac {\partial }{\partial \zeta }\{U_{x}\}^{i}_{j}=-{g^{2}\over 16\pi
^{3}}
\int dz\p[U_{z}\Tr\{U_{x}\Ud_{z}\}-N_{c}U_{x}]\frac {1}{(x-z)\p^{2}}
\nonumber\\
&\zeta \frac {\partial }{\partial \zeta }\{\Ud_{x}\}^{i}_{j}=
-{g^{2}\over 16\pi^{3}}
\int dz\p[\Ud_{z}\Tr\{U_{z}\Ud_{x}\}-N_{c}\Ud_{x}]
\frac {1}{(x-z)\p^{2}}
\label{5.21}
\end{eqnarray}
for the self-interaction diagrams of Fig.~\ref{fig:OneLoopEvol}b type. Using
eqs. (\ref{5.20},\ref{5.21}) it is easy to write down evolution equations for
arbitrary $n$-line operator. For example, the evolution equation for the
four-line operator appearing in the r.h.s. of eq.(\ref{5.2}) has the form:
\begin{eqnarray}
\lefteqn{\zeta \frac {\partial }{\partial \zeta }
\Tr\{U_{x}[x,z]_{-}\Ud_{z}[z,x]_{+}\}\Tr\{U_{z}[z,y]_{-}
\Ud_{y}[y,z]_{+}\}}\nonumber\\
&=&-{g^{2}\over 16\pi ^{3}}
\int \! dt\p \left\{\left[\Tr\{U_{x}[x,t]_{-}\Ud_{t}[t,x]_{+}\}
\Tr\{U_{t}[t,z]_{-}\Ud_{z}[z,t]_{+}\} -
N_{c}\Tr\{U_{x}[x,z]_{-}\Ud_{z}[z,x]_{+}\}\right]\right.\nonumber\\
&&\left.\Tr\{U_{z}[z,y]_{-}
\Ud_{y}[y,z]_{+}\}
\frac {(x-z)\p^{2}}{(x-t)\p^{2}(z-t)\p^{2}}\right.\nonumber\\
&+&\left.\Tr\{U_{x}[x,z]_{-}
\Ud_{z}[z,x]_{+}\}\right.\nonumber\\
&&\left.\left[\Tr\{U_{z}[z,t]_{-}\Ud_{t}[t,z]_{+}\}
\Tr\{U_{t}[t,y]_{-}\Ud_{y}[y,t]_{+}\} -
N_{c}\Tr\{U_{z}[z,y]_{-}\Ud_{y}[y,z]_{+}\}\right]
\frac {(y-z)\p^{2}}{(y-t)\p^{2}(z-t)\p^{2}}\right.\nonumber\\
&+&\left.
\left[\Tr\{U_{x}[x,z]_{-}\Ud_{z}[z,t]_{+}U_{t}[t,y]_{-}\Ud_{y}[y,z]_{+}
U_{z}[z,y]_{-}\Ud_{t}[t,x]_{+}\} +\right.\right.\nonumber\\
&&\left.\left.
\Tr\{U_{x}[x,t]_{-}\Ud_{t}[t,z]_{+}U_{z}[z,y]_{-}\Ud_{y}[y,t]_{+}
U_{t}[t,z]_{-}\Ud_{z}[z,x]_{+}\}-
2\Tr\{U_{x}[x,y]_{-}\Ud_{y}\}[y,x]_{+}\right]\right.\nonumber\\
&&\left.
\left[-\frac {(x-t,y-t)\p}{(x-t)\p^{2}(y-t)\p^{2}}-
\frac {1}{(z-t)\p^{2}}+\frac {(x-t,z-t)\p}{(x-t)\p^{2}(z-t)\p^{2}}+
\frac {(z-t,y-t)\p}{(z-t)\p^{2}(y-t)\p^{2}}\right]\right\}
\label{5.22}
\end{eqnarray}
where we have displayed the end gauge factors (\ref{U}) explicitly. Note
that each of the separate contributions (\ref{5.20}) and (\ref{5.21})
corresponding to the diagrams in Fig.~\ref{fig:OneLoopEvol}a and
\ref{fig:OneLoopEvol}b diverges at large $z$ while the
integral (over $t$) in the total answer (\ref{4.5}) is convergent. This is a
case of the usual cancellation of the IR divergent contributions between the
emission of the real
(Fig.~\ref{fig:OneLoopEvol}a) and virtual
(Fig.~\ref{fig:OneLoopEvol}b) gluons from the colorless
object (corresponding to the l.h.s. of eq (\ref{4.5})).
Another example of this
cancellation is eq.(\ref{5.2}), see the discussion above.

So, the result of the evolution of the operator in the r.h.s. of eq.(\ref{4.1})
looks like:
\begin{eqnarray}
\lefteqn{\Tr\{\Uz_{x}[x,y]_{-}\Udz_{y}[y,x]_{+}\}}\nonumber\\
&\Rightarrow& \sum
_{n=0}^{\iy}(\as\ln\frac {  \zeta }{\zeta _{0}})^{n}
\int dz^{1}dz^{2}...dz^{n}\nonumber\\
&&
\left[A_{n}(x,z^{1},z^{2},...z^{n},y)
\Tr\{\Ut_{x}[x,1]_{-}\Udt_{1}[1,x]_{+}\}\right.\nonumber\\
&&\left.\Tr\{\Ut_{1}[1,2]_{-}
\Udt_{2}[2,1]_{+}\} \dots
\Tr\{\Ut_{n}[n,y]_{-}\Udt_{y}[y,n]_{+}\}\right.\nonumber\\
&+&\left.B_{n}(x,z^{1},z^{2},...z^{n},y)\right.\nonumber\\
&&\left.\Tr\{\Ut_{x}[x,1]_{-}
\Udt_{1}[1,2]_{+}\Ut_{2}[2,3]_{-}\Udt_{3}[3,1]_{+}\Ut_{1}[1,2]_{-}
\Udt_{2}[2,x]_{+}\}\right.\nonumber\\
&&\left.\Tr\{\Ut_{3}[3,4]_{-}\Udt_{4}[4,3]_{+}\}...
\Tr\{\Ut_{n}[n,y]_{-}\Udt_{y}[y,n]_{+}\}+...\right.\nonumber\\
&+&\left.
N_{c}^{n} C_{n}(x,z^{1},z^{2},...z^{n},y;)\Tr\{\Ut_{x}[x,y]_{-}
\Udt_{y}[y,x]_{+}\}\right]
\label{5.23}
\end{eqnarray}
where $U^{(\dagger )}_{n}\equiv U^{(\dagger )}(z^{n}\p)$,
$[i,j]\equiv [x_{i},x_{j}]$ and
\newline
$A_{n}(x,z^{1},z^{2},...z^{n},y),B_{n}(x,z^{1},z^{2},...z^{n},y),
\dots C_{n}(x,z^{1},z^{2},...z^{n},y)$
are the meromorphic functions that can be obtained by using the
eqs.(\ref{5.20},\ref{5.21}) $n$ times which give us a sort of Feynman rules for
calculation of these coefficient functions. If we now evolve our operators from
$\zeta \sim \pas$ to $\zeta _{0}$ given by eq. (\ref{5.6}) we shall obtain a
series
(\ref{5.23}) of \mael
s of the operators $(U)^{n}(\Ud)^{n}$ (see eq.(\ref{4.6})) normalized at
$\zeta _{0}$.
These matrix elements correspond to small energy $\sim m^{2}$ and they can be
calculated either perturbatively (in the case the "virtual
photon" \mael\ ) or
using some model calculations such as QCD sum rules in the case of nucleon
\mael\ corresponding to small-x $\gamma ^{\ast}p$ \dis\ . It should be
mentioned that in the case of virtual photon scattering considered above we can
alculate the matrix elements of operators $U\Ud...U\Ud$ perturbatively and
therefore in the leading order in $\al_{s}$ we can replace all $U$'s
(and $\Ud$'s)
except for two by 1. (Recall that $U=1+ig\int \!A_{\mu }dx_{\mu }+...$ so
each extra $U-1$ brings at least $O(g)$). So, we return to the BFKL picture
describing the evolution of the two operators $U\Ud$ similarly to the large
$N_{c}$ case. The non-linear equation (\ref{5.2}) enters the game in the
situation like small-x \dis\ from a nucleon when the matrix elements of the
operators $U\Ud...U\Ud$ are non-perturbative so there is no reason (apart from
large $N_{c}$) to expect that extra $U$ and $\Ud$ will lead to extra smallness.
In this case, at the low "normalization point" $\zeta _{0}$ one must take into
account the whole series of the operators in the r.h.s. of eq. (\ref{5.23})
which means that we need all the coefficients $a_{n},b_{n}...c_{n}$ which must
be
obtained using the non-linear evolution equations (\ref{5.2}, \ref{5.5}) etc.

The situation may be simplified using Mueller's
dipole picture\cite{mu94}. Technically, it arises when  in each order in
$\alpha _{s}\ln(\frac {\zeta }{\zeta _{0}})$ we keep only the term
$\Tr\{\Ut_{x}\Udt_{1}\}\Tr\{\Ut_{1}\Udt_{2}\} ... \Tr\{\Ut_{n}\Udt_{y}\}
- {\rm subtractions}~$ \footnote{By "subtractions" we mean
this operator with some of
the $\Tr\{U_{k}\Ud_{k+1}\}$ substituted by $N_{c}$}  in r.h.s. of
eq.~(\ref{4.6}) --
for example, in eq.(\ref{4.5}) we keep the two first terms and disregard the
third one. In other words, we take into account
only those diagrams in Fig.~\ref{fig:OneLoopEvol}
which connect the Wilson lines belonging to the same
$\Tr\{U_{k}\Ud_{k+1}\}$. (This corresponds to the virtual photon
wavefunction in the large-$N_c$ approximation). Then the diagrams
of the corresponding effective theory are obtained by
multiple
iteration of eq.(\ref{4.2}) and give a picture  where each ``dipole"
$\Tr\{U_{k}\Ud_{k+1}\}$ can create two dipoles according to eq.(\ref{4.2}). The
motivation of this approximation is given in Ref.\cite{mu94}.

\section{Conclusion}
\label{sec:cnc}
Let us summarize our results for the operator expansion for high-energy
scattering. It has the form:
\begin{eqnarray}
\lefteqn{\int \! dx\int \! dz\delta (z\bu) T\{j_{\mu }(x+z)j_{\nu }(z)\}}
\nonumber\\
&=&\sum_{n=0}^{\iy}\al_{s}^{n}
\int
dz^{1}dz^{2}...dz^{n}\left[a_{n}(x,z^{1},z^{2},...z^{n},y;
\ln\frac {m^{2}}{s\zeta })\Tr\{\Uz_{x}[x,1]_{-}
\Udz_{1}[1,x]_{+}\}\right.\nonumber\\
&&\left.
\Tr\{\Uz_{1}[1,2]_{-}\Udz_{2}[2,1]_{+}\}
\dots\Tr\{\Uz_{n}[n,y]_{-}\Udz_{y}[y,n]_{+}\}\right.\nonumber\\
&+&\left.b_{n}(x,z^{1},z^{2},...z^{n},y;\ln\frac {m^{2}}{s\zeta })
\Tr\{\Uz_{x}[x,1]_{-}
\Udz_{1}[1,2]_{+}\Uz_{2}[2,3]_{-}\right.\nonumber\\
&&\left.\Udz_{3}[3,1]_{+}\Uz_{1}[1,2]_{-}\Udz_{2}
[2,x]_{+}\}\right.\nonumber\\
&&\left.\Tr\{\Uz_{3}[3,4]_{-}\Udz_{4}[4,3]_{+}\}...
\Tr\{\Uz_{n}[n,y]_{-}\Udz_{y}[y,n]_{+}\}+...\right.\nonumber\\
&&\left.
N_{c}^{n}
c_{n}(x,z^{1},z^{2},...z^{n},y;\ln\frac {m^{2}}{s\zeta })
\Tr\{\Uz_{x}[x,y]_{-}\Udz_{y}[y,x]_{+}\}\right]
\label{6.1}
\end{eqnarray}
where the notations are the same as in eq. (\ref{5.23}). The coefficient
functions $a_{n},b_{n},..c_{n}$ absorb all the information about the
high-energy
($\la\rightarrow\iy$) behavior of the amplitude while the matrix elements of
the Wilson-line operators, however complicated, are low-energy hadron
characteristics. In terms of functional integral representation for the
amplitude (\ref{A.fnl.int}) we make a decomposition of all the fields into
large
rapidity fields (with light-cone fractions $\al>\sqrt {m^{2}\over s\zeta }$)
and
small-rapidity ones (with $\al<\sqrt {m^{2}\over s\zeta }$). The integration
over
large Sudakov variables ($\al>\sqrt {m^{2}\over s\zeta }$) gives us the
coefficient
functions $(a_{n}...c_{n})(x_{\perp k})$ while the integrals over small
$\al<\sqrt {m^{2}\over s\zeta }$ form matrix elements of Wilson-line operators.
Coefficient functions contain logarithms of energy $\ln{s\over m^{2}\zeta }$
while
matrix elements - only $\ln\zeta $. The dependence on $\zeta $ cancels in the
final result and $\ln{s\over m^{2}}$ emerges just as in the case of usual
Wilson
expansion where the dependence of the coefficient functions and matrix elements
on the normalization point is cancelled in a similar way:
$\ln{Q^{2}\over\mu ^{2}}+\ln{\mu ^{2}\over p^{2}}=\ln{Q^{2}\over p^{2}}$.

In order to find a dependence of the amplitude on energy using this \ope\ we
must proceed as follows: first, we integrate over \lico\ fractions $\al\sim 1$
- it gives us the operator $\Tr\{U\Ud\}$ "normalized at the slope $\zeta =\ms$.
Second, using the evolutions equation we reduce the two Wilson-line operators
collinear to $p_{A}$ to the sum of the many-Wilson-line operators (almost)
collinear to $p_{B}$ times the coefficient functions containing $\ln{s\over
m^{2}}$.
In the leading logarithmic approximation, the evolution eqs.
(\ref{5.20},\ref{5.21}) are enough; beyond that, one must consider higher-order
corrections to these evolution eauations. Finally, we must compute the \mael s
of many-Wilson-line operators sandwiched between our target states. In
perturbation theory (e.g. for the \dis\ from the virtual photon) or at large
$N_{c}$ the evolution of the two-Wilson-line operator is enough - others have
the
matrix elements smaller by $g^{2}$ (or $N_{c}$)- and we have the linear BFKL
evolution described by eqs. (\ref{5.4},\ref{5.5}). If the target states are
non-perturbative (e.g. nucleons for \dis\ at small x) we must take into account
the whole non-linear evolution (\ref{5.23}) even in leading logarithmic
approximation.

There is, however, one important difference between \ope\ for \dis\ and our
expansion for high-energy scattering. In the case of Wilson's expansion the
coefficient functions were purely perturbative (up to possible contributions
from small-size vacuum fluctuations \cite{ins}) whereas all the
non-perturbative dynamics was hidden in the matrix elements. This is $not$ the
case for our \ope\ - both coefficient functions and matrix elements can have
perturbative and non-perturbative terms. For Wilson's \ope\ this perturbative
vs. non-perturbative separation was due to the fact that it corresponds to the
separation of the integrals over the transverse momenta: $p\p^{2}>\mu ^{2}$
form
coefficient functions and $p\p^{2}<\mu ^{2}$ matrix elements ( and the
characteristic scale of the coupling constant depends on the scale of
transverse momenta).  For the same reason, separation of integrals over
longitudinal variables has nothing to do with scale of $\al_{s}$ - it is
determined by scale of $p\p$ which can be either lagrge or small independent on
longitudinal momentum. So, since both \mael s and \cofu s can have the
contributions from small and large momenta - both of them do have perturbative
and non-perturbative parts. We have of course calculated only perturbative
contribution toi the \cofu s which comes from the region of large $p\p$ - the
non-perturbative contribution comes from $p\p\sim \Lambda _{QCD}$ and it
corresponds to the soft-pomeron contribution to the coefficient functions. So,
in order to separate perturbative physics from the non-perturbative one for the
high-energy scattering we must do the additional job and split the integrals
over the transverse momenta in hard and soft parts both in coefficient
functions and matrix elements. The study is in progress.

\section{Acknowledgements}
The author is grateful to John Collins for numerous discussions and help.
This work was supported by Department of Energy under
grant DE-FG02-90ER-40577 and
cooperative research agrement DE-FG02-94ER-40818.
\appendix
\section{High-energy asymptotics as a scattering from shock-wave field.}
\label{app:ExtField}
\begin{figure}[htb]
\vspace{0cm}
\mbox{
\epsfxsize=10cm
\epsfysize=9cm
\hspace{2cm}
\epsffile{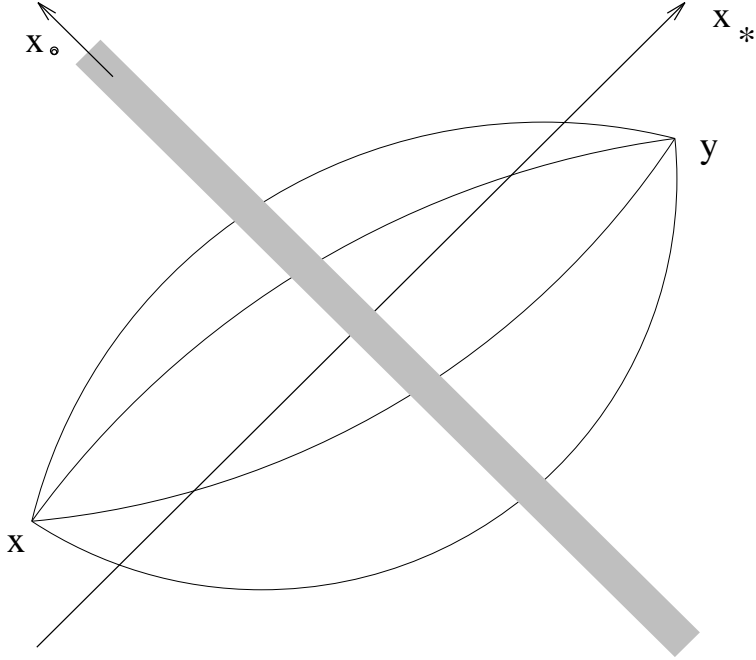}}
\vspace{0cm}
{\caption{\label{fig:pint} Quark propagator in the shock-wave field as a path
      integral.}}
\end{figure}

The structure of the answer (\ref{2.30}) for the high-energy scattering from
external field can be made transparent if instead of rescaling of the incoming
photon's momentum (\ref{lim}) one boosts instead the external field:
\begin{equation}
\int \! dx\int \! dz \delta (z\bu)e^{ip_{A}x}
\lan T\{j_{\mu }(x+z)j_{\nu }(z)\}\ran_{A}~=~
\int \! dx\int \! dz \delta (z\ci)e^{ip^{(0)}_{A}x}
\lan T\{j_{\mu }(x+z)j_{\nu }(z)\}\ran_{B}
\label{a1}
\end{equation}
where $p^{(0)}_{A}=p^{(0)}_{1}+{p_{A}^{2}\over s_{0}}p_{2}$ and the boosted
field
$B_{\mu }$ has the form:
\begin{eqnarray}
B\ci(x\ci,x\sr,x\p)&=&\la A\ci({x\ci\over \la},x\sr\la,x\p)
\nonumber\\
B\sr(x\ci,x\sr,x\p)&=&{1\over\la}A\sr({x\ci\over
\la},x\sr\la,x\p)\nonumber\\
B\p(x\ci,x\sr,x\p)&=&A\p({x\ci\over \la},x\sr\la,x\p)
\label{a2}
\end{eqnarray}
where we used the notations $x\ci\equiv x_{\mu }p_{1\mu }^{(0)},~x\sr\equiv
x_{\mu }p_{2\mu }$. The field
\begin{equation}
A_{\mu }(x\ci,x\sr,x\p)=A_{\mu }({2\over s_{0}}x\ci p_{1}^{(0)}+{2\over
s_{0}}x\sr
p_{2}+x\p)
\label{a3}
\end{equation}
is the original external field in the coordinates independent of $\la$ so we
may assume that the scales of $x\ci,x\sr$ (and $x\p$) in the function
(\ref{a3}) are $O(1)$. First, it is easy to see that at large $\la$ the field
$B_{\mu }(x)$ does not depend on $x\ci$. Moreover, in the limit of very large
$\la$ the field $B_{\mu }$ has a
form of the shock wave. It is especially clear if one writes down the field
strength tensor $G_{\mu \nu }$ for the boosted field. If we assume that
the field strength
$F_{\mu \nu }$ for the external field $A_{\mu }$ vanishes at the infinity we
get
\begin{eqnarray}
G_{\ci i}(x\ci,x\sr,x\p)&=&\la F_{\ci i}({x\ci\over\la},x\sr\la,x\p)
\rightarrow \delta (x\sr)G\p(x\p)
\nonumber\\
G_{\sr i}(x\ci,x\sr,x\p)&=&{1\over\la}F_{\sr i}({x\ci\over
\la},x\sr\la,x\p)\rightarrow 0\nonumber\\
G\csr(x\ci,x\sr,x\p)&=&F\csr({x\ci\over \la},x\sr\la,x\p)
\rightarrow 0\nonumber\\
G_{ik}(x\ci,x\sr,x\p)&=&F_{ik}({x\ci\over \la},x\sr\la,x\p)\rightarrow 0
\label{a4}
\end{eqnarray}
so the only component which survives the infinite boost is $F\cip$ and it
exists only within the thin "wall" near $x\sr=0$. In the rest of the space the
field $B_{\mu }$ is a pure gauge. Let us denote by $\Omega $ the corresponding
gauge matrix and by $B^{\Omega }$ the rotated gauge field which vanishes
everywhere except the thin wall:
\begin{equation}
B^{\Omega }\ci=lim_{\la\rightarrow\iy}\frac {\partial ^{i}}{\partial ^{2}\p}
G_{i\ci}(0,\la x\sr,x\p)\rightarrow
\delta (x\sr)\frac {\partial ^{i}}{\partial ^{2}\p}G_{i}(x\p), ~B\O\sr=B\p=0
\label{a5}
\end{equation}
Let us find the quark propagator in the $B_{\mu }$ background
(see Fig.\ref{fig:pint}). We shall at first calculate the propagator in the
external field $B^{\Omega }$ and after that make the gauge rotation back.

 We start the path-integral representation of a Green function in the external
field:
\begin{eqnarray}
&\LBB x\Big|\frac {1}{\nP}\Big| y\RBB=-i\int _{0}^{\iy}\! d\tau
\LBB x\Big|{\cal P}e^{i\tau{\cal
P}^{2}}\Big| y\RBB=
-i\int _{0}^{\iy}\! d\tau {\cal N}^{-1}
\int _{x(0)=y}^{x(\tau )=x}\!{\cal D}x(t)
\{{1\over 2}\not{\cdot x}+\not\!{B\O}(x(\tau ))\}e^{-i\int _{0}^{\tau }\!
dt{\dot{x}^{2}\over 4}}\nonumber\\
&Pexp\{ig\int _{0}^{\tau }\! dt(B\O_{\mu }(x(t))\dot{x}^{\mu }(t)+{1\over
2}\sigma ^{\mu \nu }G\O_{\mu \nu }(x(t))\}
\label{a6}
\end{eqnarray}
where $\si_{\mu \nu }\equiv {i\over2}
(\gamma _{\mu }\gamma _{\nu }-\gamma _{\nu }\gamma _{\mu })$.
First, it is easy to see that since in our external field (\ref{a4})
the only nonzero components of the field tensor is $G\O_{\ci i\p}$ only
 the first two first term of the expansion of the exponent
$\exp\{\int \! dt{i\over 2}(\sigma G\O)\}$ in powers of $(\si G)$
survive. Indeed,
$\si^{\mu \nu }G\O_{\mu \nu }={4i\over s_{0}}\np^{0}_{2}\gamma ^{i}G\O_{\ci i}$
and therefore $(\si G\O)^{2}\sim (\np_{2}\gamma ^{i})^{2}=0$ since $\np_{2}$
commutes
with $\gamma ^{i}\p$. So, the phase factor for the motion of the particle
 in the external field (\ref{a4}) has the form
\begin{equation}
Pe^{ig\int _{0}^{\tau }\! dtB\O_{\mu }(x(t))\dot{x}_{\mu }(t)}+
{2\gamma ^{i}\np_{2}\over s}\int _{0}^{\tau }\!dt'
Pe^{ig\int _{t'}^{\tau }\! dtB\O_{\mu }(x(t))\dot{x}_{\mu }(t)}
gG\O_{\ci i}(x(t'))Pe^{ig\int ^{t'}_{0}\! dtB\O_{\mu }(x(t))\dot{x}_{\mu }(t)}
\label{a7}
\end{equation}
Let us consider the case $x\sr>0,y\sr<0$ as shown in Fig.\ref{fig:pint}.
Since the  external field exists only within the infinitely thin wall at
$x\sr=0$ we can replace the gauge factor along the actual path $x_{\mu }(t)$ by
the gauge factor along the straight-line path shown in Fig. \ref{fig:pint}
which intersects the plane $x\sr=0$ at the same point  $(z\ci,z\p)$ at which
the original path does. Since the shock-wave field outside the wall vanishes we
may extend formally the limits of this segment to infinity and write the
corresponding gauge factor as $U\O(z\p)=[-\iy p_{1}+z\p,\iy p_{1}+z\p]$ where
label
$\O$ reminds that we calculate this eikonal factor in the field $B\O$. The
error brought by replacement of the original path $inside$ the wall by the
segment of straight line parallel to $p_{1}$ is ${1\over\sqrt \la}$. Indeed,
the
time of the transition of the quark through the wall is proportional to the
thickness of the wall which is $\sim{1\over\la}$ which means that it can
deviate in the perpendicular directions iside the wall only to the distances
$\sim{1\over\sqrt \la}$. Thus, if
the quark intersects this wall at some point $(z\sr,z\p)$ at the time $\tau '$
the gauge factor (\ref{a8}) reduces to:
\begin{equation}
U\O(z\p)+{\gamma ^{i}\!\np_{2}\over\dot{x}\sr(\tau ')}i\partial _{i}U\O(z\p)
\label{a8}
\end{equation}
where the last term was obtained using the identity
\begin{eqnarray}
\lefteqn{\frac {\partial }{\partial x_{i}}U(x\p)}\nonumber\\
&=&-{2i\over s_{0}}\int \! dx\sr
[\iy p^{(0)}_{1}+x\p,{2\over s_{0}}x_{*}p^{(0)}_{1}+x\p]G_{\ci i}({2\over
s_{0}}x_{*}p^{(0)}_{1}+
x\p)[{2\over s_{0}}x_{*}p^{(0)}_{1}+x\p, -\iy p^{(0)}_{1}+x\p]
\label{a9}
\end{eqnarray}
and the factor $\dot{x}\sr(\tau ')$ in eq. (\ref{a7}) comes from changing of
variable of integration from $t$ to $x\sr(t)$. Similarly,
the phase factor for the term in the r.h.s. of eq.(\ref{a6}) which
contains $\not\! B\O(x(\tau ))={2\over s_{0}}\np_{2}B\O\ci(x(\tau ))$ in front
of
the gauge factor
eq.(\ref{a6}) can be reduced to
\begin{equation}
-\np_{2}{\partial \over\partial x_{*}}[{2\over
s_{0}}x_{*}p^{(0)}_{1}+x\p,-\iy+x\p]
=-\np_{2}\delta (x_{*})[U(x\p)-1]
\label{a10}
\end{equation}
(The factor $\sim (\si G)$ is absent since it contains extra $\np_{2}$ and
$\np_{2}^{2}=0$). If we now insert the expression for the phase factors
(\ref{a7},\ref{a10})
into the path integral (\ref{a6}) we obtain
\begin{eqnarray}
&-\np_{2}\delta (x_{*})[U\O(x\p)-1]\int _{0}^{\iy}\! d\tau {\cal
N}^{-1}\int ^{x(\tau )=x}_{x(0)=y}\!{\cal D}x(t)
e^{-i\int _{\tau }^{0}\! dt{\dot{x}^{2}\over 4}}-
\nonumber\\
&{i\over 2}\int _{0}^{\iy}\! d\tau \int _{0}^{\tau }\! d\tau '\int \! dz\delta
(z_{*})
{\cal N}^{-1}\int _{x(\tau ')=z}^{x(\tau )=x}\!{\cal D}x(t)\!
\not{\dot x}(\tau )\nonumber\\
&e^{-i\int _{\tau '}^{\tau }\! dt{\dot{x}^{2}\over 4}}\{U\O(z\p)+{i\over
\dot{x}_{*}(\tau ')}\not{\partial }U\O(z\p)\np_{2}\}{\cal
N}^{-1}\int _{x(0)=y}^{x(\tau ')=z}\!{\cal D}x(t)
\dot{x}_{*}(\tau ')
e^{-i\int _{\tau '}^{\tau }\! dt{\dot{x}^{2}\over 4}}
\label{a11}
\end{eqnarray}
 The additional Jacobian factor
 $\dot{x}\sr(\tau ')$ in the numerator in the second term in r.h.s.
 of this eq. comes due to the fact that we must integrate over all
$\tau '$ from $0$ to $\tau $ and therefore we insert
$1=\int \!d\tau '\dot{x}\sr(\tau ')\delta (x\sr(\tau ')-z\sr))$ in the
functional integral (\ref{a6}).
It is convenient to make a shift of time variable $\tau '$ and to rewrite
the eq.(\ref{a10}) in the following way:
\begin{eqnarray}
&-\np_{2}\delta (x\sr)[U\O(x\p)-1]\int _{0}^{\iy}\! d\tau {\cal
N}^{-1}\int ^{x(\tau )=x}_{x(0)=y}\!{\cal D}x(t)
\!\not{\dot x}(\tau )e^{-i\int _{\tau }^{0}\! dt{\dot{x}^{2}\over 4}}
\nonumber\\
&-{i\over 2}\int _{0}^{\iy}\! d\tau \int _{0}^{\iy}\! d\tau '\int \! dz\delta
(z\sr)
{\cal N}^{-1}\int _{x(0)=z}^{x(\tau )=x}\!{\cal D}x(t)\!
\not{\dot x}(\tau )e^{-i\int _{0}^{\tau }\! dt{\dot{x}^{2}\over 4}}{\cal
N}^{-1}\int _{x(0)=y}^{x(\tau ')=z}\!{\cal D}x(t)\nonumber\\
&
\{\dot{x}\sr(\tau ')U\O(z\p)+i\!\not{\partial _{i}}U\O(z\p)\!\np_{2}\}
e^{-i\int _{\tau '}^{\tau }\! dt{\dot{x}^{2}\over 4}}
\label{a12}
\end{eqnarray}
Now, using the path-integral representations for bare propagators:
\begin{equation}
\int _{0}^{\iy}\! d\tau {\cal N}^{-1}\int ^{x(\tau )=x}_{x(0)=y}\!{\cal D}x(t)
(\tau )e^{-i\int ^{\tau }_{0}\! dt{\dot{x}^{2}\over 4}}=-\frac {1}{4\pi
^{2}(x-y)^{2}}
\label{a13}
\end{equation}
and
\begin{equation}
\int _{0}^{\iy}\! d\tau {\cal N}^{-1}\int ^{x(\tau )=x}_{x(0)=y}\!{\cal D}x(t)
\dot{x}_{\mu }(\tau )e^{-i\int ^{\tau }_{0}\! dt{\dot{x}^{2}\over 4}}=\frac {
i(x-y)_{\mu }}{\pi ^{2}(x-y)^{4}}
\label{a14}
\end{equation}
it is easy to see that the path-integral expression for the quark propagator in
the shock-wave field (\ref{a13}) reduces to
\begin{eqnarray}
 \lefteqn{\LBB x\Big|\frac {1}{\nP}\Big| y\RBB}\nonumber\\
&=&
\frac {\np_{2}}{4\pi ^{2}(x-y)^{2}}\delta (x\sr)[U\O-1](x\p)
+\int \! dz\delta (z\sr)
\frac {(\nx-\nz)\np_{2}}{2\pi ^{2}(x-z)^{4}}\{U\O(z\p)\frac {-2iy\sr}{2\pi
^{2}(z-y)^{4}}\nonumber\\
&-&i\!\not{\partial }\p U\O(z\p)\frac {\np_{2}}{4\pi  ^{2}(z-y)^{2}}\}=i\int \!
dz\delta (z\sr)
\frac {(\nx-\nz)\np_{2}}{2\pi ^{2}(x-z)^{4}}U\O(z\p)
\frac {\nz-\ny}{2\pi ^{2}(z-y)^{4}}
\label{a15}
\end{eqnarray}
(in the region $x\sr>0,y\sr<0$). It can be demonstrated that the answer for the
propagator in the region
$x\sr<0,y\sr>0$ differs from eq. (\ref{a15}) by the substitution
$U\O\leftrightarrow U^{\Omega \dagger }$. Also, the propagator outside the
shock-wave wall (at $x\sr,y\sr<0$ or $x\sr,y\sr>0$) coinside with bare
propagator so the final answer for the quark Green function in the $B\O$
background can be written down as:
\begin{eqnarray}
\lefteqn{\LBB x\Big|\frac {1}{\nP}\Big| y\RBB=}\nonumber\\
&&
-\frac {\nx-\ny}{2\pi ^{2}(x-y)^{4}}+i\int \! dz\delta (z\sr)
\frac {(\nx-\nz)\np_2}{2\pi ^{2}(x-z)^{4}}\{[U\O-1](z\p)\Theta (x\sr)\Theta
(-y\sr)-\nonumber\\
&&[U^{\Omega \dagger }-1](z\p)\Theta (y\sr)\Theta (-x\sr)\}
\frac { \nz-\ny}{2\pi
^{2}(z-y)^{4}}
\label{a16}
\end{eqnarray}
where we have used the formula
\begin{equation}
i\int \! dz \delta (z\sr)\frac {\nx-\nz}{2\pi ^{2}(x-z)^{4}}\np_{2}
\frac {\nz-\ny}{2\pi ^{2}(z-y)^{4}}=-\frac { \nx-\ny}{2\pi
^{2}(x-y)^{4}}(\Theta (x\sr)-\Theta (y\sr))
\label{a17}
\end{equation}
to separate the bare propagator.
In the momentum representation this answer (\ref{a16}) takes the form:
\begin{eqnarray}
\lefteqn{\LBB k\Big|\frac {1}{\nP}\Big| k-p\RBB}\\
\label{a18}
&=&(2\pi )^{4}\delta ^{(4)}(p)\frac
{\al^{0}_{k}\np^{0}_{1}+
\be_{k}\np_{2}+\nk\p}{\al^{0}_{k}\be_{k}s_{0}-k\p^{2}+\ie}~+\nonumber\\
&&
2\pi i\delta (\al_{p})\frac {
(\al_{k}\np^{0}_{1}+\nk\p)p_{2}}{\al^{0}_{k}\be_{k}s_{0}-k\p^{2}+\ie}{2\over
s_{0}}[\Theta (\al^{0}_{k})(U\O(p\p)-4\pi ^{2}\delta (p\p))-\nonumber\\
&&\Theta (-\al^{0}_{k})(U^{\Omega \dagger }
(p)-4\pi ^{2}\delta (p\p))]\frac {\al^{0}_{k}\np^{0}_{1}+(\nk-\np)\p}{
\al^{0}_{k}(\be_{k}-\be_{p})s_{0}-(k-p)\p^{2}+\ie}\nonumber
\end{eqnarray}
which agrees with eq.(\ref{qprop}) after integration over $\al^{0}_{p}$ and
rescaling
$\al_{p}={1\over\la}\al^{0}_{p}$ (here $\al^{0}$ is the Sudakov component along
vector
$p^{0}_{1}$).

\begin{figure}[htb]
\vspace{0cm}
\mbox{
\epsfxsize=10cm
\epsfysize=10cm
\hspace{2cm}
\epsffile{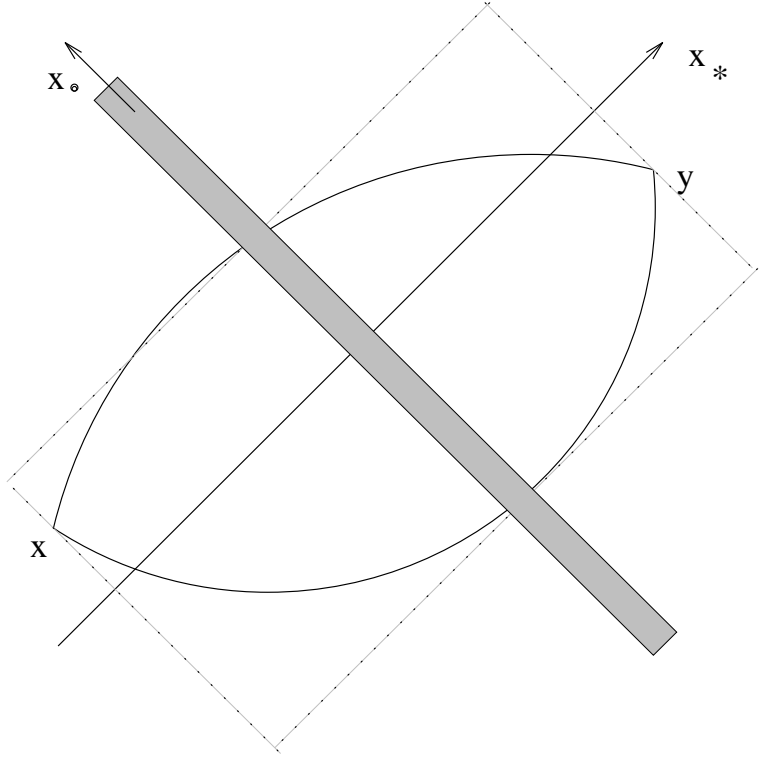}}
\vspace{0cm}
{\caption{\label{fig:14} Quark-antiquark propagation in the shock wave.}}
\end{figure}

Now, one easily obtains the quark propagator in the original field $B_{\mu }$
eq. (\ref{a2}) by making back the gauge rotation of the answer (\ref{a16}) with
matrix $\Omega ^{-1}$.
It is convenient to represent the result in the following form:
\begin{eqnarray}
\lefteqn{\LBB x\Big|\frac {1}{\nP}\Big| y\RBB~=}\\
\label{a19}
&-&
\frac {\nx-\ny}{2\pi ^{2}(x-y)^{4}}[x,y]\Theta (x\sr y\sr)\nonumber\\
&+&i\int \! dz\delta
(z\sr)
\frac {\nx-\nz}{2\pi ^{2}(x-z)^{4}}\{U(z\p;x,y)\Theta (x\sr)\Theta
(-y\sr)-\Ud(z\p;x,y)\Theta (y\sr)\Theta (-x\sr)\}\frac { \nz-\ny}{2\pi
^{2}(z-y)^{4}}\nonumber
\end{eqnarray}
where
\begin{equation}
U(z\p;x,y)=[x,z_{x}][z_{x},z_{y}][z_{y},y]~~,
{}~~z_{x}\equiv ({2\over s_{0}}z\ci p^{(0)}_{1}+{2\over s_{0}}x\sr
p_{2},z\p),~~z_{y}=z_{x}(x\sr\leftrightarrow y\sr)
\label{a20}
\end{equation}
is a gauge factor for the contour made from segments of straight lines as shown
in Fig. \ref{fig:14}. (Since the field $B_{\mu }$ outside the shock-wave
wall is a pure gauge the precise form of the contour does not matter as long as
it starts at the point $x$, intersects the wall at the point $z$ in the
direction collinear to $p_{2}$ and ends at the point $y$).

For the quark-antiquark amplitude in the shock-wave field
(see Fig \ref{fig:14})
 we get
\begin{eqnarray}
\lefteqn{\Tr\ga_{\mu }\LBB x\Big|\frac {1}{\nP}\Big| y\RBB\ga_{\nu }
\LBB y\Big|\frac {1}{\nP}\Big| x\RBB=}\\
\label{a21}
&&
\frac {Tr\ga_{\mu }(\nx-\ny)\ga_{\nu }(\ny-\nx)}{4\pi ^{4}(x-y)^{8}}
\Theta (x\sr y\sr)-\nonumber\\
&&\Theta (-x\sr y\sr)\int \! dz\delta (z\sr)
\int \! dz'\delta (z'\sr)Tr\ga_{\mu }\frac {\nx-\nz}{2\pi
^{2}(x-z)^{4}}\np_{2}\frac {\nz-\ny}{2\pi ^{2}(z-y)^{4}}\ga_{\nu }
\frac {\ny-\nz'}{2\pi ^{2}(y-z')^{4}}\np_{2}\frac { \nz'-\nx}{2\pi
^{2}(z'-x)^{4}}W(z\p;z'\p)\nonumber
\end{eqnarray}
where we can write down the gauge factor $W(z\p;z'\p)\equiv
U(z\p;x,y)\Ud(z'\p;y,x)$ as a product of two infinite Wilson-lines operators
connected by gauge segments at $\pm \iy$:
\begin{eqnarray}
\lefteqn{W(z\p;z'\p)}\\
\label{a22}
&=&\lim_{u\rightarrow\iy}[-up_{1}+z\p,up_{1}+z\p]
[up_{1}+z\p,up_{1}+z'\p][up_{1}+z'\p,-up_{1}+z'\p][-up_{1}+z'\p,-up_{1}+z\p]
\nonumber
\end{eqnarray}
The precise form of the connecting contour does not matter as long as it is
outside the shock wave. We have chosen this contour in such a way that the
gauge factor (\ref{a22}) is the same for the field $B_{\mu }$ and for the
original field $A_{\mu }$ (see eq. (\ref{a2}). Now, substituting our result for
quark-antiquark propagation (\ref{a21}) in the r.h.s of eq. (\ref{a1}) one
recovers after some algebra the eqs. (\ref{2.30}, \ref{2.31}) for the impact
factor.

\section{One-loop evolution: Wilson lines in a shock-wave background.}
\label{app:B}

\begin{figure}[htb]
\vspace{0cm}
\mbox{
\epsfxsize=15cm
\epsfysize=5cm
\hspace{0cm}
\epsffile{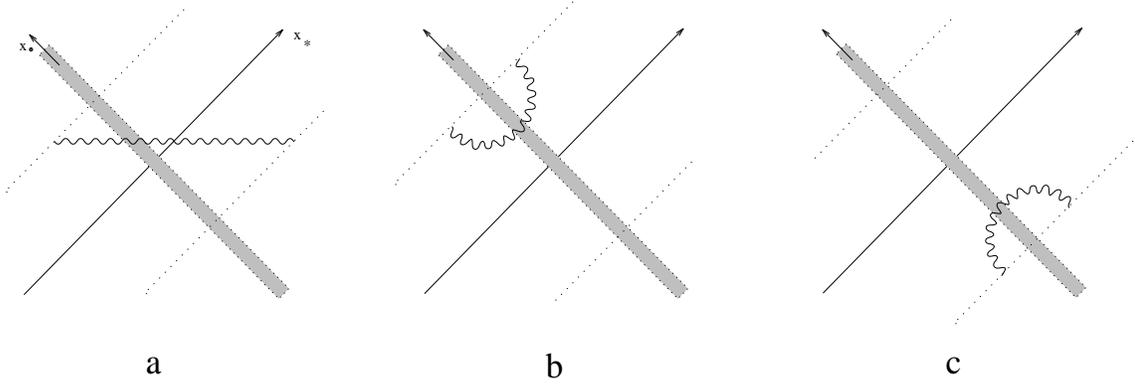}}
\vspace{1cm}
{\caption{\label{fig:pint1}Path integrals describing one-loop diagrams for
Wilson-line operators in
the shock-wave field background.}}
\end{figure}

Let us now find how the one-loop evolution of Wilson-line operators can be
obtained using the shock-wave picture of high-energy scattering. To this end,
consider the near-light-like operators $\Uhz$ and $\Uhdz$ in the external
field.
Making the rescaling (\ref{a2}) we obtain:
\begin{eqnarray}
&\lan[\iy p_{A}+x\p,-\iy p_{A}+y\p][-\iy p_{A}+x\p,\iy
p_{A}+y\p]\ran_{A}=\nonumber\\
&\lan[\iy p^{(0)}_{A}+x\p,-\iy p^{(0)}_{A}+y\p][-\iy p^{(0)}_{A}+x\p,\iy
p^{(0)}_{A}+y\p]\ran_{B}
\label{b1}
\end{eqnarray}
where the shock-wave field is given by formulas (\ref{a2}-\ref{a4})
We must find the derivative $\zeta {\partial \over\partial \zeta }$ given by
eq.
(\ref{4.2}). After rescaling according to eq. (\ref{4.2}) one obtains:
\begin{eqnarray}
\lefteqn{\zeta {\partial \over\partial \zeta
}\lan\Uh(x\p)\Uhd(y\p)\ran_{A}}\nonumber\\
&=&ig{p_{A}^{2}\over
s_{0}}\int \! udu \lan[\iy p\an+x\p, up\an+x\p]\hat{F}\sci
(up\an+x\p)[up\an+x\p,-\iy p\an+x\p]\Uhd(y\p)\ran_{B}-\nonumber\\
&&ig{p_{A}^{2}\over s_{0}}\int \! udu
\lan\Uh(x\p)[-\iy p\an+y\p, up\an+y\p]\hat{F}\sci
(up\an+y\p)[up\an+y\p,\iy p\an+y\p]\ran_{B}
\label{b2}
\end{eqnarray}
Since the ($\sci$) component of the field strength tensor vanishes for the
shock-wave field (\ref{a4}) the only nonzero contribution comes from the
diagrams with quantum gluons. In the lowest nontrivial order in $\al_{s}$ there
are three
diargams shown in Fig. \ref{fig:pint1}.

 Consider at first the diagram shown in Fig. \ref{fig:pint1}a (which
corresponds to the case $x\sr>0$, $y\sr<0$). The corresponding contribution to
r.h.s. of eq. (\ref{b2}) is:
\begin{eqnarray}
&-g^{2}\int du [\iy p\an+x\p,
up\an+x\p]t^{a}[up\an+x\p,-\iy
p\an+x\p]\nonumber\\
&\otimes\int dv [-\iy p\an+y\p,
vp\an+y\p]t^{b}[vp\an+y\p,\iy
p\an+y\p]\nonumber\\
&\LBB up^{(0)}_A+x\p\Big| up\sr\{(p^{(0)}_{A\xi }-{\cal
P}\ci\frac {p_{2\xi }}{p\cdot p_{2}})[\frac {1}{{\cal P}^{2}g_{\xi  \eta }+
2iG_{\xi \eta }}
-\frac {1}{{\cal P}^{2}g_{\xi \lambda }+2iG_{\xi \lambda
}}(D^{\al}G_{\al\lambda }
\frac {p_{2\rho }}{p\cdot p_{2}}
+\frac {p_{2\lambda }}{p\cdot p_{2}}
D^{\al}G_{\al\rho }-\nonumber\\
&\frac {p_{2\lambda }}{p\cdot p_{2}}{\cal
P}^{\be}D^{\al}G_{\al\be}\frac {p_{2\rho }}{p\cdot p_{2}})\frac {1}{{\cal
P}^{2}g_{\rho \eta }+2iF_{\rho \eta }}+...](p^{(0)}_{A\eta }
-\frac {p_{2\eta }}{p\cdot p_{2}}{\cal
P}\ci)\}-v\left\{...\right\}p\sr\Big| vp\an+y\p\RBB_{ab}
\label{b3}
\end{eqnarray}
As we discussed in Sect.4 terms in parentheses proportional to ${\cal P}\ci$
vanish after integration by parts (cf. eq. (\ref{4.4})). Further, it is easy to
check that since the only nonzero component of field strength tensor for the
shock wave is $G\O\cip$ the expression in braces in eq. (\ref{b3}) can be
reduced to ${\cal O}\O\cici$ where the operator ${\cal O}\O_{\mu \nu }$ is
given
by eq. (\ref{4.6}). Starting from this point it is convenient to perform the
calculation in the background of the rotated field $B\O$ (\ref{a5}) which is
$0$ everywhere except the shock-wave wall. (We shall make the rotation back to
field $B$ in the final answer). Then the gauge factors $[\iy,u]t^{a}[u,-\iy]$
and $[\iy,v]t^{b}[v,-\iy]$ in eq. (\ref{b3}) reduce to $t^{a}[\iy,-\iy]\otimes
t^{b}[-\iy,\iy]$ (at $x\sr>0,y\sr<0$) and we obtain:
\begin{equation}
-g^{2}t^{a}U\O\otimes t^{b}U^{\dagger \Omega }\int \! du\int \! dv
(u-v)\LBB up^{(0)}_A+x\p\Big|p\sr{\cal O}\O\cici\Big| vp\an+y\p\RBB_{ab}
\label{b4}
\end{equation}
where we have used the fact that the operator $p\sr$ commutes with ${\cal
O}\O$. Let us now derive the formula for the $(\cici)$ component of the gluon
propagator $(x|{\cal O}\O|y)$ in the shock-wave background. The path-integral
representation of $(x|{\cal O}\O\cici|y)$ has the form:
\begin{eqnarray}
\lefteqn{\LBB x\Big|4\frac {1}{{\cal P}^{2}}G^{\xi \Omega }_{\circ }
\frac { 1}{{\cal
P}^{2}}G\O_{\xi \circ }\frac {1}{{\cal P}^{2}}-\frac {1}{{\cal
P}^{2}}(D^{\al}G\O_{\al\circ }\frac {s_{0}}{2p\sr}
+\frac {s_{0}}{2p\sr}
D^{\al}G\O_{\al\nu }-\frac {s_{0}}{2p\sr}{\cal
P}^{\be}D^{\al}G\O_{\al\be}\frac {s_{0}}{2p\sr})\frac {1}{{\cal
P}^{2}}\Big| y\RBB}\nonumber\\
&=&i\int _{0}^{\iy}\! d\tau \int _{0}^{\tau }\! d\tau '\LBB x\Big|e^{i(\tau
-\tau '){\cal
P}^{2}}\{G^{\al\Omega }\ci\int _{0}^{\tau '}\! d\tau ''e^{i(\tau '-\tau
''){\cal P}^{2}}
\{G\O_{\al\circ }e^{i\tau ''{\cal P}^{2}}-{is_{0}\over 2p\sr}D^{\al}
G\O_{\al\circ
}e^{i\tau ''{\cal P}^{2}}\}\Big| y\RBB\nonumber\\
&=&
i\int _{0}^{\iy}\! d\tau {\cal N}^{-1}
\int _{x(0)=y}^{x(\tau )=x}\!{\cal D}x(t)
e^{-i\int _{0}^{\tau }\! dt{\dot{x}^{2}\over 4}}
\{4\int _{0}^{\tau }\!d\tau '\int _{0}^{\tau ' }\!d\tau ''
Pe^{ig\int _{\tau '}^{\tau }\! dtB\O_{\mu }(x(t))\dot{x}_{\mu }(t)}
gG\O_{\ci i}(x(\tau '))\nonumber\\
&&Pe^{ig\int ^{\tau '}_{\tau ''}\! dtB\O_{\mu }(x(t))\dot{x}_{\mu }(t)}
\int _{0}^{\tau '}\!d\tau ''
Pe^{ig\int _{\tau ''}^{\tau '}\! dtB\O_{\mu }(x(t))\dot{x}_{\mu }(t)}
gG\O_{\ci i}(x(\tau ''))Pe^{ig\int ^{\tau ''}_{0}\! dtB\O_{\mu
}(x(t))\dot{x}_{\mu }(t)}\nonumber\\
&&+i\int _{0}^{\tau }\!d\tau '
Pe^{ig\int _{\tau '}^{\tau }\! dtB\O_{\mu }(x(t))\dot{x}_{\mu }(t)}
{s_{0}\over \dot{x\sr}(\tau ')}gD^{\al}G\O_{\al\circ }(x(\tau '))Pe^{ig\int
^{\tau '}_{0}\! dtB\O_{\mu }(x(t))\dot{x}_{\mu }(t)}\}
\label{b5}
\end{eqnarray}
As we discussed above, the transition through the shock wave occurs in a short
time $\sim {1\over \la}$ so the gluon has no time to deviate in the transverse
directions and therefore the gauge factors in eq. (\ref{b5}) can be
approximated by segments of Wilson lines. One obtains then (cf. eq.
(\ref{a12})):
\begin{eqnarray}
\lefteqn{\LBB x\Big|{\cal O}\O\cici\Big| y\RBB}\nonumber\\
&=&
{i\over 2}s_{0}^{2}\int _{0}^{\iy}\! d\tau \int _{0}^{\tau }\! d\tau '\int \!
dz\delta
(z_{*})
{\cal N}^{-1}\int _{x(\tau ')=z}^{x(\tau )=x}\!{\cal D}x(t)\!
e^{-i\int _{\tau '}^{\tau }\! dt{\dot{x}^{2}\over 4}}\nonumber\\
&&{1\over
\dot{x}\sr(\tau ')}\{2[GG]\O(z\p)-i[DG]\O(z\p)\}{\cal
N}^{-1}\int _{x(0)=y}^{x(\tau ')=z}\!{\cal D}x(t)
e^{-i\int _{\tau '}^{\tau }\! dt{\dot{x}^{2}\over 4}}
\label{b6}
\end{eqnarray}
where $[GG]\O$ and $[DG]\O$ are the notations for the gauge factors
(\ref{4.10})
calculated for the background field $B\O_{\mu }$:
\begin{eqnarray}
&[DG]\O(x\p)=&\int du[\iy p_{1}+x\p, up_{1}+x\p]
D^{\al}G\O_{\al\circ }(up_{1}+x\p)[up_{1}+x\p,-\iy p_{1}+x\p]\nonumber\\
&[GG]\O(x\p)=&\int \! du\int \! dv \Theta (u-v) [\iy p_{1}+x\p, up_{1}+x\p]
G^{\xi \Omega }_{\circ }(up_{1}+x\p)\nonumber\\
&&[up_{1}+x\p,vp_{1}+x\p]
G\O_{\xi \circ }(vp_{1}+x\p)[vp_{1}+x\p, -\iy p_{1}+x\p]
\label{b7}
\end{eqnarray}
As we noted in Sect.4 the gauge factor $-i[DG]+2[GG]$ in braces
in eq. (\ref{b5}) is in
fact the total derivative of $U$ with respect to translations in the
perpendicular directions so we get:
\begin{eqnarray}
\lefteqn{\LBB x\Big|{\cal O}\O\cici\Big| y\RBB~=}\\
\label{b8}
&&{i\over 2}s_{0}^{2}\int _{0}^{\iy}\! d\tau
\int _{0}^{\tau }\! d\tau '\int \!
dz\delta(z_{*})
{\cal N}^{-1}\int _{x(\tau ')=z}^{x(\tau )=x}\!{\cal D}x(t)\!
e^{-i\int _{\tau '}^{\tau }\! dt{\dot{x}^{2}\over 4}}{1\over
\dot{x}\sr(\tau ')}\partial \p^{2}U\O(x\p){\cal
N}^{-1}\int _{x(0)=y}^{x(\tau ')=z}\!{\cal D}x(t)
e^{-i\int _{\tau '}^{\tau }\! dt{\dot{x}^{2}\over 4}}\nonumber
\end{eqnarray}
Using now the path-integral representation for bare propagator (\ref{a13})
and the following formula
\begin{equation}
\int _{0}^{\iy}\! d\tau {\cal N}^{-1}\int ^{x(\tau )=x}_{x(0)=y}\!{\cal D}x(t)
{1\over\dot{x}\sr(0)}e^{-i\int ^{\tau }_{0}\! dt{\dot{x}^{2}\over 4}}=
i\frac {\ln (x-y)^{2}}{16\pi ^{2}(x-y)\sr}
\label{b9}
\end{equation}
we finally obtain the $(\cici)$ component of the gluon propagator in the
shock-wave background in the form:
\begin{eqnarray}
\lefteqn{\LBB x\Big|{\cal O}\O\cici\Big| y\RBB}\\
\label{b10}
&=&
{s_{0}^{2}\over 2}\int \! dz\delta (z\sr)
\frac {\ln(x-z)^{2}}{16\pi ^{2}x\sr}[\partial ^{2}U\O(z\p)\Theta (x\sr)
\Theta (-y\sr)-\partial ^{2}U^{\dagger \Omega }(z\p)\Theta (-x\sr)
\Theta (y\sr)]\frac {1}{4\pi ^{2}(z-y)^{2}}\nonumber
\end{eqnarray}
where we have added the similar term corresponding to the case $x\sr<0,y\sr>0$.
We need also the ${\partial \over\partial x\ci}$ derivative of this propagator
(see eq. (\ref{b4}) which is:
\begin{equation}
\LBB x\Big|p\sr{\cal O}\O\cici\Big| y\RBB=
{is_{0}^{2}\over 64\pi ^{4}}\int \!
dz\frac {\delta (z\sr)}{(x-y)^{2}}
[\partial ^{2}U\O(z\p)\Theta (x\sr)\Theta (-y\sr)-\partial ^{2}
U^{\dagger \Omega }(z\p)\Theta (-x\sr)\Theta (y\sr)]\frac {1}{(z-y)^{2}}
\label{b11}
\end{equation}
In the momentum representation this eq. takes the form:
\begin{eqnarray}
\lefteqn{\LBB k\Big|{\cal O}\O\cici\Big| k-p\RBB_{ab}~=~}\\
\label{b12}
\hspace{-1cm}
&&\frac {-is_{0}}{2}2\pi \delta
(\al^{0}_{p})\frac { 1}{\al^{0}_{k}\be_{k}s-k\p^{2}+\ie}
(\Theta (\al^{0}_{k})(\partial \p^{2}U\O(p\p))-\Theta (-\al^{0}_{k})
(\partial \p^{2}\UdO(p\p))
\frac {1}{\al^{0}_{k}(\be_{k} -\be_{p})s-(k-p)\p^{2}+\ie}\nonumber
\end{eqnarray}
which agrees with eq. (\ref{4.8}) after rescaling
$\al_{p}={1\over\la}\al^{0}_{p}$.
Substituting now the eq. (\ref{b11}) into eq. (\ref{b4}) one recovers eq.
(\ref{4.10}) after some algebra
\begin{eqnarray}
&{g^{2}\over  4\pi }\LBB x\p\Big|\frac {1}{p^{2}}\partial \p^{2}U\O
\frac {1}{p^{2}}\Big| y\p\RBB_{ab}t^{a}U\O(x\p)\otimes t^{b}
U^{\dagger \Omega }(y\p)+\nonumber\\
&{g^{2}\over  4\pi }\LBB x\p\Big|\frac {1}{p^{2}}
\partial \p^{2}U^{\dagger \Omega }\frac {1}{p^{2}}\Big| y\p\RBB_{ab}
U\O(x\p)t^{a}\otimes U^{\dagger \Omega }(y\p)t^{b}
\label{b13}
\end{eqnarray}

Let us consider now the diagram shown in Fig. \ref{fig:pint1}b. The calculation
is very similar to the case of Fig. \ref{fig:pint1}a diagram considered above
so we shall only briefly outline the calculation. One starts with the
corresponding contribution to r.h.s.of eq. (\ref{b2}) which has the form (cf.
(\ref{b3}):
\begin{eqnarray}
\lefteqn{}\nonumber\\
&-&g^{2}\zeta \int \! du \int \! dv \Theta (u-v) [\iy p\an+x\p,
up\an+x\p]t^{a}[up\an+x\p,vp\an+x\p]\nonumber\\
&&t^{b}[vp\an+x\p,-\iy
p\an+x\p]\otimes \Ud(y\p)\nonumber\\
&&\LBB up^{(0)}_A+x\p\Big|up\sr\{(p^{(0)}_{A\xi }-{\cal
P}\ci\frac {p_{2\xi }}{p\cdot p_{2}})[\frac {1}{{\cal P}^{2}g_{\xi  \eta }+
2iG_{\xi \eta }}
-\frac {1}{{\cal P}^{2}g_{\xi \lambda }+2iG\O_{\xi \lambda
}}(D^{\al}G\O_{\al\lambda }
\frac {p_{2\rho }}{p\cdot p_{2}}
+\frac {p_{2\lambda }}{p\cdot p_{2}}
D^{\al}G\O_{\al\rho }-\nonumber\\
&&\frac {p_{2\lambda }}{p\cdot p_{2}}{\cal
P}^{\be}D^{\al}G_{\al\be}\frac {p_{2\rho }}{p\cdot p_{2}})\frac {1}{{\cal
P}^{2}g_{\rho \eta }+2iG_{\rho \eta }}+...](p^{(0)}_{A\eta }
-\frac {p_{2\eta }}{p\cdot p_{2}}{\cal
P}\ci)\}-v\left\{...\right\}p\sr\Big| vp\an+x\p\RBB_{ab}
\label{b14}
\end{eqnarray}
As we demonstrated in Sect.4 the terms in parentheses proportional to ${\cal
P}\bu$ vanish and after that the operator in braces reduce to ${\cal O}\cici$.
Again, it is convenient to make a gauge transformation to the rotated field
(\ref{a5}) which is 0 everywhere except the shock wave.  Then the gauge factor
$[\iy,u]t^{a}[u,v]t^{b}[v,-\iy]$ in eq. (\ref{b12}) reduces to $t^{a}[\iy,-\iy]
t^{b}$ (at $x\sr>0,y\sr<0$) and we obtain:
\begin{equation}
-g^{2}t^{a}U\O t^{b}\otimes U^{\dagger \Omega }\int \! du\int \! dv
(u-v)\LBB up^{(0)}_A+x\p\Big|p\sr{\cal O}\O\cici\Big| vp\an+x\p\RBB_{ab}
\label{b15}
\end{equation}
Using the expression (\ref{b11}) for the gluon propagator in the shock-wave
background after some algebra one obtains the answer (\ref{4.20}):
\begin{equation}
-{g^{2}\over 4\pi } t^{a}U\O(x\p)t^{b}\otimes\UdO(y\p)
\LBB x\p\Big|\frac {1}{p^{2}}
(\partial ^{2}U\O)\frac {1}{p^{2}}\Big| x\p\RBB_{ab}
\label{b16}
\end{equation}
The contribution of the diagram in Fig. \ref{fig:pint1} differs from eq.
(\ref{b16}) only in change $U\leftrightarrow\Ud, ~x\leftrightarrow y$ (see eq.
(\ref{a21})). Combining these expressions, one obtains the answer in the
rotated field (\ref{a5}) in the form:
\begin{eqnarray}
\lefteqn{{g^{2}\over 16\pi ^{3}}\int \!dz\p}\nonumber\\
&&\left\{[\{\UdO(z\p) U\O(x\p)\}^{k}_{j} \{
U\O(z\p)\UdO(y\p)\}^{i}_{l} +
\{U\O(x\p)\UdO(z\p)\}^{i}_{l}\{\UdO(y\p)U\O(z\p)\}^{k}_{j}-
\right.\nonumber\\
&&\left.
\delta ^{k}_{j}\{U\O(x\p)\UdO(y\p)\}^{i}_{l}-\delta ^{i}_{l}
\{\UdO(y\p)U\O(x\p)\}^{k}_{j}]\frac {(x-z,y-z)\p}{(x-z)\p^{2}(y-z)\p^{2}}-
\right.\nonumber\\
&&\left.
[\{U\O(z\p)\}^{i}_{j}\Tr\{U\O(x\p)\UdO(z\p)\}-N_{c}
\{U\O(x\p)\}^{i}_{j})\UdO(y\p)^{k}_{l}\frac {1}{(x-z)\p^{2}}-
\right.\nonumber\\
&&\left.
\{U\O(x\p)\}^{i}_{j}
[\UdO(z\p)^{k}_{l}\Tr\{U\O(z\p)\UdO(y\p)\}-N_{c}\{\UdO(y\p)\}^{k}_{l})]\frac {
1}{(y-z)\p^{2}}\right\}\nonumber\\
\label{b17}
\end{eqnarray}
Now we must perform the gauge rotation back to the "original" field $B_{\mu }$.
The answer is especially simple if we consider the evolution of the
gauge-invariant operator such as
$Tr\{U(x\p)[x\p,y\p]_{-}\Ud(y\p)[y\p,x\p]_{+}\}$
where the Wilson lines are connected by gauge segments at the infinity, see eq.
(\ref{endf}). We have then:
\begin{eqnarray}
\lefteqn{\zeta {\partial \over\partial \zeta }\lan
\Tr\{\Uhz(x\p)[x\p,y\p]_{-}\Uhdz(y\p)[y\p,x\p]_{+}\}\ran_{A}}\nonumber\\
&=&
{\alpha _{s}\over 4\pi } \int \! dz\p(\Tr\{U(x\p)[x\p,z\p]_{-}
\Ud(z\p )[z\p,x\p]_{+}\} \Tr\{U(z\p)[z\p,y\p]_{-}\Ud(y\p)[y\p,z\p]_{+}\}
\nonumber\\
&-&N_{c}\Tr\{U(x\p)[x\p,y\p]_{-}\Ud(y\p)[y\p,x\p]_{+}\})
\frac {(x\p-y\p)^{2}}{(x\p-z\p)^{2}(z\p-y\p^{2}}
\label{b18}
\end{eqnarray}
where we have replaced the end gauge factors like $\Omega (\iy
p_{1}+x\p)\Omega ^{\dagger }(\iy p_{1}+y\p)$ and $\Omega (-\iy
p_{1}+x\p)\Omega ^{\dagger }(-\iy p_{1}+y\p)$ by segments of gauge line
$[x\p,y\p]_{+}$
and $[x\p,y\p]_{-}$ respectively.
Since the background field $B_{\mu }$ is a pure gauge outside the shock wave
the
specific form of the contour in eq. (\ref{b18}) does not matter as long as it
has the same initial and final points. Finally, note that the gauge factors in
r.h.s. of eq. (\ref{b18}) preserve their form  after rescaling back to the
field $A_{\mu }$ so we reproduce the eq. (\ref{5.2}).

It is instructive also to see along which variable do the leading logarithmic
integration actually goes. To this end we must find matrix element of the
operator $\Uh(x\p)\Uhd(y\p)$ (see r.h.s. of eq. (\ref{b1})) in the shock-wave
background. In the first order in $\al_{s}$ one has (cf. eq. (\ref{b4}):
\begin{equation}
\lan\Uh(x\p)\Uhd(y\p)\ran_{B\O}=-ig^{2}t^{a}U\O(x\p)\otimes
t^{b}U^{\dagger \Omega }(y\p)\int _{0}^{\iy}\! du\!\int _{-\iy}^{0}\! dv
\LBB up^{(0)}_A+x\p\Big|{\cal O}\O\cici\Big| vp\an+y\p\RBB_{ab}
\label{b19}
\end{equation}
(we shall calculate only the contribution $\sim U\O$ which comes from the
region $x\sr>0, y\sr<0$ - the term $\sim \UdO$ coming from $x\sr>0, y\sr<0$ is
similar, cf. eq. (\ref{b13})). Technically it is convenient to find at first
the derivative of the integral of gluon propagator in r.h.s. of eq. (\ref{b19})
with respect to $x\p$. Using formula (\ref{b10}) for the gluon propagator
$(x|{\cal O}|y)$ we obtain:
\begin{eqnarray}
\lefteqn{-ig^{2}\int \! du\!\int \! dv \LBB up^{(0)}_A+x\p\Big|
p_{i}{\cal O}\O\cici\Big| vp\an+y\p\RBB_{ab}}~=\\
\label{b20}
&&{g^{2}\over 16\pi ^{4}}\int \! dz\p\!\int _{0}^{\iy} \frac {du}{u}dv\int \!
dz\ci\frac {(x\p-z\p)_{i}[\partial ^{2}\p U\O(z\p)]_{ab}}{[u(u\zeta
s_{0}-2z\ci)-(x-z)\p^{2}-\ie][v(v\zeta s_{0}+2z\ci)-(x-z)\p^{2}-\ie]}\nonumber
\end{eqnarray}
The integration over $z\ci$ can be performed by taking the residue and the
result is
\begin{equation}
-i{g^{2}\over 16\pi ^{3}}\int \! dz\p\int _{0}^{\iy} \frac {du}{u}
dv\frac {(x\p-z\p)_{i}[\partial ^{2}\p
U(z\p)]_{ab}}{[(x-z)\p^{2}v+(y-z)\p^{2}v-uv(u+v)\zeta s_{0}+\ie]}
\label{b21}
\end{equation}
This integral diverges logarithmically when $u\rightarrow O$ - in other words,
when the emission of quantum gluon occurs in the vicinity of the shock wave.
(note that if we had done integration by parts, the divergence would be at
$v\rightarrow 0$ so there is no asymmetry between $u$ and $v$). The size of the
shock wave $z\sr\sim m^{-1}/\la$ (where $1/m$ is the characteristic transverse
size) serves as the lower cutoff for this integration and we obtain
\begin{equation}
-i{g^{2}\over 16\pi ^{3}}\ln\la\int \! dz\p\int _{0}^{1} \frac {d\al}{\al}
\frac {(x\p-z\p)_{i}[\partial ^{2}\p
U(z\p)]_{ab}}{[(x-z)\p^{2}\bar{\al}+(y-z)\p^{2}\al]}=
-{g^{2}\over 16\pi ^{3}}\ln\la\LBB x\p\Big|{p_{i}\over p^{2}}
(\partial \p^{2}U){1\over
p^{2}}\Big| y\p\RBB_{ab}
\label{b22}
\end{equation}
(recall that $\bar{\al}\equiv 1-\al$). Thus, we have the contribution of the
diagram in Fig. \ref{fig:pint1} in leading logarithmic approximation in the
following form:
\begin{equation}
\lan\Uh(x\p)\Uhd(y\p)\ran_{B\O}=-({g^{2}\over 2\pi }\ln\la)
t^{a}U\O(x\p)\otimes U^{\dagger \Omega }(y\p)
\LBB x\p\Big|{1\over p^{2}}(\partial \p^{2}U){1\over p^{2}}
\Big| y\p\RBB_{ab}
\label{b23}
\end{equation}
which agrees with the first term in eq. (\ref{b13}) (recall that
$\zeta {\partial \over\partial \zeta }=-\half\la{\partial \over \partial
\la}$).

\section{Gluon propagator in the axial gauge.}
\label{app:GlueProp}

Our aim here is to derive the expression for the gluon propagator in the
external field in the
axial gauge. The propagator of the "quantum" gauge field $A^{q}$ in the
external
"classical" field $A^{cl}$ in the axial gauge $e_{\mu }A_{\mu }=0$ can be
represented as the following functional integral:
\begin{equation}
G_{\mu \nu }^{ab}(x,y)=\lim_{w\rightarrow 0}N^{-1}\int DA
A^{qa}_{\mu }(x)A^{qb}_{\nu }(y)e^{i\int dz\Tr\{ A^{q}_{\al}(z)
(D^{2}g^{\al\be}-D^{\al}D^{\be}-2igF_{cl}^{\al\be}-{1\over
w}e^{\al}e^{\be})A^{q}_{\be}(z)\}}
\label{c1}
\end{equation}
where $D_{\mu }=\partial _{\mu }-igA^{cl}_{\mu }$. Hereafter we shall omit the
label "cl" from the external field. This propagator can be formally written
down as
\begin{equation}
iG^{ab}_{\mu \nu }(x,y)=\LBB x\Big|\frac {1}{\Box^{\mu \nu }-
{\cal P}^{\mu }{\cal
P}^{\nu }+\frac {1}{w}e^{\mu }e^{\nu }}\Big| y\RBB^{ab}
\label{c2}
\end{equation}
where $\Box^{\mu \nu }={\cal P}^{2}g^{\mu \nu }+2igF^{\mu \nu }$. It is easy to
check
that the operator in r.h.s. of eq.(\ref{c2}) satisfies the recursion formula
\begin{eqnarray}
\lefteqn{\frac {1}{\Box^{\mu \nu }-{\cal P}^{\mu }{\cal P}^{\nu }+
\frac {1}{w}e^{\mu
}e^{\nu }}}\nonumber\\
\label{c3}
&=&(\delta _{\mu }^{\xi }
-{\cal P}_{\mu }\frac {e^{\xi }}{{\cal P}e})\frac {1}{\Box^{\xi \eta }}(\delta
_{\nu }^{\eta }-
\frac {e^{\eta }}{{\cal P}e}{\cal P}_{\nu })+{\cal P}_{\mu }
\frac {w}{({\cal P}e)^{2}}
{\cal P}_{\nu }-\\
& & \hspace{-1in}
\frac {1}{\Box^{\mu \al}-{\cal P}^{\mu }{\cal P}^{\al} +\frac {1}{w}e^{\mu
}e^{\al}}(D_{\la}F^{\la\al}\frac {e^{\xi }}{{\cal P}e}
-{\cal P}^{\al}\frac {1}{{\cal P}^{2}}D_{\la}F^{\la\xi })\frac {1}{\Box^{\xi
\eta }}(\delta _{\nu }^{\eta }-
\frac {e^{\eta }}{{\cal P}e}{\cal P}_{\nu })\nonumber
\end{eqnarray}
which gives the propagator as an expansion in powers of the operator
$D_{\la}F^{a}_{\la\al}=-g\bar{\psi }t^{a}\gamma _{\al}\psi $. We shall see
below that
in the leading logarithmic approximation we need the terms not higher than the
first nontrivial order in this operator. With this accuracy
 \begin{eqnarray}
\lefteqn{\frac {1}{\Box^{\mu \nu }-{\cal P}^{\mu }{\cal P}^{\nu }+
\frac {1}{w}e^{\mu
}e^{\nu }}}\nonumber\\
&=&(\delta _{\mu }^{\xi }
-{\cal P}_{\mu }\frac {e^{\xi }}{{\cal P}e})\frac {1}{\Box^{\xi \eta }}(\delta
_{\nu }^{\eta }-
\frac {e^{\eta }}{{\cal P}e}{\cal P}_{\nu })+{\cal P}_{\mu }
\frac {w}{({\cal P}e)^{2}}
{\cal P}_{\nu }-\nonumber\\
&&(\delta _{\mu }^{\xi }
-{\cal P}_{\mu }\frac {e^{\xi }}{{\cal P}e})\frac {1}{\Box^{\xi \eta
}}(D_{\la}F^{\la\eta }
\frac {e^{\rho }}{{\cal P}e}
+\frac {e^{\eta }}{{\cal P}e}D_{\la}F^{\la\rho }-
\frac {e^{\eta }}{{\cal P}e}{\cal P}^{\be}D_{\al}F^{\al\be}
\frac {e^{\rho }}{{\cal P}e})\frac {1}{\Box^{\rho \si}}(\delta _{\nu }^{\si}-
\frac {e^{\si}}{{\cal P}e}{\cal P}_{\nu })
\label{c4}
\end{eqnarray}
Taking now $w\rightarrow 0$ we obtain the propagator in external field in axial
gauge in the form:
 \begin{eqnarray}
\lefteqn{iG_{\mu \nu }^{ab}(x,y)}\nonumber\\
&=&(\delta _{\mu }^{\xi }
-{\cal P}_{\mu }\frac {e^{\xi }}{{\cal P}e})\frac {1}{\Box^{\xi \eta }}(\delta
_{\nu }^{\eta }-
\frac {e^{\eta }}{{\cal P}e}{\cal P}_{\nu })\nonumber\\
&-&(\delta _{\mu }^{\xi }
-{\cal P}_{\mu }\frac {e^{\xi }}{{\cal P}e})\frac {1}{\Box^{\xi \eta }}
(D_{\la}F^{\la\eta }
\frac {e^{\rho }}{{\cal P}e}
+\frac {e^{\eta }}{{\cal P}e}D_{\la}F^{\la\rho }-
\frac {e^{\eta }}{{\cal P}e}{\cal P}^{\be}D_{\al}F^{\al\be}
\frac {e^{\rho }}{{\cal P}e})\frac {1}{\Box^{\rho \si}}(\delta _{\nu }^{\si}-
\frac {e^{\si}}{{\cal P}e}{\cal P}_{\nu })+...
\label{c5}
\end{eqnarray}
where the dots stand for the terms of second (and higher) order in
$D^{\la}F_{\la\rho }$. It can be demonstrated that for our purposes
a first few terms of the expansion of operators ${1\over\Box}$ in
powers of $F_{\xi\eta}$ are enough, namely
\begin{eqnarray}
\lefteqn{iG_{\mu \nu }^{ab}(x,y)}\nonumber\\
&=&(\delta _{\mu }^{\xi }
-{\cal P}_{\mu }\frac {e^{\xi }}{{\cal P}e})\nonumber\\
&&\left[\frac {\delta^{\xi\eta}}{{\cal P}^2}-2\frac {1}{{\cal P^2}}
F^{\xi\eta}\frac {1}{{\cal P^2}}+4\frac {1}{{\cal P^2}}F^{\xi\eta}
\frac {1}{{\cal P}^2}F^{\xi\eta}\frac {1}{{\cal P}^2} -\frac {1}{{\cal P}^2}
(D_{\la}F^{\la\xi }
\frac {e^{\eta }}{{\cal P}e}
+\frac {e^{\xi }}{{\cal P}e}D_{\la}F^{\la\eta }-
\frac {e^{\xi }}{{\cal P}e}{\cal P}^{\be}D_{\al}F^{\al\be}
\frac {e^{\eta }}{{\cal P}e})\frac {1}{{\cal P}^2}\right]\nonumber\\
&&(\delta _{\nu }^{\eta}-
\frac {e^{\eta}}{{\cal P}e}{\cal P}_{\nu })+...
\label{c6}
\end{eqnarray}

\end{document}